\documentclass[3p,authoryear]{elsarticle}

\usepackage{float}
\usepackage{graphicx,ifthen,color,algorithm,palatino}
\usepackage{amsmath,amsthm,amssymb,amsfonts,verbatim}
\usepackage{mathrsfs,accents,setspace}
\usepackage{subfigure,caption}
\usepackage{multirow,makecell}
\usepackage{soul}
\usepackage{natbib}
\newboolean{ShowFigures}
\newboolean{UnBlinded}
\newboolean{DoubleSpaced}
\setboolean{ShowFigures}{true}
\setboolean{UnBlinded}{true}
\setboolean{DoubleSpaced}{false}

\def\ba{\boldsymbol{a}}
\def\bb{\boldsymbol{b}}
\def\bc{\boldsymbol{c}}

\def\bk{\boldsymbol{k}}

\def\bm{\boldsymbol{m}}

\def\br{\boldsymbol{r}}
\def\bs{\boldsymbol{s}}
\def\bt{\boldsymbol{t}}
\def\bu{\boldsymbol{u}}
\def\bv{\boldsymbol{v}}
\def\bw{\boldsymbol{w}}
\def\bx{\boldsymbol{x}}
\def\by{\boldsymbol{y}}
\def\bz{\boldsymbol{z}}
\def\bA{\boldsymbol{A}}

\def\bC{\boldsymbol{C}}

\def\bI{\boldsymbol{I}}

\def\bK{\boldsymbol{K}}
\def\bL{\boldsymbol{L}}
\def\bM{\boldsymbol{M}}
\def\bN{\boldsymbol{N}}

\def\bQ{\boldsymbol{Q}}
\def\bR{\boldsymbol{R}}

\def\bT{\boldsymbol{T}}

\def\bV{\boldsymbol{V}}

\def\bX{\boldsymbol{X}}
\def\bY{\boldsymbol{Y}}

\def\bbeta{\boldsymbol{\beta}}

\def\bvarepsilon{\boldsymbol{\varepsilon}}
\def\bzeta{\boldsymbol{\zeta}}
\def\bdeta{\boldsymbol{\eta}}
\def\btheta{\boldsymbol{\theta}}

\def\blambda{\boldsymbol{\lambda}}
\def\bmu{\boldsymbol{\mu}}

\def\btau{\boldsymbol{\tau}}

\def\bomega{\boldsymbol{\omega}}

\def\bLambda{\boldsymbol{\Lambda}}

\def\bSigma{\boldsymbol{\Sigma}}

\def\bOmega{\boldsymbol{\Omega}}
\def\bzero{\boldsymbol{0}}
\def\bone{\boldsymbol{1}}
\def\vecof{\mathop{\mbox{\rm vec}}}

\def\dv{d_{\mbox{\tiny $\bv$}}}
\def\dw{d_{\mbox{\tiny $\bw$}}}
\def\dH{d^{\mbox{\tiny H}}}
\def\dV{d^{\mbox{\tiny V}}}
\def\LH{\bL^{\mbox{\tiny H}}}
\def\LV{\bL^{\mbox{\tiny V}}}
\def\LzeroH{\bL_0^{\mbox{\tiny H}}}
\def\LzeroV{\bL_0^{\mbox{\tiny V}}}
\def\LoneD{\bL_{\mbox{\tiny 1D}}}
\def\LtwoD{\bL_{\mbox{\tiny 2D}}}
\def\dM{d_{\mbox{\tiny $\bM$}}}
\def\rH{r^{\mbox{\tiny H}}}
\def\rV{r^{\mbox{\tiny V}}}

\def\brH{\br^{\mbox{\tiny H}}}
\def\brV{\br^{\mbox{\tiny V}}}

\def\szeroH{\bs_0^{\mbox{\tiny H}}}

\def\sH{\bs^{\mbox{\tiny H}}}
\def\sV{\bs^{\mbox{\tiny V}}}
\def\tH{\bt^{\mbox{\tiny H}}}
\def\tV{\bt^{\mbox{\tiny V}}}

\def\bI{\boldsymbol{I}}

\def\diagonal{\mathop{\mbox{\rm diagonal}}}
\def\diag{\mathop{\mbox{\rm diag}}}

\def\simind{\stackrel{{\tiny \mbox{ind.}}}{\sim}}
\newtheorem{lemma}{\textbf{Lemma}}
\newtheorem{result}{\textbf{Result}}
\newtheorem{definition}{\textbf{Definition}}

\def\biggerbdeta{\mbox{\large $\bdeta$}}
\def\GVMP{G_{\tiny{\mbox{VMP}}}}
\def\etaSUBpyxsigsqepsilonTOx
{\biggerbdeta_{\mbox{\footnotesize$p(\by|\,\bx,\sigma^{2}_{\varepsilon})\rightarrow\bx$}}}
\def\etaSUBpyxsigsqepsFROMTOx
{\biggerbdeta_{\mbox{\footnotesize$p(\by|\,\bx,\sigma^{2}_{\varepsilon})\leftrightarrow\bx$}}}
\def\etaSUBpyxsigsqepsilonTOsigsqepsilon
{\biggerbdeta_{\mbox{\footnotesize$p(\by|\,\bx,\sigma^{2}_{\varepsilon})\rightarrow\sigma^2_{\varepsilon}$}}}
\def\etaSUBpyxsigsqepsFROMTOsigsqeps
{\biggerbdeta_{\mbox{\footnotesize$p(\by|\,\bx,\sigma^{2}_{\varepsilon})\leftrightarrow\sigma^2_{\varepsilon}$}}}
\def\etaSUBpxbsigsqxTOx
{\biggerbdeta_{\mbox{\footnotesize$p(\bx|\,\bb,\sigma^{2}_{x})\rightarrow\bx$}}}
\def\etaSUBxTOpxbsigsqx
{\biggerbdeta_{\mbox{\footnotesize$\bx\rightarrow p(\bx|\,\bb,\sigma^{2}_{x})$}}}
\def\etaSUBpxbsigsqxFROMTOx
{\biggerbdeta_{\mbox{\footnotesize$p(\bx|\,\bb,\sigma^{2}_{x})\leftrightarrow\bx$}}}
\def\etaSUBpxbsigsqxTOsigsqx
{\biggerbdeta_{\mbox{\footnotesize$p(\bx|\,\bb,\sigma^{2}_{x})\rightarrow\sigma^{2}_{x}$}}}

\def\etaSUBsigsqxTOpxbsigsqx
{\biggerbdeta_{\mbox{\footnotesize$\sigma^{2}_{x}\rightarrow p(\bx|\,\bb,\sigma^{2}_{x})$}}}
\def\etaSUBpxbsigsqxFROMTOsigsqx
{\biggerbdeta_{\mbox{\footnotesize$p(\bx|\,\bb,\sigma^{2}_{x})\leftrightarrow\sigma^{2}_{x}$}}}
\def\etaSUBpaxTOax
{\biggerbdeta_{\mbox{\footnotesize$p(a_x)\rightarrow a_x$}}}
\def\etaSUBpaepsilonTOaepsilon
{\biggerbdeta_{\mbox{\footnotesize$p(a_\varepsilon)\rightarrow a_\varepsilon$}}}

\def\etaSUBpsigsqxaxTOsigsqx
{\biggerbdeta_{\mbox{\footnotesize$p(\sigma^2_x|a_x)\rightarrow \sigma^2_x$}}}
\def\etaSUBpsigsqxaxTOax
{\biggerbdeta_{\mbox{\footnotesize$p(\sigma^2_x|a_x)\rightarrow a_x$}}}
\def\etaSUBpxbsigsqxTOx
{\biggerbdeta_{\mbox{\footnotesize$p(\bx|\bb,\sigma^2_x)\rightarrow \bx$}}}
\def\etaSUBpyxsigsqepsilonTOx
{\biggerbdeta_{\mbox{\footnotesize$p(\by|\bx,\sigma^2_\varepsilon)\rightarrow \bx$}}}
\def\etaSUBpyxsigsqepsilonTOsigsqepsilon
{\biggerbdeta_{\mbox{\footnotesize$p(\by|\bx,\sigma^2_\varepsilon)\rightarrow \sigma^2_\varepsilon$}}}
\def\etaSUBaxTOpsigsqxax
{\biggerbdeta_{\mbox{\footnotesize$a_x\rightarrow p(\sigma^2_x|a_x)$}}}

\def\etaSUBaepsilonTOpsigsqepsilonaepsilon
{\biggerbdeta_{\mbox{\footnotesize$a_\varepsilon\rightarrow p(\sigma^2_\varepsilon|a_\varepsilon)$}}}
\def\etaSUBxTOpyxsigsqepsilon
{\biggerbdeta_{\mbox{\footnotesize$\bx\rightarrow p(\by|\bx,\sigma^2_\varepsilon)$}}}
\def\etaSUBsigsqepsilonTOpyxsigsqepsilon
{\biggerbdeta_{\mbox{\footnotesize$\sigma^2_\varepsilon\rightarrow p(\by|\bx,\sigma^2_\varepsilon)$}}}
\def\etaSUBsigsqxTOpsigsqxax
{\biggerbdeta_{\mbox{\footnotesize$\sigma^{2}_{x}\rightarrow p(\sigma^{2}_{x}|a_x)$}}}
\def\etaSUBaxTOpax
{\biggerbdeta_{\mbox{\footnotesize$a_x\rightarrow p(a_x)$}}}

\def\etaSUBsigsqepsilonTOpsigsqepsilonaepsilon
{\biggerbdeta_{\mbox{\footnotesize$\sigma^2_\varepsilon\rightarrow p(\sigma^2_\varepsilon|a_\varepsilon)$}}}
\def\etaSUBpsigsqepsilonaepsilonTOsigsqepsilon
{\biggerbdeta_{\mbox{\footnotesize$p(\sigma^2_\varepsilon|a_\varepsilon)\rightarrow\sigma^2_\varepsilon$}}} 
\def\etaSUBpsigsqepsilonaepsilonTOaepsilon
{\biggerbdeta_{\mbox{\footnotesize$p(\sigma^2_\varepsilon|a_\varepsilon)\rightarrow a_\varepsilon$}}}
\def\etaSUBaepsilonTOpaepsilon
{\biggerbdeta_{\mbox{\footnotesize$a_\varepsilon\rightarrow p(a_\varepsilon)$}}}
\def\etaSUBqx
{\biggerbdeta_{\mbox{\footnotesize$q(\bx)$}}}
\def\etaSUBqsigsqx
{\biggerbdeta_{\mbox{\footnotesize$q(\sigma^2_x)$}}}
\def\etaSUBqsigsqepsilon
{\biggerbdeta_{\mbox{\footnotesize$q(\sigma^2_\epsilon)$}}}
\def\etaSUBqlambda
{\biggerbdeta_{\mbox{\footnotesize$q(\lambda)$}}}
\def\mSUBpxbsigsqxTOx
{m_{\mbox{\footnotesize$p(\bx|\,\bb,\sigma^{2}_{x})\rightarrow\bx$}}\left(\bx\right)}
\def\mSUBxTOpxbsigsqx
{m_{\mbox{\footnotesize$\bx\rightarrow p(\bx|\,\bb,\sigma^{2}_{x})$}}\left(\bx\right)}
\def\mSUBpxbsigsqxTOsigsqx
{m_{\mbox{\footnotesize$p(\bx|\,\bb,\sigma^{2}_{x})\rightarrow\sigma^2_x$}}\left(\sigma^2_x\right)}
\def\mSUBsigsqxTOpxbsigsqx
{m_{\mbox{\footnotesize$\sigma^2_x\rightarrow p(\bx|\,\bb,\sigma^{2}_{x})$}}\left(\sigma^2_x\right)}
\def\mSUBpxbsigsqxTOb
{m_{\mbox{\footnotesize$p(\bx|\,\bb,\sigma^{2}_{x})\rightarrow\bb$}}\left(\bb\right)}
\def\mSUBbTOpxbsigsqx
{m_{\mbox{\footnotesize$\bb\rightarrow p(\bx|\,\bb,\sigma^{2}_{x})$}}\left(\bb\right)}
\def\mSUBpbTOb
{m_{\mbox{\footnotesize$p(\bb)\rightarrow\bb$}}\left(\bb\right)}
\def\GSUBpaxTOax
{G_{\mbox{\footnotesize$p(a_x)\rightarrow a_x$}}}
\def\GSUBpaepsilonTOaepsilon
{G_{\mbox{\footnotesize$p(a_\varepsilon)\rightarrow a_\varepsilon$}}}
\def\GSUBsigsqxTOpsigsqxax
{G_{\mbox{\footnotesize$\sigma^{2}_{x}\rightarrow p(\sigma^{2}_{x}|a_x)$}}}
\def\GSUBpsigsqxaxTOsigsqx
{G_{\mbox{\footnotesize$p(\sigma^{2}_{x}|a_x)\rightarrow\sigma^{2}_{x}$}}}
\def\GSUBsigsqepsilonTOpsigsqepsilonaepsilon
{G_{\mbox{\footnotesize$\sigma^{2}_{\varepsilon}\rightarrow p(\sigma^{2}_{\varepsilon}|a_\varepsilon)$}}}
\def\GSUBpsigsqepsilonaepsilonTOsigsqepsilon
{G_{\mbox{\footnotesize$p(\sigma^{2}_{\varepsilon}|a_\varepsilon)\rightarrow\sigma^{2}_{\varepsilon}$}}}
\def\GSUBaxTOpsigsqxax
{G_{\mbox{\footnotesize$a_{x}\rightarrow p(\sigma^{2}_{x}|a_x)$}}}
\def\GSUBpsigsqxaxTOax
{G_{\mbox{\footnotesize$p(\sigma^{2}_{x}|a_x)\rightarrow a_{x}$}}}
\def\GSUBaepsilonTOpsigsqepsilonaepsilon
{G_{\mbox{\footnotesize$a_{\varepsilon}\rightarrow p(\sigma^{2}_{\varepsilon}|a_\varepsilon)$}}}
\def\GSUBpsigsqepsilonaepsilonTOaepsilon
{G_{\mbox{\footnotesize$p(\sigma^{2}_{\varepsilon}|a_\varepsilon)\rightarrow a_{\varepsilon}$}}}
\def\tridiag{\mathop{\mbox{\rm tridiag}}}
\def\sparsetridiag{\mathop{\mbox{\rm sparsetridiag}}}

\def\muqx
{\bmu_{q\left(\bx\right)}}
\def\Sigmaqx
{\bSigma_{q\left(\bx\right)}}
\def\muqrecipsigsqx
{\mu_{q\left(1/\sigma^2_x\right)}}
\def\kappaqsigsqx
{\kappa_{q\left(\sigma^2_x\right)}}
\def\lambdaqsigsqx
{\lambda_{q\left(\sigma^2_x\right)}}
\def\muqrecipsigsqepsilon
{\mu_{q\left(1/\sigma^2_\varepsilon\right)}}
\def\kappaqsigsqepsilon
{\kappa_{q\left(\sigma^2_\varepsilon\right)}}
\def\lambdaqsigsqepsilon
{\lambda_{q\left(\sigma^2_\varepsilon\right)}}
\def\kappaqax
{\kappa_{q\left(a_x\right)}}
\def\lambdaqax
{\lambda_{q\left(a_x\right)}}
\def\kappaqaepsilon
{\kappa_{q\left(a_\varepsilon\right)}}
\def\lambdaqaepsilon
{\lambda_{q\left(a_\varepsilon\right)}}
\def\xDelta
{\bx_\Delta}

\def\tr{\mbox{tr}}
\def\stackdum{\mathop{\mbox{\rm stack}}}
\def\stack#1{\stackdum_{#1}}

\makeatother

\journal{Computational Statistics and Data Analysis}

\begin{document}

\begin{frontmatter}



\title{A variational inference framework for inverse problems}

\author[1]{Luca Maestrini}
\ead{luca.maestrini@anu.edu.au}
\author[2]{Robert G. Aykroyd}
\author[3]{Matt P. Wand}

\affiliation[1]{organization={Research School of Finance, Actuarial Studies and Statistics, Australian National University},
            addressline={Building 26C Kingsley Street}, 
            city={Canberra},
            postcode={Acton 2601}, 
            state={Australian Capital Territory},
            country={Australia}}

\affiliation[2]{organization={Department of Statistics, School of Mathematics, University of Leeds},
            city={Leeds},
            postcode={LS2 9JT}, 
            country={United Kingdom}}
            
\affiliation[3]{organization={School of Mathematical and Physical Sciences, University of Technology Sydney},
            addressline={P.O. Box 123 Broadway}, 
            city={Sydney},
            postcode={Ultimo 2007}, 
            state={New South Wales},
            country={Australia}}

\begin{abstract}
A framework is presented for fitting inverse problem models via variational Bayes approximations. This methodology guarantees flexibility to statistical model specification for a broad range of applications, good accuracy and reduced model fitting times. 
The message passing and factor graph fragment approach to variational Bayes that is also described facilitates streamlined implementation of approximate inference algorithms and allows for supple inclusion of numerous response distributions and penalizations into the inverse problem model. 
Models for one- and two-dimensional response variables are examined and an infrastructure is laid down where efficient algorithm updates based on nullifying weak interactions between variables can also be derived for inverse problems in higher dimensions. 
An image processing application and a simulation exercise motivated by biomedical problems reveal the computational advantage offered by efficient implementation of variational Bayes over Markov chain Monte Carlo. 
\end{abstract}


\end{frontmatter}

\section{Introduction}\label{sec:intro}

\noindent
Inverse problems are essentially statistical regression problems where a response depending on a number of parameters is measured and the goal is to interpret the parameter estimates, rather then predict the outcome. Stable fitting of inverse problems is crucial but this is generally hindered by a large number of parameters and the presence of predictors which are highly correlated. 

Let $\by$ denote a vector of data and suppose this data is related to a vector of unknown parameters $\bx$ to estimate by a linear regression problem $E(\by) = \bK\bx$, where $\bK$ is given, or a nonlinear one such that $E(\by) = g(\bx)$, where $g$ is a known function. From a Bayesian perspective, model fitting can be performed by placing a prior on $\bx$ to then find the maximum a posteriori estimate $\hat{\bx}=\mbox{argmax}_{\bx} p(\bx\vert \by)$. This appears straightforward in principle, however, in typical applications, inverse problems may be ill-posed in a sense that either the solution does not exist, is not unique or does not depend smoothly on the data, as small noise variations can produce significantly different estimates \citep{hadamard1902problemes}. 
A remedy is to introduce a penalization in the model formulation and use Bayesian hierarchical models, but these can be slow to fit via standard Markov chain Monte Carlo methods. To overcome this issue, we propose and study variational Bayes methods. The direct and message passing approaches to variational Bayes we examine facilitate inverse problem fitting in Bayesian settings with reduced computational times.

The use of variational Bayesian methods for inverse problems has been shown in the literature concerning neural source reconstruction, including \cite{sato2004hierarchical}, \cite{kiebel2008variational}, \cite{wipf2009unified} and \cite{nathoo2014variational}.
Approximate inference methods motivated by a broader class of inverse problem applications are in their infancy. A small, growing, literature includes \cite{mcgrory2009variational}, \cite{gehre2014expectation} and \cite{guha2015variational}. \cite{arridge2018variational} and \cite{zhang2019expectation} respectively study usage of Gaussian variational approximations and expectation propagation to fit inverse problems models with Poisson responses. \cite{tonolini2020variational} propose a framework to train variational inference for imaging inverse problems exploiting existing image data. \cite{agrawal2022variational} study variational inference for inverse problems with gamma hyperpriors. \cite{povala2022variational} present a stochastic variational Bayes approach based on sparse precision matrices. 

The state-of-the-art in approximate inference for inverse problems is to derive and code algorithm updates from scratch each time a model is modified. The message passing on factor graph fragment approach to variational Bayes we suggest in this work overcomes the issue. \cite{wand2017fast} has spearheaded adoption of this approach to fast approximation inference in regression-type models via variational message passing (VMP). In the same spirit, we lay down similar infrastructure for inverse problems and propose VMP as an alternative to the more common mean field variational Bayes (MFVB). We show how to perform approximate inference by combining algorithms for single factor graph components, or \textit{fragments}, that arise from inverse problem models. The resultant factor graph fragments facilitate streamlined implementation of fast approximate algorithms and form the basis for software development for use in applications. The factor graph fragment paradigm allows for easy incorporation of different penalization structures in the model or changes to the distribution of the outcome variable. In fact, VMP on factor graph fragments is such that the corresponding algorithms only need to be derived once for a particular fragment and can be used for any arbitrarily complex model including such a fragment. Hence dramatically reducing set-up overheads as well as providing fast implementation.

Motivated by a real biomedical problem, we identify a base inverse problem model and describe how to efficiently perform MFVB and VMP. The application we show concerns medical positron emission tomography imaging where the raw data is processed for image enhancement. The data were collected to illustrate a small animal imaging system which can be used in biotechnology and pre-clinical medical research to help detect tumors or organ dysfunctions.
An application to two-dimensional deconvolution problems motivated by an archaeological exploration is also embarked upon the base framework by varying the response and penalization distributional assumptions. This is illustrated in the supplementary material.

\subsection{Overview of the article}

Section \ref{sec:InvProbModel} defines a reference inverse problem model for illustrating the methodology in use and our computational developments. The variational approximation engine for inverse problem fitting and inference is introduced in Section \ref{sec:varInf}. Section \ref{sec:stremVarInf} examines strategies to streamline variational inference algorithms. An application to real biomedical data is treated in Section \ref{sec:biomed}. The same section reports results from a study involving simulations which resemble the analyzed biomedical dataset. The supplementary material provides an illustration concerning archaeological data performed via VMP, where the Normal response and Laplace penalization of the base model are replaced by a Skew Normal distribution for the outcome variable and a Horseshoe penalization. Concluding remarks and extensions are discussed in Section \ref{sec:discussion}.

Before setting up our reference linear inverse problem model and presenting variational algorithms for approximate model fitting we introduce some useful notation.

\subsection{Useful notation}\label{sec:notation}

For a matrix $\bA$ of size $d_1\times d_2$, $\vecof(\bA)$ is the $d_1d_2\times 1$ vector obtained by stacking the columns of $\bA$ underneath each other in order from left to right. If $\ba$ is a $(d_1d_2)\times 1$ vector then $\vecof^{-1}_{d_1\times d_2}(\ba)$ is the $d_1\times d_2$ matrix such that $\vecof\big\{\vecof^{-1}_{d_1\times d_2}(\ba)\big\}=\ba$; when the $\vecof$ operator inverse produces a square matrix the subscript is omitted. Vectors of $d$ zeros or ones are respectively denoted by $\bzero_d$ and $\bone_d$.

\section{Base inverse problem model\label{sec:InvProbModel}}

We consider linear inverse problems having the following formulation:
\begin{equation}
\by=\bK\bx+\bvarepsilon,\quad\bvarepsilon\sim N(\bzero,\sigma^2\bI),
\label{eq:baseModel}
\end{equation}
where $\by$ is an $m\times 1$ vector of observed data, $\bK$ is a matrix acting as a linear operator of size $m\times m$, $\bx$ is a $m\times 1$ vector of unknown parameters and $\bvarepsilon$ is a Normal error vector of length $m$. For ease of illustration, we focus on the case where the vectors $\by$ and $\bx$ have equal length $m$ and therefore $\bK$ is a square matrix. Nevertheless, the methodology presented here can be adapted to the situation in which $\by$ has length $n$ different from and typically smaller than the length $m$ of $\bx$. Motivated by our biomedical application, we focus on a particular type of forward problem where $\bK$ is a kernel matrix. Another formulation of $\bK$ is discussed in the application to archaeological data treated in the supplementary material.

We assume that the the vector of observations $\by$ has a one-to-one correspondence with the vector of parameters $\bx=(x_1,\ldots,x_m)$ to be estimated. For simplicity, here we only model first nearest neighbor interactions, or differences, between elements of $\bx$. If these elements are identified by a system of coordinates in two dimensions, then first nearest neighbor interactions are the differences between one parameter and those in adjacent locations on the horizontal and vertical coordinates of the parameter.

\begin{figure}[!ht]
	\centering
	{\includegraphics[width=0.3\textwidth]{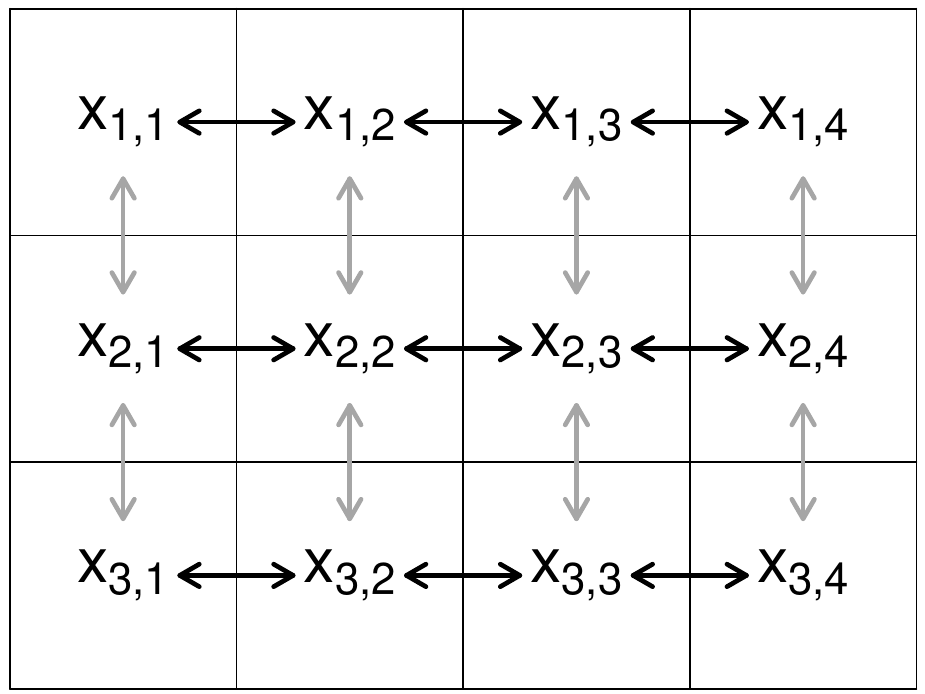}}
	\caption{\it Example of a $3\times 4$ grid of pixels with 9 horizontal differences marked by black arrows and 8 vertical differences in grey. The total number of first order nearest neighbor differences is 17.}
	\label{fig:pixelGrid} 
\end{figure}

Suppose the aim is to study a linear inverse problem in a Bayesian framework according to the model
\begin{equation}
\begin{array}{c}
y_i\vert \bx,\sigma^2_{\varepsilon}\simind N\left(\left(\bK\bx\right)_i,\sigma^2_{\varepsilon}\right),\quad i=1,\ldots,m,\\[1ex]
(\xDelta)_j\vert b_j,\sigma^2_x\simind N\left(0,\sigma^2_x/b_j\right),\quad b_j\simind\mbox{Inverse-$\chi^2$}\left(2,1\right),\quad j=1,\ldots,d,\\[1ex]
\sigma^2_\varepsilon\vert a_\varepsilon\sim \mbox{Inverse-$\chi^2$}\left(1,1/a_\varepsilon\right),\quad a_\varepsilon\sim\mbox{Inverse-$\chi^2$}\left(1,1/A^2_\varepsilon\right),\\[1ex]
\sigma^2_x\vert a_x\sim \mbox{Inverse-$\chi^2$}\left(1,1/a_x\right),\quad a_x\sim\mbox{Inverse-$\chi^2$}\left(1,1/A^2_x\right),
\end{array}
\label{eq:linInvProbModel}
\end{equation}
where $\bK$ is a matrix of size $m\times m$, $\xDelta$ is a vector of $d$ differences between pairs of elements in $\bx$ and $A_\varepsilon,A_x>0$ are user-specified hyperparameters. The auxiliary variables $a_\varepsilon$ and $a_x$ generate Half-Cauchy$\left(A_\varepsilon\right)$ and Half-Cauchy$\left(A_x\right)$ priors on the scale parameters $\sigma_\varepsilon$ and $\sigma_x$, respectively. Specifically, the density function of a random variable $\sigma>0$ having a Half-Cauchy$\left(A\right)$ distribution is $p(\sigma)=2/[A\pi\{1+(\sigma/A)^2\}]$, with $A>0$.
For problems that only contemplate first nearest neighbor differences, the scalar $d$ coincides with the number of unique up to sign differences between pairs of elements of $\bx$ coming from adjacent locations. In the one-dimensional case, $\bx$ can be interpreted as a vector matching $m$ spatial locations on a line and the number of differences between adjacent locations will be $d=m-1$. Model \eqref{eq:linInvProbModel} also encompasses higher-dimensional problems. In bidimensional settings, $\bx$ can be conveniently expressed as the vectorization of a grid, or matrix, of pixels $\bX$ by setting $\bx=\vecof\left(\bX\right)$. If $\bX$ has size $m_1\times m_2$, the first nearest neighbor differences are $d=m_1\left(m_2-1\right) + m_2\left(m_1-1\right)$. In the simple example of Figure \ref{fig:pixelGrid} where $\bX$ is of size $3\times 4$, the number of horizontal and vertical differences are respectively 9 and 8, giving $d=17$ differences in total. In a similar vein, the model can be applied to three-dimensional problems by letting $\bx$ be the vectorization of voxel-type data. 

The distributional assumption on $\xDelta$ in model \eqref{eq:linInvProbModel} can be conveniently re-expressed as
\begin{equation}
\bL\bx\vert\bb,\sigma^2_x\sim N\left(\bzero_d,\sigma^2_x\diag\left(\bb\right)^{-1}\right),
\label{eq:xFactor}
\end{equation}
where $\bL$ is some $d\times m$ \textit{contrast matrix} such that $\xDelta\equiv\bL\bx$ and $\bb=(b_1,\ldots,b_d)$. For instance in one-dimensional problems contemplating first nearest neighbor interactions, the contrast matrix can be defined as
\begin{equation}
	\LoneD\equiv\left[\begin{array}{ccccc}
		-1 & 1 & 0 & \cdots & 0\\
		0 & -1 & 1 & \ddots & \vdots\\
		\vdots & \ddots & \ddots & \ddots & 0\\
		0 & \cdots & 0 & -1 & 1
	\end{array}\right],
	\label{eq:L1D}
\end{equation}
i.e. as the $(m-1)\times m$ matrix such that
\begin{equation*}
\xDelta=\LoneD\bx=\left[x_{2}-x_{1},\,x_{3}-x_{2},\,\ldots,\,x_{m}-x_{m-1}\right]^T,
\end{equation*}
where $m-1$ is the number of unique up to sign differences between adjacent elements of $\bx$.
In practice there is no need to compute a contrast matrix, although for deriving variational algorithms it is useful to carry $\bL$ around. As for the matrix defined in \eqref{eq:L1D}, it is convenient to design $\bL$ as a matrix whose number of rows and columns are respectively equal to the number of differences $d$ and the length of $\bx$, $m$, also in higher-dimensional problems. 
Model \eqref{eq:linInvProbModel} also incorporates a Laplace penalization that originates from the auxiliary variables $b_j>0$, $j=1,\ldots,d$, and the following result.
\begin{result}
	Let $x$ and $b$ be random variables such that
	\begin{equation*}
	x\,\vert\, b\sim N\left(0,\sigma^2/b\right)\quad\mbox{and}\quad b\sim\mbox{Inverse-$\chi^2$}\left(2,1\right),\quad\mbox{with}\quad\sigma>0.
	\end{equation*}
	Then $x\sim\mbox{Laplace}\left(0,\sigma\right)$.
	\label{res:laplaceRes}
\end{result} 

The choice of imposing priors on the difference  $\xDelta$ is mainly motivated by the two-dimensional applications on biomedical and archaeological imaging considered in this work. Such a choice gives adequate smoothing for flat regions (i.e., it increases image deblurring), but it may oversmooth discontinuities (see Section 3.2 of \citeauthor{aykroyd2001bayesian}, \citeyear{aykroyd2001bayesian}, for a detailed discussion on this and remedies). We focus on first-neighbor differences to illustrate efficient computation of the variational algorithm updates through removal of the contrast matrix $\boldsymbol{L}$ as per Section \ref{sec:stremVarInf}. Another widely used neighborhood system is the second order one, which is based on the eight nearest neighbors (e.g., \citeauthor{green1990bayesian}, \citeyear{green1990bayesian}) and also provides ground for efficient algorithm implementations.

If the dimension of $\bx$ increases, the number of first neighbor differences increases and reduces computational efficiency. This issue is often solved in practice by partitioning the surface where the observations are collected into smaller regions. Sometimes this is also done to facilitate the application of the inverse problem to irregular (e.g., non-rectangular) surfaces. For example, when large archaeological fields are explored it is standard practice to divide the area into grids and examine each grid as soon as it has been surveyed, and the inverse problem reconstruction of each grid is typically extended to part of the neighboring grids in each direction to get a smoother reconstruction (\citeauthor{aykroyd2001bayesian}, \citeyear{aykroyd2001bayesian}, Section 5).

The design of the matrix $\bK$ varies according to the inverse problem characteristics. As illustration and for later use on real biomedical data we consider a Gaussian kernel matrix $\bK$ and show its application to a simple unidimensional problem.
If $\by$ is a vector of $m$ recordings from a unidimensional space having one-to-one correspondence with $\bx$, the $(i,j)$th entry of a Gaussian kernel matrix $\bK$ is given as follows:
\begin{equation}
K_{ij}=(2\pi\delta^2)^{-1/2}\exp\{-(i-j)^2/(2\delta^2)\},\quad i=1,\ldots,m,\quad j=1,\ldots,m,
\label{eq:Kmatrix}
\end{equation}
where $\delta>0$ is a parameter that governs the amount of blur. Here $\delta$ is assumed to be fixed, but it can be estimated within the variational framework by imposing a prior and selected using, for example, a posterior mean estimate. Since this parameter is required to be positive, sensible choices of prior are gamma and inverse gamma distributions \citep{weir1997fully,aykroyd2001bayesian}. Alternatively, to limit computational costs one could first obtain a discrete approximation of the posterior of $\delta$ for a discrete grid of $\delta$ values.

\begin{figure}[!ht]
	\centering
	{\includegraphics[width=1\textwidth]{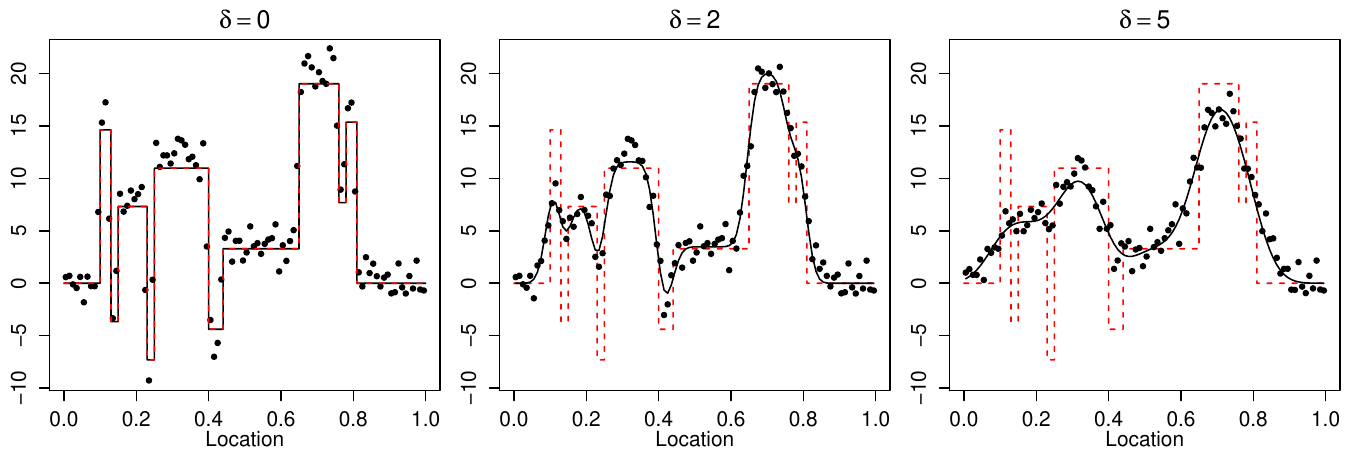}}
	\caption{\it  Data points obtained through the Blocks test function (red dashed line) along with blurred test function (solid line) for different levels of blur: no blur ($\delta=0$), moderate blur ($\delta=2$) and large blur ($\delta=5$).}
	\label{fig:1dExampPlots} 
\end{figure}

To illustrate the effect of variations in $\delta$, consider the \emph{Blocks} test function \citep{donoho1994ideal,nason2008wavelet} from
the \texttt{wavethresh} package \citep{nason2022wavethresh} available in \textsf{R} \citep{R2023}. The piecewise constant nature of this function makes estimation a very challenging problem especially when tackled as an inverse problem, but it is well motivated by stratigraphy problems in archaeology \citep{allum1999empirical,aykroyd2001bayesian}. 
Figure \ref{fig:1dExampPlots} shows three examples produced through \eqref{eq:baseModel} and the Gaussian kernel \eqref{eq:Kmatrix}, with $m=100$, $\sigma=1$ and $\delta=0,2,5$. In each plot the red dashed line shows the true Blocks function to estimate. In the plot corresponding to $\delta=0$ blurring is not added to the generation process and  the data points are randomly scattered around the true function. Blurring is introduced when $\delta>0$ and points scatter around a rounded solid line. The cases where $\delta=2$ and $\delta=5$ respectively correspond to moderate and large blurring of the underlying true function and hence moderate and difficult inverse problems.

In two dimensions, assuming the observed data is stored in a matrix $\bY$, each element of a Gaussian kernel matrix $\bK$ links an element of $\by=\vecof(\bY)$ to one of $\bx=\vecof(\bX)$. The entry of $\bK$ corresponding to a pair $(Y_{ij},X_{i'j'})$ is given by
\begin{equation}
(2\pi\delta^2)^{-1}\exp[-\{(i-i')^2+(j-j')^2\}/(2\delta^2)],\quad\mbox{for }\delta>0.
\label{eq:Kbiomed}
\end{equation}
The effect of $\delta$ on blurring is analogous to the one described for unidimensional problems.

\section{Model fitting via variational methods \label{sec:varInf}}

In this section, we study variational Bayes approximations, specifically MFVB and VMP, for fitting the base model \eqref{eq:linInvProbModel}. Emphasis is placed onto the message passing on factor graph fragment prescription, which provides the infrastructure for compartmentalized and scalable algorithm implementations.

For studying MFVB and VMP note that the joint density function of all the random variables and random vectors in model \eqref{eq:linInvProbModel} admits the following factorization: 
\begin{equation}
p\left(\by,\bx,\bb,\sigma^2_\varepsilon,\sigma^2_x,a_\varepsilon,a_x\right)=p\left(\by\vert\bx,\sigma^2_\varepsilon\right)p\left(\bx\vert\bb,\sigma^2_x\right)p\left(\bb\right)p\left(\sigma^2_\varepsilon\vert a_\varepsilon\right)p\left(\sigma^2_x\vert a_x\right)p\left(a_\varepsilon\right)p\left(a_x\right).
\label{eq:modelFactoriz}
\end{equation}
Both variational inference procedures are based upon approximating the joint posterior density function through a product of approximating density functions. A possible mean field restriction on the joint posterior density function of all parameters in \eqref{eq:linInvProbModel} is
\begin{equation}
p\left(\bx,\bb,\sigma^2_\varepsilon,\sigma_x^2,a_\varepsilon,a_x\vert\by\right)\approx q\left(\bx\right)q\left(\sigma^2_\varepsilon\right)q\left(\sigma^2_x\right)q\left(a_\varepsilon\right)q\left(a_x\right)\prod_{j=1}^d q\left(b_j\right).
\label{eq:mfRest}
\end{equation}
Typically, more restrictive density products facilitate the derivation of a variational algorithm but also negatively impact on the quality of the approximation. On the other hand, less severe restrictions may increase algebraic and computational complexity of the variational algorithms. The one used above provides a good trade-off between tractability and accuracy of the approximation.

The scope of MFVB and VMP is to provide expressions for the optimal $q$-densities that minimize the Kullback–Leibler divergence between the approximating densities themselves and the left-hand side of \eqref{eq:mfRest}. The former is based upon a directed acyclic graph interpretation of the model, whereas the latter benefits from a factor graph representation and its subsequent division into factor graph fragments. While the two variational inference procedures lead to ostensibly different iterative algorithms, they converge to the identical posterior density function approximations in the case where the inputs of the algorithms are the same and parameters are updated in the same sequence, since they are each founded upon the same optimization problem (Wand, 2017).
Details for fitting model \eqref{eq:linInvProbModel} under restriction \eqref{eq:mfRest} via MFVB and VMP are shown in Sections \ref{sec:MFVB} and \ref{sec:VMP}.

\subsection{Mean field variational Bayes}\label{sec:MFVB}

Mean field variational Bayes is a well established approximate Bayesian inference technique where tractability is achieved through factorization of the approximating density. Here we provide a short introduction to MFVB and we refer the reader to Section 3 of \cite{wand2011mean} for fuller details on its derivation.

Let $\bz$ be a vector of observed data and $\btheta\in\Theta$ represent all model parameters. The logarithm of the model marginal likelihood, $\log p(\bz)$, satisfies
\begin{equation*}
\log p(\bz)\geq\log\underline{p}(\bz;q)\equiv\int q(\btheta)\log\left\{\frac{p(\bz,\btheta)}{q(\btheta)}\right\}d\btheta,
\label{eq:LowerBound}
\end{equation*}
where $\underline{p}(\bz;q)$ is a lower-bound depending on a density function $q$ defined over $\Theta$ and the joint density function $p(\bz,\btheta)$. It can be shown that maximizing the above lower-bound is equivalent to minimizing the Kullback-Leibler divergence between $q(\btheta)$ and the joint posterior density function $p(\btheta\vert\bz)$,
\begin{equation*}
\mbox{KL}(q(\btheta)\,\Vert\, p(\btheta\vert\bz))=\int q(\btheta)\log\left\{\frac{q(\btheta)}{p(\btheta\vert\bz)}\right\}d\btheta.
\end{equation*}
If the approximating density is factorized according to a partition $(\btheta_1,\ldots,\btheta_s)$ of $\btheta$ such that $q(\btheta)=\prod_{k=1}^sq(\btheta_k)$, as on the right side of \eqref{eq:mfRest}, then the optimal approximating densities satisfy
\begin{equation}
	q^*(\btheta_k)\propto\exp\big[E_{q(\btheta\backslash \btheta_k)}\{\log p(\btheta_k\vert\bz,\btheta\backslash \btheta_k )\}\big],\quad k=1,\ldots,s,
\label{eq:qkupdate}
\end{equation}
where $E_{q(\btheta\backslash \btheta_k)}$ denotes the expectation with respect to all the approximating densities except $q(\btheta_k)$ and $\btheta\backslash \btheta_k$ represents the entries of $\btheta$ with $\btheta_k$ omitted.
Under mild regularity conditions it can be shown that optimization of the lower bound can be performed via a coordinate ascent scheme converging to a local maximizer.

\begin{figure}[!ht]
	\centering
	{\includegraphics[width=0.5\textwidth]{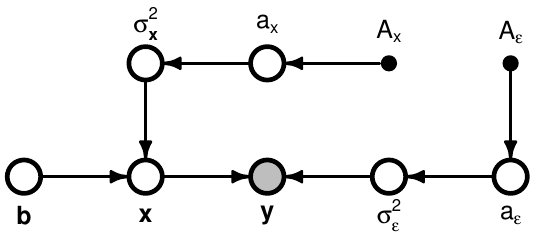}}
	\caption{\it Directed acyclic graph representation of the Normal response model with Laplace penalization in \eqref{eq:linInvProbModel}. The shaded circle corresponds to the observed data. The unshaded circles correspond to model parameters and auxiliary variables. The small solid circles correspond to hyperparameters.}
	\label{fig:InvProbDAG} 
\end{figure}

Figure \ref{fig:InvProbDAG} is a directed acyclic graph representation of model \eqref{eq:linInvProbModel}, which forms the basis for deriving an MFVB algorithm. The shaded circle corresponds to the observed data, the empty circles correspond to model parameters and auxiliary variables, and the small solid circles are used for hyperparameters.
According to \eqref{eq:qkupdate}, the full conditional density functions of the nodes in the directed acyclic graph provide expressions for the optimal approximating densities. For instance, the full conditional density of $\bx$ arising from model \eqref{eq:linInvProbModel} and \eqref{eq:xFactor} is
\begin{equation*}
p\left(\bx\vert\mbox{rest}\right)\propto\exp\left[-\frac{1}{2}\left\{\bx^T\bL^T\left(\frac{1}{\sigma^2_\varepsilon}\bK^T\bK+\frac{1}{\sigma^2_x}\diag\left(\bb\right)\right)\bL\bx-\frac{2}{\sigma^2_\varepsilon}\bx^T\bK^T\by\right\}\right].
\end{equation*}
From application of \eqref{eq:qkupdate} it follows that
\begin{equation}
    q^*(\bx)\,\text{is a}\, N\big(\muqx,\Sigmaqx\big)\,\text{density function}\label{eq:qxDens}
\end{equation}
with
\begin{equation*}
\muqx\equiv\mu_{q(1/\sigma_{\varepsilon}^2)}\Sigmaqx\bK^T\by\,\,\,\mbox{and}\,\,\,\Sigmaqx\equiv\left(\mu_{q(1/\sigma^2_\varepsilon)}\bK^T\bK+\mu_{q(1/\sigma^2_x)}\bL^T\diag(\mu_{q(\bb)})\bL\right)^{-1},
\end{equation*}
where $\mu_{q(1/\sigma_{\varepsilon}^2)}$, $\mu_{q(1/\sigma^2_x)}$ and $\mu_{q(\bb)}$ respectively denote the expectations of $1/\sigma_{\varepsilon}^2$, $1/\sigma^2_x$ and $\bb$ computed with respect to the optimal approximating densities $q^*(\sigma^2_\varepsilon)$, $q^*(\sigma^2_x)$ and $q^*(\bb)=\prod_{j=1}^d q^*\left(b_j\right)$ arising from the mean field restriction \eqref{eq:mfRest}.
Referring to restriction \eqref{eq:mfRest}, the other MFVB approximations to the posterior density functions of the parameters and auxiliary variables depicted in the directed acyclic graph have the following optimal forms:
\begin{align}
&q^*(b_j)\,\text{is an}\, \text{Inverse-Gaussian}\big(\mu_{q(b_j)},\lambda_{q(b_j)}\big)\,\text{density function, for}\, j=1,\ldots,d,\label{eq:qbDens}\\
&q^*(\sigma^2_\varepsilon)\,\mbox{is an}\, \mbox{Inverse-}\chi^2\left(\kappaqsigsqepsilon,\lambdaqsigsqepsilon\right)\,\text{density function},\label{eq:qsigsqepsDens}\\
&q^*(\sigma^2_x)\,\mbox{is an}\,\mbox{Inverse-}\chi^2\big(\kappaqsigsqx,\lambdaqsigsqx\big)\,\text{density function},\label{eq:qsigsqxDens}\\
&q^*(a_\varepsilon)\,\mbox{is an}\,\mbox{Inverse-}\chi^2\big(\kappaqaepsilon,\lambdaqaepsilon\big)\,\text{density function}\label{eq:qaepsDens}\\
\text{and }&q^*(a_x)\,\mbox{is an}\,\mbox{Inverse-}\chi^2\big(\kappaqax,\lambdaqax\big)\,\text{density function}.\label{eq:qaxDens}
\end{align}
Details about the optimal density function parameters are given in the supplement.
Such parameters can be obtained through the MFVB iterative scheme listed as Algorithm \ref{alg:MFVBalgo}.

\begin{algorithm}[!th]
	\begin{center}
		\begin{minipage}[t]{154mm}
			\begin{small}
				\begin{itemize}
					
					\setlength\itemsep{4pt}
					
					\vspace{0.2cm}
					
					\item[] \textbf{Data Inputs:} $\by$ $(m\times 1)$, $\bK$ $(m\times m)$. 
					
					\item[] \textbf{Hyperparameter Inputs:} $A_\varepsilon>0$, $A_x>0$.
					
					\item[] \textbf{Initialize:} $\mu_{q\left(1/\sigma^2_\varepsilon\right)}>0$, $\mu_{q\left(1/\sigma^2_x\right)}>0$, $\mu_{q\left(1/a_\varepsilon\right)}>0$, $\mu_{q\left(1/a_x\right)}>0$, $\bmu_{q(\bb)}$ $(d\times 1)$ vector of\\
					\null$\qquad\qquad$ positive elements.
					
					\item[] $\kappaqsigsqepsilon\longleftarrow m+1\quad; \quad\kappaqsigsqx\longleftarrow d+1\quad;\quad\kappaqaepsilon\longleftarrow 2\quad;\quad\kappaqax\longleftarrow 2$.
					
					\item[] \textbf{Cycle:}
					\begin{itemize}
						
						\setlength\itemsep{4pt}
						
						\item[] $\Sigmaqx\longleftarrow\left(\mu_{q(1/\sigma^2_\varepsilon)}\bK^T\bK+\mu_{q(1/\sigma^2_x)}\bL^T\diag(\bmu_{q(\bb)})\bL\right)^{-1}$
						\item[] $\muqx\longleftarrow\mu_{q(1/\sigma_{\varepsilon}^2)}\Sigmaqx\bK^T\by$
						\item[] $\lambdaqsigsqepsilon\longleftarrow\mu_{q(1/a_\varepsilon)}+\Vert\by-\bK\muqx\Vert^2+\tr\left(\bK^T\bK\Sigmaqx\right)$
						\item[] $\muqrecipsigsqepsilon\longleftarrow\kappaqsigsqepsilon/\lambdaqsigsqepsilon$
						\item[] $\lambdaqaepsilon\longleftarrow\mu_{q(1/\sigma^2_\varepsilon)}+1/A_\varepsilon^2\quad;\quad\mu_{q(1/a_\varepsilon)}\longleftarrow\kappaqaepsilon/\lambdaqaepsilon$
						\item[]
						$\btau_1 \longleftarrow \big(\bL\muqx\big)^2+\diagonal(\bL\Sigmaqx\bL^T)$
						\item[]
						$\lambdaqsigsqx\longleftarrow\mu_{q(1/a_x)}+\mu_{q\left(\bb\right)}^T\btau_1$
						\item[] $\muqrecipsigsqx\longleftarrow\kappaqsigsqx/\lambdaqsigsqx$
						\item[] $\lambdaqax\longleftarrow\mu_{q(1/\sigma^2_x)}+1/A_x^2\quad;\quad\mu_{q(1/a_x)}\longleftarrow\kappaqax/\lambdaqax$
						\item[] $\btau_2 \longleftarrow \muqrecipsigsqx\btau_1\quad;\quad\bmu_{q\left(\bb\right)} \longleftarrow 1/\sqrt{\btau_2}$ 						
					\end{itemize}
					
					\item[] \textbf{Output:} $\muqx$, $\Sigmaqx$, $\bmu_{q\left(\bb\right)}$, $\blambda_{q\left(\bb\right)}\equiv\bone_d$, $\kappaqsigsqepsilon$, $\lambdaqsigsqepsilon$, $\kappaqsigsqx$, $\lambdaqsigsqx$, $\kappaqaepsilon$, $\lambdaqaepsilon$, $\kappaqax$, $\lambdaqax$.
				\end{itemize}
			\end{small}
		\end{minipage}
	\end{center}
	\caption{\textit{Mean field variational Bayes scheme for fitting the inverse problem model \eqref{eq:linInvProbModel}, using product density restriction \eqref{eq:mfRest}.}}
	\label{alg:MFVBalgo} 
\end{algorithm}

%

\subsection{Variational message passing}\label{sec:VMP}

The idea behind variational message passing as presented in \cite{wand2017fast} is that the same approximations in \eqref{eq:qkupdate} can be achieved by exploiting a convenient factor graph representation of the model.
A detailed description of VMP as a method for fitting statistical models that have a factor graph representation is provided in Sections 2--4 of \cite{wand2017fast} and briefly summarized here. The same notational conventions of \cite{wand2017fast} concerning message passing are used in this work. 

A factor graph is an ensemble of factors connected to stochastic nodes by edges. Let $f_{h}$, $h=1,\ldots,r$, denote a generic factor and function of one or more stochastic nodes and $\btheta_k$, $k=1,\ldots,s$, be a generic stochastic variable represented by a node. If $\btheta_k$ is a neighbor of $f_{h}$ in the factor graph, then the messages passed from $f_{h}$ to $\btheta_k$ and from $\btheta_k$ to $f_{h}$ are functions of $\btheta_k$ and are denoted by $\bm_{f_h\rightarrow\btheta_k}(\btheta_k)$ and $\bm_{\btheta_k\rightarrow f_h}(\btheta_k)$, respectively. 

In the optic of deriving an algorithm, the VMP stochastic node to factor message updates are given by
\begin{equation}
    \bm_{\btheta_k\rightarrow f_h}(\btheta_k)\longleftarrow\propto\prod_{h'\neq h\,:\,k\in\text{neighbors}(h')}\bm_{f_{h'}\rightarrow\btheta_k(\btheta_k)}
    \label{eq:nodeToFacUpdate}
\end{equation}
and the factor to stochastic node message updates are
\begin{equation}
    \bm_{ f_h\rightarrow\btheta_k}(\btheta_k)\longleftarrow\propto\exp\left[E_{f_h\rightarrow\btheta_k}\left\{\log f_h(\btheta_{\text{neighbors}(h)})\right\}\right],
    \label{eq:facToNodeUpdate}
\end{equation}
where $E_{f_h\rightarrow\btheta_k}$ denotes expectation with respect to the density function
\begin{equation}
    \frac{\prod_{k'\in\text{neighbors}(h)\backslash\{k\}}\bm_{ f_h\rightarrow\btheta_{k'}}(\btheta_{k'})\bm_{ \btheta_{k'}\rightarrow f_h}(\btheta_{k'})}{\prod_{k'\in\text{neighbors}(h)\backslash\{k\}}\int\bm_{ f_h\rightarrow\btheta_{k'}}(\btheta_{k'})\bm_{ \btheta_{k'}\rightarrow f_h}(\btheta_{k'})d\btheta_{k'}},
    \label{eq:VMPexpectation}
\end{equation}
$\text{neighbors}(h)\equiv\{k=1,\ldots,s\,:\,\btheta_k\text{ is a neighbor of }f_h\}$ and the $\longleftarrow\propto$ symbol means that the function of $\btheta_k$ on the left-hand side is updated according to the expression on the right-hand side but that multiplicative factors not depending on $\btheta_k$ can be ignored. Upon convergence of the messages, the optimal $q$-densities are obtained via
\begin{equation}
    q^*(\btheta_k)\propto\prod_{h\,:\,k\in\text{neighbors}(h)}\bm_{f_h\rightarrow\btheta_k}(\btheta_k).
    \label{eq:qStarVMP}
\end{equation}

The messages above are typically proportional to an exponential family density function and so are such that
\begin{equation}
\bm_{f_h\rightarrow\btheta_k}(\btheta_k)\propto\exp\big\{\bT(\btheta_k)^T\biggerbdeta_{ {f_h\rightarrow\btheta_k}}\big\}\quad\mbox{and}\quad \bm_{\btheta_k\rightarrow f_h}(\btheta_k)\propto\exp\big\{\bT(\btheta_k)^T\biggerbdeta_{\btheta_k\rightarrow f_h}\big\},
\label{eq:expFamMsgs}
\end{equation}
where $\bT(\btheta_k)$ is a sufficient statistic vector, and $\biggerbdeta_{f_h\rightarrow\btheta_k}$ and $\biggerbdeta_{\btheta_k\rightarrow f_h}$ are the message natural parameter vectors. Then for each parameter $\btheta_k$, the optimal approximating density $q^*(\btheta_k)$ belongs to an exponential family with natural parameter vector
\begin{equation}
    \biggerbdeta_{{\tiny q}(\btheta_k})\equiv\prod_{h\,:\,k\in\text{neighbors}(h)}\biggerbdeta_{f_h\rightarrow\btheta_k}(\btheta_k)
    \label{eq:etaVMP}
\end{equation}
that can be computed at convergence of a VMP algorithm.

\begin{figure}[!ht]
	\centering
	{\includegraphics[width=.6\textwidth]{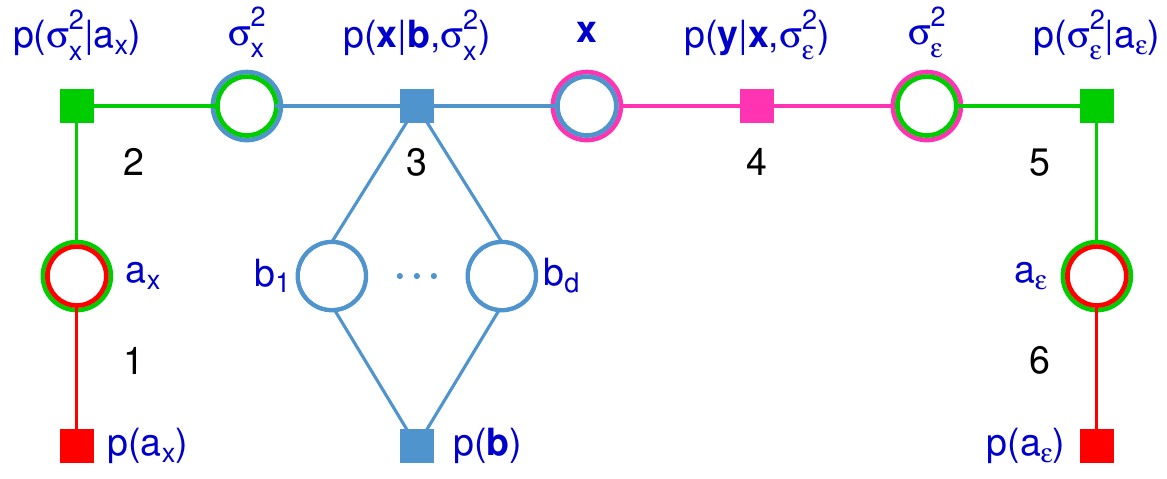}}
	\caption{\it Factor graph representation of the Normal response model with Laplace penalization in \eqref{eq:linInvProbModel}, where the square nodes correspond to the density functions, or factors, at the right-hand side of \eqref{eq:modelFactoriz}. The circular nodes correspond to stochastic nodes of the $q$-density factorization in \eqref{eq:mfRest}. The fragments are numbered 1 to 6, whereas colors identify different fragments types.}
	\label{fig:InvProbFacGraph} 
\end{figure}

The notion of a factor graph fragment, or simply \emph{fragment}, allows for compartmentalization of algebra and computer code and can be exploited to fit inverse problems via VMP without resorting to calculations involving \eqref{eq:nodeToFacUpdate} and \eqref{eq:facToNodeUpdate} each time a VMP algorithm is derived for a new model.
The corresponding factor graph representation of \eqref{eq:modelFactoriz} given the density product restriction \eqref{eq:mfRest} appears in Figure \ref{fig:InvProbFacGraph}. Colors mark different fragment types, in accordance with the nomenclature presented in \cite{wand2017fast} for variational message passing on factor graph fragments and numbers label seven factor graph fragments. Some of these have been studied in previous works.
Those numbered 1, 2, 5 and 6 are already catalogued in \cite{maestrini2021inverse} as \emph{Inverse G-Wishart prior fragments} (1 and 6) and \emph{iterated Inverse G-Wishart fragments} (2 and 5). Fragment number 4 corresponds to the \emph{Gaussian likelihood fragment} treated in Section 4.1.5 of \cite{wand2017fast}, whose notation can be aligned with that of model \eqref{eq:linInvProbModel} by settings $\bA$, $\btheta_1$ and $\theta_2$ equal to the current $\bK$, $\bx$ and $\sigma^2_\varepsilon$, respectively.
In the view of VMP, we can just read off from equations (38) and (39) of \cite{wand2017fast}, which themselves originate from \eqref{eq:facToNodeUpdate}, and get the following updates for fragment 4 involving factor $p(\by|\,\bx,\sigma^{2}_{\varepsilon})$ and stochastic nodes $\bx$ and $\sigma^2_{\varepsilon}$:
\begin{equation*}
\etaSUBpyxsigsqepsilonTOx \longleftarrow \left[\begin{array}{c} 
\bK^{T}\by\\[1ex]
-\frac{1}{2}\vecof\left(\bK^{T}\bK\right)
\end{array}\right]\Big\{\left(\etaSUBpyxsigsqepsFROMTOsigsqeps\right)_1 + 1\Big\}\Big/\Big\{\left(\etaSUBpyxsigsqepsFROMTOsigsqeps\right)_2\Big\}
\end{equation*}
and
\begin{equation*}
\etaSUBpyxsigsqepsilonTOsigsqepsilon\longleftarrow \left[\begin{array}{c}
-n/2\\[1ex]
\GVMP\left(\etaSUBpyxsigsqepsFROMTOx;\bK^T\bK,\bK^T\by,\by^T\by\right)
\end{array}\right],
\end{equation*}
where, according to the definition in Section 2.7 of \cite{wand2017fast}, $\GVMP\left([\begin{array}{cc}
    \bv_1^T\,\,
    \bv_2^T
    \end{array}]^T;\bQ,\br,s\right)\equiv-(1/8)\tr\left(\bQ\{\vecof^{-1}(\bv_2)\}^{-1}[\bv_1\bv_1^T\{\vecof^{-1}(\bv_2)\}^{-1}-2\bI]\right)$\\ $-(1/2)\br^T\{\vecof^{-1}(\bv_2)\}^{-1}\bv_1-(1/2)s$, for a $d\times 1$ vector $\bv_1$, $d^2\times 1$ vector $\bv_2$ such that $\vecof^{-1}(\bv_2)$ is symmetric, $d\times d$ matrix $\bQ$, $d\times 1$ vector $\br$ and $s\in\mathbb{R}$.

New calculations are needed only for fragment 3 to compose a VMP algorithm for the whole model. In fragment 3, factor $p\left(\bb\right)$ symbolizes the specification
\begin{equation}
b_j\simind\mbox{Inverse-$\chi^2$}\left(2,1\right),\quad j=1,\ldots,d,
\label{eq:bFactor}
\end{equation} 
but may also represent other penalizations. Examples of alternative shrinkage distributions are displayed in Table \ref{tab:bDistrib} and discussed in Section \ref{sec:alternPenaliz}.
The factor $p\left(\bx\vert\bb,\sigma^2_x\right)$ of fragment 3 corresponds to the possibly improper specification \eqref{eq:xFactor}. The logarithm of this likelihood factor is
\begin{equation}
\log p\left(\bx\vert\bb,\sigma^2_x\right)=-\frac{d}{2}\log\left(\sigma^2_x\right)-\frac{1}{2\sigma^2_x}\bx^T\bL^T\diag\left(b\right)\bL\bx + \frac{1}{2}\sum_{i=1}^{d}\log\left(b_i\right) + \mbox{const}.
\label{eq:xLogLik}
\end{equation}
Combining \eqref{eq:xLogLik} with the auxiliary variable distributional assumption in \eqref{eq:bFactor} we get the VMP scheme listed as Algorithm \ref{alg:VMPalgo} for fitting the penalization part of model \eqref{eq:linInvProbModel}. This originates from first noticing that, as a function of $\bx$, \eqref{eq:xLogLik} can be written as
\begin{equation*}
\log p\left(\bx\vert\bb,\sigma^2_x\right) = \frac{1}{\sigma_{x}^{2}}\left[\begin{array}{c}
\bx\\
\vecof\left(\bx\bx^{T}\right)
\end{array}\right]^{T}\left[\begin{array}{c}
\boldsymbol{0}\\
-\frac{1}{2}\vecof\left(\bL^{T}\diag\left(\bb\right)\bL\right)
\end{array}\right] + \mbox{const}, 
\end{equation*}
which in light of \eqref{eq:facToNodeUpdate} and \eqref{eq:expFamMsgs} originates the message sent from factor $\left(\bx\vert\bb,\sigma^2_x\right)$ to node $\bx$
\begin{equation*}
\mSUBpxbsigsqxTOx = \exp\left\{\left[\begin{array}{c}
\bx\\
\vecof\left(\bx\bx^{T}\right)
\end{array}\right]^{T} \etaSUBpxbsigsqxTOx\right\}.
\end{equation*}
This message is within the Multivariate Normal family, with
\begin{equation*}
\etaSUBpxbsigsqxTOx \equiv E_{\boxtimes}\left(1/\sigma_{x}^{2}\right)\left[\begin{array}{c}
\boldsymbol{0}\\
-\frac{1}{2}\vecof\left(\bL^{T}\diag\left\{E_{\oplus}\left(\bb\right)\right\}\bL\right)
\end{array}\right],
\end{equation*}
where, following \eqref{eq:VMPexpectation}, $E_{\boxtimes}$ denotes expectation with respect to the normalization of
\begin{equation*}
\mSUBpxbsigsqxTOsigsqx \mSUBsigsqxTOpxbsigsqx
\end{equation*}
and $E_{\oplus}$ denotes expectation with respect to the normalization of
\begin{equation*}
\mSUBpxbsigsqxTOb \mSUBbTOpxbsigsqx .
\end{equation*}
Rewriting \eqref{eq:xLogLik} as a function of $\sigma^2_x$ and $\bbeta$ and applying similar considerations we get to Algorithm \ref{alg:VMPalgo} (see the supplement for full derivations).

\begin{algorithm}[!th]
	\begin{center}
		\begin{minipage}[t]{154mm}
			\begin{small}
				\begin{itemize}
					
					\setlength\itemsep{4pt}
					
					\vspace{0.2cm}
					
					\item[] \textbf{Inputs:} $\etaSUBpxbsigsqxTOx$, $\etaSUBxTOpxbsigsqx$, $\etaSUBpxbsigsqxTOsigsqx$, $\etaSUBsigsqxTOpxbsigsqx$.
					
					\item[] \textbf{Updates:}
					
					\begin{itemize}
						
						\setlength\itemsep{4pt}
						
						\item[] $\mu_{q\left(1/\sigma^2_x\right)} \longleftarrow \Big\{\left(\etaSUBpxbsigsqxFROMTOsigsqx\right)_1+1\Big\}\Big/\Big\{\left(\etaSUBpxbsigsqxFROMTOsigsqx\right)_2\Big\}$
						\item[] $\bOmega_1 \longleftarrow -\frac{1}{2}\left\{\vecof^{-1}\left(\left(\etaSUBpxbsigsqxFROMTOx\right)_2\right)\right\}^{-1}$
						\item[] $\bomega_2 \longleftarrow \bOmega_1\left(\etaSUBpxbsigsqxFROMTOsigsqx\right)_1\quad;\quad\bomega_3 \longleftarrow \bL\bomega_2\quad;\quad\bomega_4 \longleftarrow \diagonal\left(\bL\bOmega_1\bL^T\right)$
						\item[] $\bomega_5 \longleftarrow \bomega_3^2 + \bomega_4\quad;\quad\bomega_6 \longleftarrow \mu_{q\left(1/\sigma^2_x\right)}\bomega_5$
						\item[] $\bmu_{q\left(\bb\right)} \longleftarrow 1/\sqrt{\bomega_6}\quad;\quad\bOmega_7 \longleftarrow \bL^T\diag\left(\mu_{q\left(\bb\right)}\right)\bL$
					\end{itemize}
					
					\item[] until convergence.
					
					\item[] \textbf{Outputs:} 
					
					$\quad\etaSUBpxbsigsqxTOx\longleftarrow\mu_{q\left(1/\sigma^2_x\right)}\left[\begin{array}{c}
					\bzero\\[1ex]
					-\frac{1}{2}\vecof\left(\bOmega_{7}\right)
					\end{array}\right]\quad;\quad\etaSUBpxbsigsqxTOsigsqx\longleftarrow-\frac{1}{2}\left[\begin{array}{c}
					m_{L}\\[1ex]
					\mu_{q\left(\bb\right)}^T\bomega_{5}
					\end{array}\right].$
				\end{itemize}
			\end{small}
		\end{minipage}
	\end{center}
	\caption{\textit{Variational message passing inputs, updates and outputs of the penalization likelihood fragment given by \eqref{eq:xFactor} and \eqref{eq:bFactor}, and corresponding to factor graph fragment 3 of Figure \ref{fig:InvProbFacGraph}.}}
	\label{alg:VMPalgo} 
\end{algorithm}

The combination of Algorithm \ref{alg:VMPalgo} with the Inverse G-Wishart fragment algorithms of \cite{maestrini2021inverse} and the Gaussian likelihood fragment algorithm of \cite{wand2017fast} gives rise to a full VMP procedure for fitting and approximate inference on model \eqref{eq:linInvProbModel}. Complete details about the implementation of VMP for fitting the inverse problem model \eqref{eq:linInvProbModel} are provided in the supplement. At convergence of the VMP procedure, \eqref{eq:qStarVMP} and \eqref{eq:etaVMP} can be applied to obtain the approximating densities in explicit form. For instance, we have
\begin{equation*}
\begin{array}{c}
q^*(\bx)\propto\exp\left\{\left[\begin{array}{c}
\bx\\[1ex]
\vecof\big(\bx\bx^T\big)
\end{array}\right]^T\etaSUBqx\right\},\,\mbox{with}\,\,\etaSUBqx\equiv\etaSUBpxbsigsqxTOx + \etaSUBpyxsigsqepsilonTOx,
\end{array}
\end{equation*}
which can be rewritten in terms of the common parameters of the approximating density \eqref{eq:qxDens} of MFVB by setting
\begin{equation*}
    \begin{array}{c}		\muqx=\Sigmaqx\left(\etaSUBqx\right)_{1:m}\quad\mbox{and}\quad\Sigmaqx=-\frac{1}{2}\vecof^{-1}\left\{\left(\etaSUBqx\right)_{(m+1):m^2}\right\}.
    \end{array}
\end{equation*}
Both MFVB and VMP can achieve the same approximations \eqref{eq:qxDens}--\eqref{eq:qaxDens} since it is possible to establish a direct correspondence between the initialisation values of the two variational procedures. For example, using the hyperparameter input $A_x$ in Algorithm \ref{alg:MFVBalgo} for MFVB corresponds to a call of VMP to the Inverse G-Wishart Prior Fragment \cite[Algorithm 1]{maestrini2021inverse} with inputs $G_{\Theta}=G_{\tiny\mbox{diag}}$, $\xi_{\Theta}=1$, $\bLambda_{\Theta}=A_x^{-2}$ and $G_{\tiny\mbox{diag}}$ being a graph corresponding to a diagonal matrix, to initialise the natural parameter vector $\etaSUBpaxTOax$ of the message sent from factor $p(a_x)$ to stochastic node $a_x$.
Further details about the initialisation of VMP are provided in the supplement.

\subsection{Alternative response and penalization distributions}\label{sec:alternPenaliz}

The message passing on factor graph fragments paradigm allows for flexible imposition of non-Normal response distributions. For instance, fragment 4 of Figure \ref{fig:InvProbFacGraph} can be replaced by one of the likelihood fragments identified in \cite{nolan2017accurate}, \cite{maestrini2018variational} or \cite{mclean2019variational} to accommodate a variety of response distributions such as, for instance, binary-logistic, Poisson, Negative Binomial, $t$, Asymmetric Laplace, Skew Normal, Skew $t$ and Finite Normal Mixtures.

Other penalization structures can be easily incorporated by varying the distributional assumption on the vector of auxiliary variables $\bb$. \cite{neville2014mean} studied MFVB inference for three continuous sparse signal shrinkage distributions, namely the Horseshoe, Negative-Exponential-Gamma and Generalized Double Pareto distributions, that can replace the Laplace penalization employed in model \eqref{eq:linInvProbModel}. References for the development of these sparse shrinkage priors are respectively \cite{carvalho2010horseshoe}, \cite{griffin2011bayesian} and \cite{armagan2013generalized}.  

The derivation of variational algorithms for models containing these shrinkage distributions can be quite challenging. In fact, the variational algorithm algebraic complexity and inference performance rely upon an accurate choice of their auxiliary variable representations. \cite{neville2014mean} propose two alternative auxiliary variable representations for each of the aforementioned shrinkage distributions by making use of either one or two sets of auxiliary variables. Their empirical and theoretical evidence show the supremacy of representations based on a single set of auxiliary variables in terms of posterior density approximation and computational complexity for all three cases. If this auxiliary variable representation is chosen, then the three penalizations can be easily imposed in model \eqref{eq:linInvProbModel} as a replacement to the Laplace distribution by simply modifying the distributional assumption on the auxiliary vector $\bb$. Algorithms \ref{alg:MFVBalgo} and \ref{alg:VMPalgo} can still be used by simply replacing the update for $\bmu_{q\left(\bb\right)}$ with one of those listed in the last column of Table \ref{tab:bDistrib}. Some expressions in Table \ref{tab:bDistrib} related to the Negative-Exponential-Gamma and Generalized Double Pareto cases depend on the parabolic cylinder function of order $\nu$, $\mathcal{D}_\nu(x)$, and $\mathcal{R}_\nu(x)\equiv\mathcal{D}_{-\nu-2}(x)/\mathcal{D}_{-\nu-1}(x)$, for $\nu>0$ and $x>0$. Efficient computation of function $\mathcal{R}_\nu(x)$ is discussed in Appendix A of \cite{neville2014mean}.

Several other penalizations can be imposed on the base inverse problem model. The penalization in model \eqref{eq:linInvProbModel} is a particular case of the Bayesian lasso of \cite{park2008bayesian} that makes use of a Gamma prior on the Laplace squared scale parameter. \cite{tung2019bayesian} show the use of MFVB for variable selection in generalized linear mixed models via Bayesian adaptive lasso. \cite{ormerod2017variational} develop a MFVB approximation to a linear model with a spike-and-slab prior on the regression coefficients. A detailed discussion on variable selection priors and variational inference fitting goes beyond the scope of this article. However, for the analysis of archaeological data provided in the supplement we show how the Laplace penalization can be easily replaced by a Horseshoe prior without deriving a VMP algorithm from scratch. In the same real data analysis we replace the Normal response assumption with a Skew Normal one. 
 
\begin{table}
	\begin{center}
	\begin{tabular}{cccc}
	\Xhline{3\arrayrulewidth}\\[-2.2ex]
	\textbf{Penalization}      & \textbf{Density function} & \textbf{MFVB or VMP update} \\
	\Xhline{3\arrayrulewidth}\\[-2ex]
	Horseshoe & $p\left(b_j\right)=\pi^{-1}b_j^{-1/2}\left(1+b_j\right)^{-1}$ & $\bmu_{q\left(\bb\right)} \longleftarrow\frac{2}{\bzeta\odot\exp\left(\frac{1}{2}\bzeta\right)\odot E_1\left(\frac{1}{2}\bzeta\right)}-1$ \\[1.5ex]
	\hline\\[-2.2ex]
	Negative- & \multirow{2}{*}{$p\left(b_j\right)=\lambda b_j^{\lambda-1}\left(1+b_j\right)^{-\lambda-1}$} & \multirow{2}{*}{$\bmu_{q\left(\bb\right)} \longleftarrow(2\lambda + 1)\mathcal{R}_{2\lambda}\left(\sqrt{\bzeta}\right)\odot\sqrt{\bzeta}$} \\
	Exponential-Gamma & & \\
	\hline\\[-2ex] 
	Generalized & $p\left(b_j\right)=\frac{1}{2}\left(1+\lambda\right)\lambda^{1+\lambda}$ & \multirow{2}{*}{$\bmu_{q\left(\bb\right)} \longleftarrow\frac{\sqrt{2}\left(\lambda + 1\right)}{\sqrt{\bzeta}\odot(\sqrt{2}\lambda + \bzeta)}$}\\
	Double Pareto& $\times b_j^{\left(\lambda-2\right)/2}e^{\lambda^2 b_j/4}\mathcal{D}_{-\lambda-2}\left(\lambda\sqrt{b_j}\right)$ & \\[0.6ex]
	\hline     
	\end{tabular}
	\end{center}
	\caption{\it Distributions of the auxiliary variables $b_j>0$, $j=1\ldots d$, that produce the penalizations analyzed in \cite{neville2014mean} when introduced in model \eqref{eq:linInvProbModel} in lieu of the Laplace penalization. Here the shape parameter $\lambda>0$ is fixed. 		
	For each distribution, the MFVB or VMP update for $\bmu_{q\left(\bb\right)}$ is displayed in the last column. When Algorithm \ref{alg:MFVBalgo} is used to perform MFVB, $\bzeta\equiv\btau_2$; when Algorithm \ref{alg:VMPalgo} is used for running VMP, $\bzeta\equiv\bomega_6$.}
	\label{tab:bDistrib} 
\end{table}

%

\section{Streamlined variational algorithm updates \label{sec:stremVarInf}}

In typical inverse problem applications the number of observations can be very high and a na\"{i}ve implementation of Algorithms \ref{alg:VMPalgo} and \ref{alg:MFVBalgo} may lead to a bottleneck due to operations involving a large contrast matrix $\bL$ and big matrix inversion steps related to $\bK$. However, the structure of matrices $\bL$ and $\bK$ is such that computationally expensive algorithm updates may be efficiently performed. In this section we propose solutions to simplify algorithm updates and reduce their computational complexity. The results shown here are designed for one- and two-dimensional problems but are applicable to extensions to higher dimensions.

\subsection{Removal of the contrast matrix \label{sec:Lremoval}}

The contrast matrix $\bL$ is a potentially massive sparse matrix that one does not want to compute or store. The number of rows of $\bL$ is given by the $d$ differences between the elements of $\bx$ considered, whereas the number of columns is given by the length $m$ of $\bx$. For one-dimensional problems with a first nearest neighbor structure only the two longest diagonals of $\bL$ have non-zero elements, as shown in \eqref{eq:L1D}. The contrast matrix of two-dimensional problems under the same assumptions is sparse and has number of non-zero elements equal to twice the number of differences between elements of $\bx$, that is $2d$. 

The updates of the variational Algorithms \ref{alg:MFVBalgo} and \ref{alg:VMPalgo} that make use of matrix $\bL$ have the following form:
\begin{align}
	& \bL\bv,\mbox{ where }\bv\mbox{ is a vector}; \label{eq:exp1}\\
	& \diagonal(\bL\bM\bL^T),\mbox{ where }\bM\mbox{ is a symmetric positive definite matrix};\label{eq:exp2}\\
	& \bL^T\diag(\bw)\bL,\mbox{ where }\bw\mbox{ is a vector}.\label{eq:exp3}
\end{align}
The results presented in the supplementary material allow efficient computation of the MFVB and VMP updates that utilise matrix $\bL$ in the forms described in \eqref{eq:exp1}--\eqref{eq:exp3}. One- and two-dimensional problems are considered and the results can be potentially extended to higher dimensions, e.g. for voxel-type data in three-dimensions.

\subsection{Sparsification of $\bK$}\label{sec:Ksparsification}

The matrix $\bK$ is a potentially big matrix whose size depends on the number of observations and the length of $\bx$.
It is easy to notice from the MFVB scheme presented as Algorithm \ref{alg:MFVBalgo} that the algebraic operations where matrix $\bK$ appears have the following generic forms:
\begin{align}
	& \left(s_1\bK^T\bK + s_2\bL^T\diag(\bomega)\bL\right)^{-1},\mbox{ where }s_1\mbox{ and }s_2\mbox{ are scalar positive numbers};\label{eq:Kform1}\\
	&\bM\bK^T\bv,\mbox{ where }\bM\mbox{ is a symmetric positive definite matrix and }\bv\mbox{ is a vector};\label{eq:Kform2}\\
	& \Vert\bu-\bK\bt\Vert^2,\mbox{ where }\bu\mbox{ and }\bt\mbox{ are vectors};\label{eq:Kform3}\\
	& \tr(\bK^T\bK\bN),\mbox{ where }\bN\mbox{ is a matrix}\label{eq:Kform4}.
\end{align}
For what concerns VMP, the expression in \eqref{eq:Kform1} appears in the update for $\bOmega_1$ of Algorithm \ref{alg:VMPalgo}, whereas those in \eqref{eq:Kform2}--\eqref{eq:Kform4} arise in the Gaussian likelihood fragment numbered as fragment 5 in the Figure \ref{fig:InvProbFacGraph} factor graph.
Visual inspection of \eqref{eq:Kform1}--\eqref{eq:Kform4} suggests it is worth studying the structure of $\bK$ and $\bK^T\bK$ for a computationally efficient implementation of the variational algorithm updates.

The focus of this section is placed on two-dimensional inverse problems. Again, we restrict our attention to the case where $\bX$ and $\bY$ are both $m_1\times m_2$ matrices and each element of $\bX$ has a one-to-one correspondence with an element having the same position in $\bY$. Under these conditions $\bK$ is a square matrix of size $m\times m$, with $m=m_1m_2$. If model \eqref{eq:linInvProbModel} is used, setting $\bx=\vecof(\bX)$ and $\by=\vecof(\bY)$, the matrix $\bK$ has the following structure:
\begin{equation*}
\bK=\left[\begin{array}{cccc}
\bK_{1} & \bK_{2} & \cdots & \bK_{m_2} \\
\bK_{2} & \bK_1 & \ddots & \vdots \\
\vdots & \ddots & \ddots & \bK_2\\
\bK_{m_2} & \cdots & \bK_{2} & \bK_1 
\end{array}\right],\mbox{ with }
\bK_i=\left[\begin{array}{cccc}
K_{i,1} & K_{i,2} & \cdots & K_{i,m_1} \\
K_{i,2} & K_{i,1} & \ddots & \vdots \\
\vdots & \ddots & \ddots & K_{i,2}\\
K_{i,m_1} & \cdots & K_{i,2} & K_{i,1}
\end{array}\right],\, i=1,\ldots,m_2.
\end{equation*}
Therefore $\bK$ is a symmetric block-Toeplitz matrix with $m_2$ unique sub-blocks, each being $m_1\times m_1$ symmetric Toeplitz matrices. For simple unidimensional problems $\bK$ is a symmetric Toeplitz matrix.

\begin{figure}[!h]
	\centering
	{\includegraphics[width=0.8\textwidth]{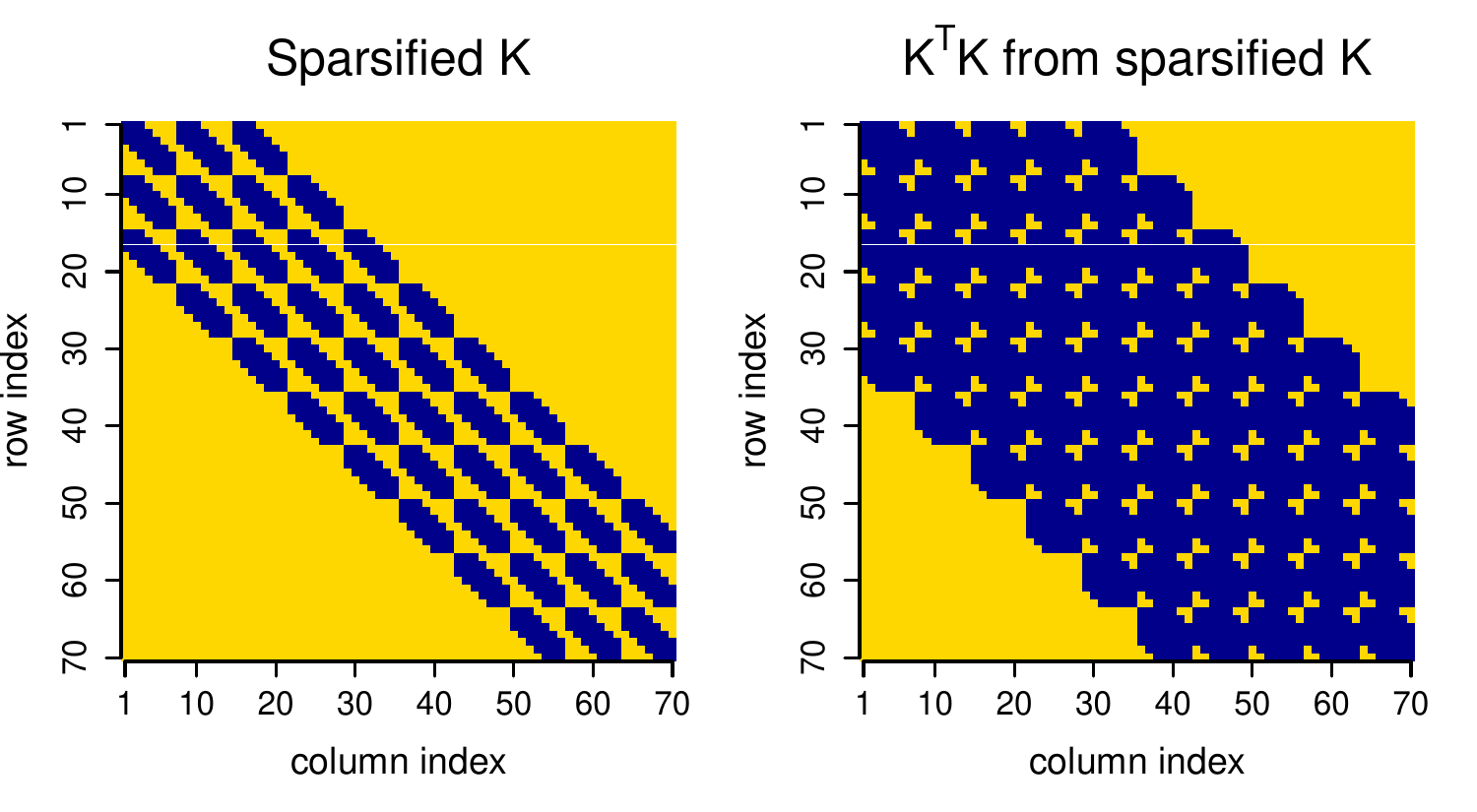}}
	\caption{\it Representation of a sparse $\bK$ matrix of size $70\times70$ for an inverse problem over a pixel grid of size $7\times10$ (left panel) and the corresponding $\bK^T\bK$ matrix (right panel). These matrices are obtained by using a truncation number $\ell=2$. Non-zero values are depicted in blue.}
	\label{fig:sparseKandKTK} 
\end{figure}

Both the MFVB and VMP updates
\begin{align}
&\Sigmaqx\longleftarrow\left(\mu_{q(1/\sigma^2_\varepsilon)}\bK^T\bK+\mu_{q(1/\sigma^2_x)}\bL^T\diag(\mu_{q(\bb)})\bL\right)^{-1}\,\,\mbox{of Algorithm \ref{alg:MFVBalgo} and}\label{eq:SigmaUpdate}\\
&\bOmega_1 \longleftarrow -\frac{1}{2}\left\{\vecof^{-1}\left(\left(\etaSUBpxbsigsqxFROMTOx\right)_2\right)\right\}^{-1}\,\,\mbox{of Algorithm \ref{alg:VMPalgo}}\label{eq:OmegaUpdate}
\end{align}
require inversion of a matrix of size $m\times m$. From \eqref{eq:SigmaUpdate} it is easy to notice that the structure of the matrix being inverted is influenced by $\bK$ through $\bK^T\bK$. 
A possible idea to reduce computational burden induced by these updates is to sparsify $\bK$ in such a way that also $\bK^T\bK$ and the final matrix to invert are sparse. 
Since $\bK$ linearly links elements in $\bX$ with those in $\bY$ and given the one-to-one correspondence between $\bX$ and $\bY$, it is reasonable to set to zero the matrix $\boldsymbol{K}$ elements which correspond to interactions between pairs of locations whose distance exceeds a certain \emph{truncation} value $\ell\in\mathbb{N}$, with $\ell<\min(m_1,m_2)$. This strategy can find application, for instance, both in the context of biomedical imaging discussed in Section \ref{sec:biomed} and archaeological field survey treated in the supplementary material. More formally, this consists in setting to zero the entries of $\bK$ that model dependence between pairs of elements of $\bX$ and $\bY$, $(X_{ij},Y_{i'j'})$, such that
\begin{equation*}
\max\left(\vert i-i'\vert,\vert j-j'\vert\right)>\ell.
\end{equation*}
The resultant $\bK$ is an $\ell$-block-banded matrix whose sub-blocks are $\ell$-banded matrices. Also $\bK^T\bK$ may result in a sparse matrix for particular choices of $\ell$, as stated in the following result. 
\begin{result}
	Let $\bA$ be an $\ell$-block-banded matrix of size $m\times m$ with $\ell$-banded sub-blocks and such that $0<\ell<(m-1)/2$. Then $\bA^T\bA$ is a symmetric $2\ell$-block-banded matrix whose sub-blocks are $2\ell$-banded matrices.
	\label{res:Ksparsity}
\end{result} 
Figure \ref{fig:sparseKandKTK} depicts the $\bK$ matrix and $\bK^T\bK$ block for an inverse problem where $\bX$ and $\bY$ have size $7\times 10$ and $\bK$ is sparsified by applying a truncation value $\ell=2$. The blue color indicates non-zero matrix entries. In this case, matrix $\bK$ is $2$-block-banded matrix with $2$-banded sub-blocks, whereas $\bK^T\bK$ is a $4$-block-banded matrix with $4$-banded sub-blocks.

It is easy to check through Lemma \ref{lem:lemmaC2d} by applying such a sparsification strategy to $\bK^T\bK$ that the same sparsity structure is imposed to the matrices being inverted in updates \eqref{eq:SigmaUpdate} and \eqref{eq:OmegaUpdate}. Hence, for appropriate choices of $\ell$, the updates involve inversion of a sparse matrix having a block-banded structure and banded matrices in the main block-diagonals.

The suggested sparsification strategy has a physical interpretation. In the two-dimensional problems under examination each element of $\bY$ linearly depends on a subset of the elements of $\bX$ through $\bK$. If the elements of $\bK$ are set to zero according to a truncation number $\ell$, then such subset is given by those elements of $\bX$ that fall inside of a circle of diameter $2\ell+1$ around the element of $\bX$ having one-to-one correspondence with the $\bY$ entry.
Sparsifying the expression to be inverted in \eqref{eq:SigmaUpdate} means setting to zero some elements of the precision matrix $\bOmega_{q(\bx)}=\Sigmaqx^{-1}$, where $\Sigmaqx$ is the covariance matrix of the optimal approximating density function $q^*(\bx)$ in \eqref{eq:qxDens}. Since $q^*(\bx)$ is a Multivariate Normal density function, $\big(\bOmega_{q(\bx)}\big)_{ij}=0$, for $i,j=1,\ldots,m$,  if and only if $x_i$ and $x_j$ are conditionally independent given all the other elements of $\bx$. 

Note that differently from the algebraic results proposed for removal of $\bL$, the sparsification applied to $\bK$ comes from nullifying interactions between elements of $\bX$ and $\bY$, and therefore it introduces another level of approximation to the variational fitting procedure.

\subsubsection{Block-banded matrix algebra}

\cite{asif2005block} propose two algorithms to invert block-banded matrices whose inverses are positive definite matrices: one resorting to Cholesky factors and an alternative implementation that avoids Cholesky factorizations. These algorithms require the inversion of smaller matrices having the size of the block-banded matrix sub-blocks. In the two-dimensional inverse problems under examination these sub-blocks have a banded structure. An approach for inverting banded matrices is described in \cite{kilicc2013inverse}. These algebraic approaches for handling block-banded and banded matrices may allow for stable computation of variational algorithm steps involving sparse matrix inversions such as \eqref{eq:Kform1}. However, the simplest way to perform efficient sparse matrix inversion and computations is to employ software for sparse matrix algebra. The functions contained in the \textsf{R} package \texttt{spam} \citep{furrer2010spam} allow efficient management of sparse matrices and implement matrix operations. This package can be used in combination with package \texttt{spam64} \cite{gerber2017extending} to speed up such functions in 64 bit machines. Well established software is also available for lower-level languages such as the linear algebra libraries Armadillo and Eigen for \textsf{C++} coding.

In general, matrices having banded or block-banded inverses are full matrices. Nonetheless, the inverse of a banded matrix may be referred to as \textit{band-dominated matrix} \citep{bickel2012approximating}, since the entries of its inverse exponentially decay with the distance from the main diagonal \citep[Theorem 2.4]{demko1984decay}. This property can be generalized to block-banded matrices and the inverse of a block-banded matrix can be approximated by a block-banded matrix with the same sparsity structure, i.e. with zero blocks off the main block band \citep{wijewardhana2016bayesian}.
The blocks outside the $\ell$-block band of a symmetric positive definite matrix having an $\ell$-block-banded inverse are called \textit{nonsignificant} blocks and those in the $\ell$-block band are called \textit{significant} blocks. Theorem 3 of \cite{asif2005block} states that nonsignificant blocks can be obtained from the significant ones. 
Then, a possible way to further speed-up algebra involving these matrices and reduce memory usage is to impose a block-banded structure to the precision matrix $\bOmega_{q(\bx)}$ and approximate $\Sigmaqx$ with a block-banded matrix having the same structure of $\bOmega_{q(\bx)}$. In this case only the significant blocks of the covariance matrix $\Sigmaqx$, which solely depend on the significant blocks of $\bOmega_{q(\bx)}$, need to be computed.

\section{Biomedical data study}\label{sec:biomed}

This section demonstrates the use of variational inference for two-dimensional inverse problems motivated by a biomedical application. An illustration on a real dataset and on simulations that mimic the real data are provided.

We assess the performances of variational inference through comparison with MCMC and computation of accuracy. 
For a generic univariate parameter $\theta$, the approximation accuracy of a density $q(\theta)$ to a posterior density $p(\theta\vert\by)$ is measured through
\begin{equation}
\mbox{accuracy}\equiv 100\left\{1-\frac{1}{2}\int_{-\infty}^{\infty}\big\vert q(\theta)-p(\theta\vert\by)\big\vert d\theta\right\}\%,
\label{eq:accExpr}
\end{equation}
so that $0\%\leq\mbox{accuracy}\leq 100\%$, with $100\%$ indicating perfect matching between the approximating and posterior density functions. 
We compute accuracy using Markov chain Monte Carlo as a benchmark. 
A standard Metropolis--Hastings algorithm \cite{metropolis1953equation,hastings1970monte} is used to produce approximate samples from the posterior distribution. The Markov chains are started at feasible points in the parameter space and the retained samples are used to approximate the corresponding posterior density functions via kernel density estimation. Accuracy is then obtained from \eqref{eq:accExpr} with replacement of $p(\theta\vert\by)$ by MCMC density estimates of the posterior density functions.
Variational inference is performed removing the contrast matrix $\bL$ through Lemmas \ref{lem:lemmaA2d}--\ref{lem:lemmaC2d} and sparsifying the matrix $\bK$ via truncation of interactions, as explained in Section \ref{sec:stremVarInf}. Also our MCMC implementation does not make direct use of matrices $\bL$ and $\bK$ to reduce computational burden. 
The simulation study was run on a personal computer with a 64 bit Windows 10 operating system, an Intel i7-7500U central processing unit at 2.7 gigahertz and 16 gigabytes of random access memory. Variational inference was fully performed in \textsf{R}, whereas MCMC was run in \textsf{R} with subroutines replacing $\bL$ and $\bK$ matrix operations implemented in \textsf{C++}.

\subsection{Real biomedical data}

We test the performance of our variational inference approach on a real biomedical application from the realm of tomographic data. 
Tomography aims to display cross-sections through human and animal bodies, or other solid objects, using data collected around the body. 

The data, kindly provided by BioEmission Technology Solutions, Athens, Greece, were collected to illustrate a small animal imaging system, gamma-eye, which can be used in biotechnology and pre-clinical medical research \citep{georgiou2017characterization}. A technetium radioisotope (99m Tc-MIBI) was injected via the tail vein of a mouse and mainly absorbed by organs such as heart, liver and kidneys, and then excreted. In humans such techniques are used to monitor heart function and mice are often used in pre-clinical studies.  A single plane gamma-camera image of an adult mouse was collected with the camera at a distance of 5mm from the nearest point of the mouse and 35mm from the support bed. The $29\times58$, pixels of the data image are 1.7mm apart giving a field of view of about 5cm by 10cm. The mouse was anesthetized and so this corresponds to the “at rest” part of a human scan which would also involve a “stress test”. The total data recording time was 3 hours.

\begin{figure}
	\centering
	{\includegraphics[width=1\textwidth]{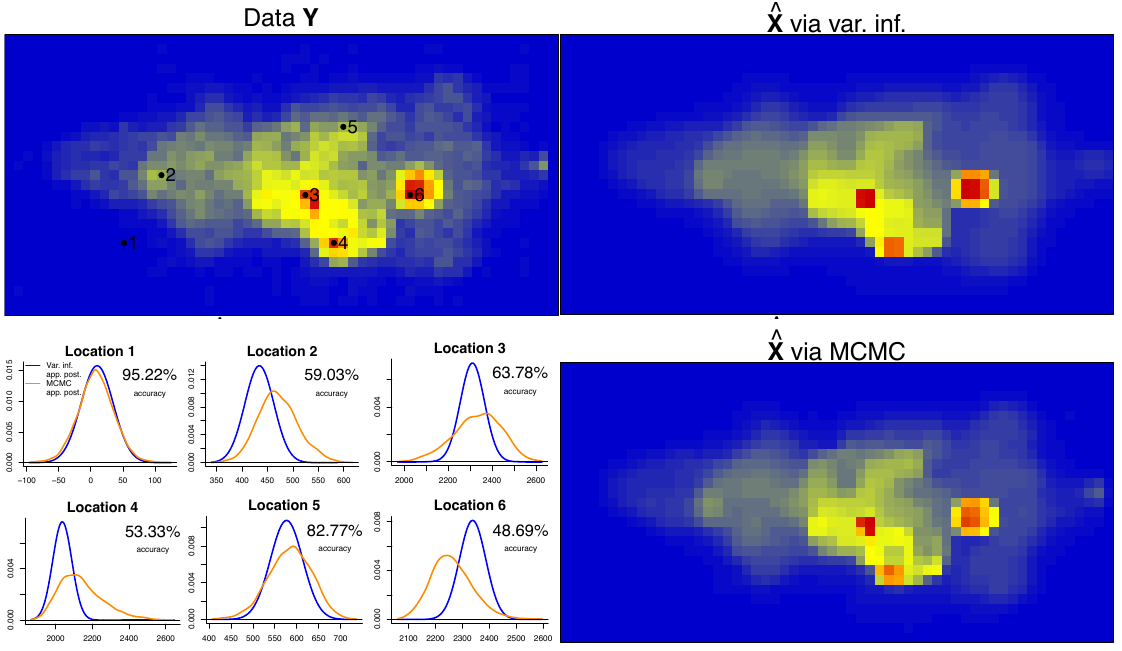}
	\caption{\it Analysis of the data provided by BioEmission Technology Solutions, Athens, Greece, performed by fitting model \eqref{eq:linInvProbModel} via variational inference and MCMC. The panel on the upper-left side displays the observed data $\bY$. The two panels on the left-hand side show, from top to bottom, the reconstruction $\widehat{\bX}$ of $\bX$ obtained via variational inference and MCMC. The plots on the lower-left side show the approximate marginal posterior densities produces through variational inference (blue lines) and MCMC (orange lines) for a selection of pixels.}
	\label{fig:mouseResults}}
\end{figure}

The objective is to reconstruct an image by removing blur from the observed scan of the mouse shown and denoted as $\bY$ in Figure \ref{fig:mouseResults}. We adopt model \eqref{eq:linInvProbModel} setting $\by=\vecof(\bY)$ and using a Gaussian kernel matrix $\bK$ obtained through \eqref{eq:Kbiomed} with $\delta=0.7$. Hence, $\by$ and $\bx$ are vectors of length equal to the number of pixels, 1,682, and $\bK$ has size $1,682\times 1,682$. We set to zero the elements of $\bK$ expressing interactions between locations that have 10 or more pixels between each other using $\ell=10$. 
The random walk Metropolis--Hastings algorithm is used to produce approximate samples from the posterior distribution by simulating a Markov chain with burn-in of 1000 followed by 100,000, then thinned by a factor of 20. Hyperparameters are set to values that give rise to diffuse priors. Specifically, $A_{\varepsilon}=A_x=10^{5}$. Estimates of $\bX$ obtained via variational inference and MCMC are included in Figure \ref{fig:mouseResults}. Here the term variational inference refers to both MFVB and VMP, as they both provide the same results by construction. The estimate of $\bX$ obtained via variational inference corresponds to the inverse vectorization of $\bmu_{q(\bx)}$ from \eqref{eq:qxDens}, whereas that of MCMC is given by the mean of the sampled chains. 

Figure \ref{fig:mouseResults} also displays approximate posterior density plots for a selection of six representative pixels.
Five of the six selected pixels correspond to targeted organs of the mouse, namely thyroid, liver, kidneys and bladder, while the remaining pixel is located outside but near the mouse body. Overall variational approximations provide good image reconstruction and facilitate visualisation of the mouse body shape and organs. The posterior density approximations are also satisfactory in terms of accuracy, that is, area overlap between MCMC. As typical of variational approximations based on mean field restrictions, variational inference underestimates the variance of the approximate posterior densities. Plots of some representative bivariate posterior densities are also provided in the supplement. These show that in all cases the variational approximation covers a smaller area, indicating that it underestimates uncertainty relative to MCMC.

\subsection{Simulated biomedical data}

We employ the real biomedical data image processed through MCMC, the one corresponding to the lower-right panel of Figure \ref{fig:mouseResults}, to simulate datasets and study the performance of variational inference in comparison with MCMC. In this simulation study we also keep track of computational times and calculate percentages of coverage. For a given parameter, the percentage of coverage corresponds to the proportion of simulations where the true parameter falls inside its 95\% credible interval obtained through variational inference.
 
Let $\widehat{\bX}_{\tiny\mbox{MCMC}}$ be the inverse vectorization of the estimate of $\bx$ obtained as the sample mean of the corresponding MCMC chains.
We simulate data through:
\begin{equation*}
\by=\bK\vecof(\widehat{\bX}_{\tiny\mbox{MCMC}})+\bvarepsilon,\quad\bvarepsilon\sim N(\bzero,\sigma^2_{\varepsilon}\bI),
\end{equation*}
with $\sigma_{\varepsilon}=50$ and $\bK$ generated according to \eqref{eq:Kbiomed}, using $\delta$ values from the set $(0.7,0.8,0.9)$ and without truncating interactions ($\ell=\infty$). The example plots provided in the supplement show how the blur increases for higher $\delta$. For each $\delta$ value we generate 100 datasets and fit model \eqref{eq:linInvProbModel} via both MFVB and MCMC. For fitting we apply the same $\bK$ matrix used to generate the datasets, i.e. without truncation of interactions, but also fit the model again setting to zero the elements of $\bK$ associated with pairs of observations whose distance exceeds a truncation value $\ell=5$.
The MFVB algorithm is stopped when the relative difference of estimates of $\bX$ goes below $10^{-2}$ between two iterations. We  generate Markov chains of length 6,000 and retain 5,000 of them after discarding 1,000 warm-up samples. 

Table \ref{tab:SimStudRes} summarizes the results of the simulation study including: accuracy of the variational inference estimates of $\bX$ versus MCMC; variational inference percentage of coverage for $\bX$, i.e. the number of times the entries of $\widehat{\bX}_{\tiny\mbox{MCMC}}$ falls inside their 95\% variational inference credible intervals; variational inference and MCMC computational times. Average and standard deviations (in brackets) of each indicator are displayed for each combination of $\delta$ and $\ell$ values. 

For each pixel of each simulated dataset we compute the accuracy of the variational approximation using \eqref{eq:accExpr} and then average over the 100 replicates. We then calculate average and standard deviation over all the entries of $\bX$ and display these in Table \ref{tab:SimStudRes}. We repeat the same procedure for measuring the percentage of coverage performances. The mean accuracy values range between 88.07 and 87.10, whereas the mean percentages of coverage are between 95.09 and 92.75 and therefore close to the nominal 95\% level. Both accuracy and coverage performances slightly degrade for higher values of $\delta$ and blur. 

The variational inference and MCMC computational times are displayed in minutes and show that variational inference is around 100 times faster for the $\delta=0.7$ setting. Imposing a truncation $\ell=5$ to $\bK$ reduces the MCMC computational times by about six to seven minutes on average for the three $\delta$ values under examination as opposed to not applying truncation. The time performances of variational inference are not particularly affected by truncation as the \textsf{R} package \texttt{spam} efficiently manages the algorithm updates involving $\bK$ in both cases with and without truncation. Once again, variational inference has been fully performed in \textsf{R}, while the main MCMC function has been implemented in \textsf{R} and uses \textsf{C++} subroutines that replace the $\bK$ and $\bL$ matrix operations. Therefore, further computational advantages may be achieved implementing variational inference in \textsf{C++}. 

\begin{table}[!h]
	\centering
	\footnotesize
	\begin{tabular}{l l c c }
        \Xhline{3\arrayrulewidth}
		& & $\ell=\infty$ (no truncation) & $\ell=5$ \\
		\Xhline{3\arrayrulewidth}
		\multirow{4}{*}{$\delta=0.7$} 
		& Var. inf. vs MCMC accuracy for $\bX$ & 88.07 (4.23) & 88.04 (4.23) \\
		& Var. inf. percentage of coverage for $\bX$ & 95.09 (16.36) & 95.09 (16.36)\\
		& Var. inf. comp. time (minutes) & 0.85 (0.24) & 0.84 (0.24) \\
		& MCMC comp. time (minutes) & 87.67 (0.29) & 81.61 (0.39) \\
		\hline
		\multirow{4}{*}{$\delta=0.8$} 
		& Var. inf. vs MCMC accuracy for $\bX$ & 87.54 (5.17) & 87.55 (5.20) \\ 
		& Var. inf. percentage of coverage for $\bX$ & 94.01 (19.12) & 94.01 (19.12) \\
		& Var. inf. comp. time (minutes) & 1.20 (0.28) & 1.19 (0.28) \\
		& MCMC comp. time (minutes) & 88.08 (0.57) & 82.12 (0.58) \\
		\hline
		\multirow{4}{*}{$\delta=0.9$} 
		& Var. inf. vs MCMC accuracy for $\bX$ & 87.10 (5.58) & 87.12 (5.65) \\
		& Var. inf. percentage of coverage for $\bX$ & 92.75 (22.02) & 92.75 (22.02) \\
		& Var. inf. comp. time (minutes) & 1.43 (0.43) & 1.43 (0.43) \\
		& MCMC comp. time (minutes) & 88.64 (0.30) & 81.70 (0.44) \\
		\hline
	\end{tabular}
	\caption{Results of the simulation study based on 100 replicates per $\delta$ value generated using the kernel matrix $\bK$ defined in \eqref{eq:Kbiomed} without truncation. Fitting is performed via variational inference and MCMC using the same $\bK$ matrix employed to generate the data (no truncation) and also imposing a truncation threshold $\ell=5$ to the interactions expressed by $\bK$.  Higher $\delta$ values correspond to more blur. The smaller $\ell$, the sparser the $\bK$ matrix used to fit the model is.}
	\label{tab:SimStudRes}
\end{table}

\section{Discussion \label{sec:discussion}}

We have laid down the infrastructure for performing variational approximate inference on applications that can be studied through statistical inverse problems models. Our variational algorithms allow fast approximate fitting for these models, with satisfactory accuracy compared with standard MCMC. 

The run-time of MCMC estimation for inverse problems is usually excessive on a standard personal computer. This means that parameter tuning, model diagnostics and sensitivity analyses are rarely performed. The use of variational inference methods, which are quick by comparison, means that multiple parameter settings and multiple models can be considered in a reasonable length of time. This opens-up the possibility of model diagnostics, such as influence and leverage, to become a routine part of applied inverse problems solution. Further, there is a subjective element, as always, to the choice of model components and in particular the hierarchical prior components. It would be a great advance for applications areas, such as medicine and archaeology, if rogue measurements could be identified and their influence on estimation could be quantified. Also, if any arbitrary parts could be shown to have insignificant effect on results, this would lead to far greater confidence and hence a wider acceptability of advanced statistical modelling approaches. 

Hence, the implications of our work are not limited to the numerical results presented, but we provide a framework for other researchers to develop a richer set of model exploration methods for inverse problems. The flexibility of our approach is such that non-Normal likelihood distributions and other penalizations can be incorporated at will. Furthermore, this research sets the basis for several future directions to explore such as the study of settings with more than two dimensions, number of observations not matching that of data recording locations or more complex neighbor dependence structures.

\ifthenelse{\boolean{UnBlinded}}{
	
\section*{Acknowledgments}

We are grateful to Alan Welsh for advice related to this research. This research was supported by the Australian Research Council Discovery Project DP180100597 and the Australian Research Council Centre of Excellence for Mathematical and Statistical Frontiers.

}{}

\bibliographystyle{apalike}
\bibliography{ref}

\vfill\eject

%
%
\renewcommand{\theequation}{S.\arabic{equation}}
\renewcommand{\thesection}{S.\arabic{section}}
\renewcommand{\thetable}{S.\arabic{table}}
\renewcommand{\thefigure}{S.\arabic{figure}}
\setcounter{equation}{0}
\setcounter{table}{0}
\setcounter{section}{0}
\setcounter{page}{1}
\setcounter{footnote}{0}
\setcounter{figure}{0}

\begin{center}

{\Large Supplement for:}
\vskip3mm

\centerline{\Large\bf A Variational inference framework for inverse problems}
\vskip7mm
\ifthenelse{\boolean{UnBlinded}}{
\centerline{\normalsize\sc By L. Maestrini$\null^1$, R.G. Aykroyd$\null^2$ and M.P. Wand$\null^3$}
\vskip5mm
\centerline{\textit{$\null^1$Australian National University, $\null^2$University of Leeds and $\null^3$University of Technology Sydney}}}{}
\vskip20mm

\end{center}

\section{Algorithm derivations}

This section provides full justification of Algorithms \ref{alg:MFVBalgo} and \ref{alg:VMPalgo}.

\subsection{Derivation of Algorithm \ref{alg:MFVBalgo}}\label{sec:MFVBalgoDerivn}

The optimal approximating densities satisfy expression \eqref{eq:qkupdate}.
The full conditional density functions of each hidden node in the directed acyclic graph in Figure \ref{fig:InvProbDAG} allow to derive expressions for the optimal approximating densities.
First,
\begin{align*}
p\left(\bx\vert\mbox{rest}\right)&\propto p\left(\by\vert\bx,\sigma^2_\varepsilon\right)p\left(\bx\vert\bb,\sigma^2_x\right)\\
&\propto\exp\left[-\frac{1}{2}\left\{\bx^T\bL^T\left(\frac{1}{\sigma^2_\varepsilon}\bK^T\bK+\frac{1}{\sigma^2_x}\diag\left(\bb\right)\right)\bL\bx-\frac{2}{\sigma^2_\varepsilon}\bx^T\bK^T\by\right\}\right],
\end{align*}
that is,
\begin{equation*}
\bx\vert\mbox{rest}\sim N\left(\left(\frac{1}{\sigma^2_\varepsilon}\bK^T\bK+\frac{1}{\sigma^2_x}\bL^T\diag\left(\bb\right)\bL\right)^{-1}\frac{1}{\sigma^2_\varepsilon}\bK^T\by,\left(\frac{1}{\sigma^2_{\varepsilon}}\bK^T\bK+\frac{1}{\sigma^2_x}\bL^T\diag\left(b\right)\bL\right)^{-1}\right).
\end{equation*}
Hence $q^*(\bx)$ is $N\big(\muqx,\Sigmaqx\big)$ with
\begin{equation*}
\muqx\equiv\mu_{q(1/\sigma_{\varepsilon}^2)}\Sigmaqx\bK^T\by\,\,\,\mbox{and}\,\,\,\Sigmaqx\equiv\left(\mu_{q(1/\sigma^2_\varepsilon)}\bK^T\bK+\mu_{q(1/\sigma^2_x)}\bL^T\diag(\mu_{q(\bb)})\bL\right)^{-1}.
\end{equation*}

Given that
\begin{equation}
p\left(\bb\right)\propto \prod_{j=1}^{d} b_j^{-2}\exp\left\{-1/(2b_j)\right\},
\label{eq:pbDensFunc}
\end{equation}
then, for $1\leq j\leq d$,
\begin{equation*}
p\left(b_j\vert\mbox{rest}\right)\propto p\left(\bx\vert b_j,\sigma^2_x\right)p\left(b_j\right)\propto b_j^{-3/2}\exp\left[-\frac{1}{2}\left\{\frac{b_j\left(\bL\bx\right)_j^2}{\sigma^2_x} + \frac{1}{b_j}\right\}\right]
\end{equation*}
and so
\begin{equation*}
b_j\vert\mbox{rest}\sim \mbox{Inverse-Gaussian}\Big(\sqrt{\sigma_x^2\big/\left(\bL\bx\right)_j^2},1\Big).
\end{equation*}
It follows that $q^*\left(b_j\right)$ is $\mbox{Inverse-Gaussian}\big(\mu_{q(b_j)},\lambda_{q(b_j)}\big)$, with
\begin{equation*}
\mu_{q(b_j)}\equiv\left[\muqrecipsigsqx\left\{\big(\bL\muqx\big)^2_j+\big(\diagonal(\bL\Sigmaqx\bL^T)\big)_j\right\} \right]^{-1/2}\,\,\,\mbox{and}\,\,\,\lambda_{q(b_j)}\equiv 1.
\end{equation*}

Next,
\begin{equation*}
p\left(\sigma^2_\varepsilon\vert\mbox{rest}\right)\propto p\left(\by\vert\bx,\sigma^2_\varepsilon\right)p\left(\sigma^2_\varepsilon\vert a_\varepsilon\right)\propto \left(\sigma_\varepsilon^2\right)^{-\frac{m+1}{2}-1}\exp\left\{-\frac{1}{2\sigma^2_\varepsilon}\left(\frac{1}{a_\varepsilon}+\Vert\by-\bK\bx\Vert^2\right)\right\},
\end{equation*}
that provides
\begin{equation*}
\sigma^2_\varepsilon\vert\mbox{rest}\sim\mbox{Inverse-}\chi^2\left(m+1,1/a_\varepsilon+\Vert\by-\bK\bx\Vert^2\right).
\end{equation*}
Then $q^*\left(\sigma^2_\varepsilon\right)$ is Inverse-$\chi^2$$\left(\kappaqsigsqepsilon,\lambdaqsigsqepsilon\right)$ with
\begin{equation*}
\kappaqsigsqepsilon\equiv m+1\,\,\,\mbox{and}\,\,\,
\lambdaqsigsqepsilon\equiv\mu_{q(1/a_\varepsilon)}+\Vert\by-\bK\muqx\Vert^2+\tr\left(\bK^T\bK\Sigmaqx\right),
\end{equation*}
which provide $\muqrecipsigsqepsilon\equiv\kappaqsigsqepsilon/\lambdaqsigsqepsilon$.

Also,
\begin{equation*}
p\left(\sigma^2_x\vert\mbox{rest}\right)\propto p\left(\bx\vert\bb,\sigma^2_x\right)p\left(\sigma^2_x\vert a_x\right)\propto \left(\sigma_x\right)^{-\frac{d+1}{2}-1}\exp\left\{-\frac{1}{2\sigma^2_x}\left(\frac{1}{a_x}+\bx^T\bL^T\diag\left(\bb\right)\bL\bx\right)\right\},
\end{equation*}
implying that
\begin{equation*}
\sigma^2_x\vert\mbox{rest}\sim\mbox{Inverse-}\chi^2\Bigg(d+1,1/a_x+\sum_{j=1}^d b_j \left(\bL\bx\right)^2_j\Bigg).
\end{equation*}
This leads to $q^*\left(\sigma^2_x\right)$ being Inverse-$\chi^2\left(\kappaqsigsqx,\lambdaqsigsqx\right)$ with
\begin{equation*}
\kappaqsigsqx\equiv d+1\,\,\,\mbox{and}\,\,\,
\lambdaqsigsqx\equiv\mu_{q(1/a_x)}+\sum_{j=1}^{d}\mu_{q\left(b_j\right)}\left\{\big(\bL\muqx\big)^2_j+\big(\diagonal(\bL\Sigmaqx\bL^T)\big)_j\right\}.
\end{equation*}
The moment expression $\muqrecipsigsqx\equiv\kappaqsigsqx/\lambdaqsigsqx$ follows.

As regards auxiliary variable $a_\varepsilon$,
\begin{equation*}
p(a_\varepsilon\vert\mbox{rest})\propto p\left(\sigma^2_\varepsilon\vert a_\varepsilon\right)p\left(a_\varepsilon\right)\propto\exp\left\{-2\log(\alpha_\varepsilon)-\frac{1}{2\alpha_\varepsilon}\left(\frac{1}{\sigma^2_\varepsilon}+\frac{1}{A^2_\varepsilon}\right)\right\},
\end{equation*}
which implies
\begin{equation*}
a_\varepsilon\vert\mbox{rest}\sim\mbox{Inverse-}\chi^2\left(2,1/\sigma^2_\varepsilon+1/A_\varepsilon^2\right).
\end{equation*}
It follows that $q^*(a_\varepsilon)$ is Inverse-$\chi^2\left(\kappaqaepsilon,\lambdaqaepsilon\right)$ with 
\begin{equation*}
\kappa_{q(a_\varepsilon)}\equiv2\,\,\,\mbox{and}\,\,\,\lambda_{q(a_\varepsilon)}\equiv\mu_{q(1/\sigma^2_\varepsilon)}+1/A_\varepsilon^2.
\end{equation*}
This gives $\mu_{q(1/a_\varepsilon)}\equiv2/\lambda_{q(a_x)}$.

Analogously to $a_\varepsilon$, $q^*(a_x)$ is Inverse-$\chi^2\left(\kappaqax,\lambdaqax\right)$ with 
\begin{equation*}
\kappa_{q(a_x)}\equiv 2\,\,\,\mbox{and}\,\,\,\lambda_{q(a_x)}\equiv\mu_{q(1/\sigma^2_x)}+1/A_x^2.
\end{equation*}
Expression $\mu_{q(1/a_x)}\equiv2/\lambda_{q(a_x)}$ follows. 

\subsection{Derivation of Algorithm \ref{alg:VMPalgo}}\label{sec:VMPalgoDerivn}

Consider expression \eqref{eq:xLogLik} for $\log p\left(\bx\vert\bb,\sigma^2_x\right)$ as a function of the single components $\bx$, $\sigma^2_x$ and $\bb$.
As a function of $\bx$, we get
\begin{equation*}
\log p\left(\bx\vert\bb,\sigma^2_x\right) = \frac{1}{\sigma_{x}^{2}}\left[\begin{array}{c}
\bx\\
\vecof\left(\bx\bx^{T}\right)
\end{array}\right]^{T}\left[\begin{array}{c}
\boldsymbol{0}\\
-\frac{1}{2}\vecof\left(\bL^{T}\diag\left(\bb\right)\bL\right)
\end{array}\right] + \mbox{const}. 
\end{equation*}
Hence
\begin{equation*}
\mSUBpxbsigsqxTOx = \exp\left\{\left[\begin{array}{c}
\bx\\
\vecof\left(\bx\bx^{T}\right)
\end{array}\right]^{T} \etaSUBpxbsigsqxTOx\right\},
\end{equation*}
which is within the Multivariate Normal family, with
\begin{equation*}
\etaSUBpxbsigsqxTOx \equiv E_{\boxtimes}\left(1/\sigma_{x}^{2}\right)\left[\begin{array}{c}
\boldsymbol{0}\\
-\frac{1}{2}\vecof\left(\bL^{T}\diag\left\{E_{\oplus}\left(\bb\right)\right\}\bL\right)
\end{array}\right],
\end{equation*}
where $E_{\boxtimes}$ denotes expectation with respect to the normalization of
\begin{equation*}
\mSUBpxbsigsqxTOsigsqx \mSUBsigsqxTOpxbsigsqx
\end{equation*}
and $E_{\oplus}$ denotes expectation with respect to the normalization of
\begin{equation*}
\mSUBpxbsigsqxTOb \mSUBbTOpxbsigsqx .
\end{equation*}

As a function of $\sigma^2_x$,
\begin{equation*}
\log p\left(x\vert\bb,\sigma^2_x\right) = \left[\begin{array}{c}
\log\left(\sigma^2_x\right)\\
1/\sigma^2_x
\end{array}\right]^{T}\left[\begin{array}{c}
-d/2\\
-\frac{1}{2}\bx^T\bL^{T}\diag\left(\bb\right)\bL\bx
\end{array}\right] + \mbox{const}. 
\end{equation*}
It follows that
\begin{equation*}
\mSUBpxbsigsqxTOsigsqx = \exp\left\{\left[\begin{array}{c}
\log\left(\sigma^2_x\right)\\
1/\sigma^2_x
\end{array}\right]^{T} \etaSUBpxbsigsqxTOsigsqx\right\},
\end{equation*}
which is within the Inverse Chi-Squared family, where
\begin{equation*}
\etaSUBpxbsigsqxTOsigsqx \equiv \left[\begin{array}{c}
-d/2\\
-\frac{1}{2}E_{\otimes}\left[\bx^T\bL^{T}\diag\left\{E_{\oplus}\left(\bb\right)\right\}\bL\bx\right]
\end{array}\right],
\end{equation*}
with $E_{\otimes}$ denoting expectation with respect to the normalization of
\begin{equation*}
\mSUBpxbsigsqxTOx \mSUBxTOpxbsigsqx .
\end{equation*}

As a function of $\bb$,
\begin{equation*}
\log p\left(\bx\vert\bb,\sigma^2_x\right) = \frac{1}{2} \sum_{j=1}^{d} \left\{\log\left(b_j\right) - \left(\bL\bx\right)_j^2 b_j/\sigma^2_x \right\} + \mbox{const}.
\end{equation*}
Therefore
\begin{equation*}
\mSUBpxbsigsqxTOb \propto \prod_{j=1}^{d} b_j^{1/2}\exp\left[-\frac{1}{2}E_{\boxtimes}\left(1/\sigma_{x}^{2}\right)E_{\otimes}\left\{ \left(\bL\bx\right)_{j}^{2}\right\}\right].
\end{equation*}
It is simple to show that
\begin{equation}
\mSUBpbTOb\propto p\left(\bb\right)\propto \prod_{j=1}^{d} b_j^{-2}\exp\left\{-1/(2b_j)\right\}.
\label{eq:pbTObmessage}
\end{equation}
From \eqref{eq:nodeToFacUpdate},
\begin{equation*}
\mSUBbTOpxbsigsqx = \mSUBpbTOb
\end{equation*}
and so
\begin{align*}
&\mSUBpxbsigsqxTOb \mSUBbTOpxbsigsqx \\
&\qquad\qquad\qquad= \prod_{j=1}^{d} b_j ^{-3/2}\exp\left\{ \left[\begin{array}{c}
b_{j}\\
1/b_{j}
\end{array}\right]^{T}\left[\begin{array}{c}
-\frac{1}{2}E_{\boxtimes}\left(1/\sigma_{x}^{2}\right)E_{\otimes}\left\{ \left(\bL\bx\right)_{j}^{2}\right\}\\
-1/2
\end{array}\right] \right\},
\end{align*}
which is a product of Inverse Gaussian density functions with natural parameter vector
\begin{equation*}
\left[-\frac{1}{2}E_{\boxtimes}\left(1/\sigma_{x}^{2}\right)E_{\otimes}\left\{ \left(\bL\bx\right)_{j}^{2}\right\},\quad -\frac{1}{2}\right]^T_{1\leq j\leq d}.
\end{equation*}
It follows from Table S.1 of the supplementary material of \cite{wand2017fast} that
\begin{equation*}
E_{\odot}\left(\bb \right) = \left[\left[E_{\boxtimes}\left(1/\sigma_{x}^{2}\right)E_{\otimes}\left\{ \left(\bL\bx\right)_{j}^{2}\right\} \right]^{-1/2}\right]_{1\leq j\leq d}.
\end{equation*}
Again from Table S.1 of of \cite{wand2017fast}, since $E_{\boxtimes}$ corresponds to the expectation of an Inverse Chi-Squared distribution with natural parameter vector $\etaSUBpxbsigsqxFROMTOsigsqx$,
\begin{equation*}
E_{\boxtimes} \left(1/\sigma^2_x\right)= \Big\{\left(\etaSUBpxbsigsqxFROMTOsigsqx\right)_1 + 1\Big\}\Big/\Big\{\left(\etaSUBpxbsigsqxFROMTOsigsqx\right)_2\Big\}.
\end{equation*}
Next,
\begin{equation*}
E_{\otimes}\left\{\left(\bL\bx\right)_j^2\right\} = Var_{\otimes} \left\{\left(\bL\bx\right)_j\right\} + \left[E_{\otimes}\left\{\left(\bL\bx\right)_j\right\}\right]^2.
\end{equation*}
Making use of Table S.1 of \cite{wand2017fast}, we get
\begin{align*}
\left[Var_{\otimes} \left\{\left(\bL\bx\right)_j\right\}\right]_{1\leq j \leq d} &= -\frac{1}{2}\diagonal\left[\bL \left\{\vecof^{-1}\left(\left(\etaSUBpxbsigsqxFROMTOx\right)_2\right)\right\}^{-1}\bL^T\right]\\
\left[\left[E_{\otimes}\left\{\left(\bL\bx\right)_j\right\}\right]^2\right]_{1\leq j \leq d} &= \frac{1}{4}\left[\bL \left\{\vecof^{-1}\left(\left(\etaSUBpxbsigsqxFROMTOx\right)_2\right)\right\}^{-1}\left(\etaSUBpxbsigsqxFROMTOx\right)_1\right]^2
\end{align*}
and
\begin{equation*}
E_{\otimes}\left\{\left(\bL\bx\right)_j^2\right\}=\bomega_3\odot\bomega_3+\bomega_4,
\end{equation*}
where the expressions for $\bomega_3$ and $\bomega_4$ are provided in Algorithm \ref{alg:VMPalgo}. 
Similarly,
\begin{align*}
E_{\otimes}\left[\bx^T\bL^T\diag\left\{E_{\oplus}\left(\bb\right)\right\}\bL\bx\right]=& E_{\oplus}\left(\bb\right)^T\Big[\left\{\bL E_{\oplus}\left(\bx\right)\right\} \odot \left\{\bL E_{\oplus}\left(\bx\right)\right\}\\
&+\diagonal\left\{\bL Cov_{\otimes}\left(\bx\right) \bL^T \right\}\Big]=\mu_{q(\bb)}^T\left(\bomega_3^2+\bomega_4\right).
\end{align*}
The expressions for $\etaSUBpxbsigsqxTOx$ and $\etaSUBpxbsigsqxTOsigsqx$ outputted by Algorithm \ref{alg:VMPalgo} follow.

\subsection{Introducing alternative penalizations}

One of the continuous distributions for sparse signal shrinkage of \cite{neville2014mean} can be used in lieu of the Laplace penalization. This entails replacement of \eqref{eq:pbDensFunc} for MFVB and \eqref{eq:pbTObmessage} for VMP with one of the following:
\begin{equation*}
\begin{array}{ll}
\mSUBpbTOb\propto\prod_{j=1}^{d} b_j^{-1/2}\left(1+b_j\right)^{-1} & \mbox{(Horseshoe),}\\[2ex]
\mSUBpbTOb\propto\prod_{j=1}^{d} b_j^{\lambda-1}\left(1+b_j\right)^{-1-\lambda} & \mbox{(Negative-Exponential-Gamma),}\\[2ex]
\mSUBpbTOb\propto\prod_{j=1}^{d} b_j^{\left(\lambda-2\right)/2}e^{\lambda^2 b_j/4}\mathcal{D}_{-\lambda-2}\left(\lambda\sqrt{b_j}\right) & \mbox{(Generalized-Double-Pareto).}
\end{array}
\end{equation*}
The adjustments in Algorithms \ref{alg:MFVBalgo} and \ref{alg:VMPalgo} that allow for inclusion of such penalizations are summarized in Table \ref{tab:bDistrib}.

\section{Implementation of variational message passing}

This section demonstrates how to fully implement VMP on the base model, making use of Algorithm $\ref{alg:VMPalgo}$ and other relevant VMP algorithms for fragments that have already been studied in previous works.

The VMP approach to fit model \eqref{eq:linInvProbModel} under restriction \eqref{eq:mfRest} takes a response vector $\by$ of length $m$ and a $\bK$ matrix of size $(m\times m)$ as data inputs, and $A_\varepsilon,A_x>0$ as hyperparameter inputs. At convergence, VMP provides the optimal posterior density function approximations \eqref{eq:qxDens}--\eqref{eq:qaxDens}.

\subsection{Initialization}

The message natural parameters arising from the factor graph in Figure \ref{fig:InvProbFacGraph} have to be initialised at feasible points in the parameter space.

The natural parameter vector $\etaSUBpaxTOax$ can be initialized through the Inverse G-Wishart Prior Fragment \cite[Algorithm 1]{maestrini2021inverse} with inputs:
\begin{equation*}
G_{\Theta}=G_{\tiny\mbox{diag}},\,\,\,\xi_{\Theta}=1,\,\,\,\mbox{and}\,\,\,\bLambda_{\Theta}=A_x^{-2}.
\end{equation*}
The algorithm also provides the graph $\GSUBpaxTOax$ as an output. An analogous call to the same algorithm provides $\etaSUBpaepsilonTOaepsilon$ and $\GSUBpaepsilonTOaepsilon$. 
The remaining factor to stochastic node message natural parameters can be initialized, for example, as follows:
\begin{equation*}
\begin{array}{c}
	\etaSUBpsigsqxaxTOsigsqx\longleftarrow\left[\begin{array}{c}
	-3/2\\
	-1
	\end{array}\right],\,\,\, \etaSUBpsigsqxaxTOax\longleftarrow\left[\begin{array}{c}
	-3/2\\
	-1
	\end{array}\right],\\[2.5ex]
	\etaSUBpxbsigsqxTOx\longleftarrow\left[\begin{array}{c}
	\bzero_m\\
	-\frac{1}{2}\vecof(\bI_m)
	\end{array}\right],\,\,\,
	\etaSUBpxbsigsqxTOsigsqx\longleftarrow\left[\begin{array}{c}
	-3/2\\
	-1
	\end{array}\right],\\[2.5ex]
	\etaSUBpyxsigsqepsilonTOx\longleftarrow\left[\begin{array}{c}
	\bzero_m\\
	-\frac{1}{2}\vecof(\bI_m)
	\end{array}\right],\,\,\, \etaSUBpyxsigsqepsilonTOsigsqepsilon\longleftarrow\left[\begin{array}{c}
	-3/2\\
	-1
	\end{array}\right],\\[2.5ex]
	\etaSUBpsigsqepsilonaepsilonTOsigsqepsilon\longleftarrow\left[\begin{array}{c}
	-3/2\\
	-1
	\end{array}\right],\,\,\, \etaSUBpsigsqepsilonaepsilonTOaepsilon\longleftarrow\left[\begin{array}{c}
	-3/2\\
	-1
	\end{array}\right].
    \end{array}
\end{equation*}

One way to initialize the stochastic node to factor message natural parameters is the following:
\begin{equation*}
\begin{array}{c}
\etaSUBaxTOpsigsqxax\longleftarrow\etaSUBpaxTOax,\,\,\,\etaSUBaepsilonTOpsigsqepsilonaepsilon\longleftarrow\etaSUBpaepsilonTOaepsilon,\\[2.5ex] \etaSUBsigsqepsilonTOpyxsigsqepsilon\longleftarrow\etaSUBpsigsqepsilonaepsilonTOsigsqepsilon,\\[2.5ex]
\etaSUBsigsqepsilonTOpsigsqepsilonaepsilon\longleftarrow\left[\begin{array}{c}
-3/2\\
-1
\end{array}\right],\,\,\,
\etaSUBsigsqxTOpsigsqxax\longleftarrow\left[\begin{array}{c}
-3/2\\
-1
\end{array}\right],\\[2.5ex]
\etaSUBxTOpxbsigsqx\longleftarrow\left[\begin{array}{c}
\bK^T\by\\
-\frac{1}{2}\vecof(\bK^T\bK)
\end{array}\right],\,\,\,\etaSUBxTOpyxsigsqepsilon\longleftarrow\left[\begin{array}{c}
\bzero_m\\
-\frac{1}{2}\vecof(\bI_m)
\end{array}\right].
\end{array}
\end{equation*}

\subsection{Variational message passing iterations}

Once the natural parameter vector initializations are carried out, the stochastic node to factor and factor to stochastic node message parameters are updated in cycle until convergence. A possible way to assess convergence is monitoring the relative difference of parameter estimates from subsequent iterations. The updates for factor to stochastic node message parameters are performed via Algorithm \ref{alg:VMPalgo} and other VMP schemes proposed in the existing literature.

\subsubsection{Stochastic node to factor message parameter updates}

The stochastic node to factor message updates follow from \eqref{eq:nodeToFacUpdate}. For the factor graph of Figure \eqref{fig:InvProbFacGraph} these updates are:

\begin{equation*}
\begin{array}{c}	
	\etaSUBxTOpyxsigsqepsilon\longleftarrow\etaSUBpxbsigsqxTOx,\,\,\,	\etaSUBxTOpxbsigsqx\longleftarrow\etaSUBpyxsigsqepsilonTOx, \\
	\etaSUBsigsqepsilonTOpyxsigsqepsilon\longleftarrow\etaSUBpsigsqepsilonaepsilonTOsigsqepsilon,\,\,\,\etaSUBsigsqepsilonTOpsigsqepsilonaepsilon\longleftarrow\etaSUBpyxsigsqepsilonTOsigsqepsilon,\\
	\GSUBsigsqepsilonTOpsigsqepsilonaepsilon\longleftarrow G_{\tiny\mbox{full}},\,\,\,\etaSUBsigsqxTOpxbsigsqx\longleftarrow\etaSUBpsigsqxaxTOsigsqx,\\
	\GSUBsigsqxTOpsigsqxax\longleftarrow G_{\tiny\mbox{full}},\,\,\, \etaSUBsigsqxTOpsigsqxax\longleftarrow\etaSUBpxbsigsqxTOsigsqx,\\
	\GSUBaepsilonTOpsigsqepsilonaepsilon\longleftarrow\GSUBpaepsilonTOaepsilon,\,\,\,\etaSUBaepsilonTOpsigsqepsilonaepsilon\longleftarrow\etaSUBpaepsilonTOaepsilon,\\
	\GSUBaxTOpsigsqxax\longleftarrow\GSUBpaxTOax,\,\,\,\etaSUBaxTOpsigsqxax\longleftarrow\etaSUBpaxTOax,\\
	\etaSUBaepsilonTOpaepsilon\longleftarrow\etaSUBpsigsqepsilonaepsilonTOaepsilon,\,\,\,\etaSUBaxTOpax\longleftarrow\etaSUBpsigsqxaxTOax.
\end{array}
\end{equation*}
Note that the updates for $\etaSUBaepsilonTOpsigsqepsilonaepsilon$ and $\etaSUBaxTOpsigsqxax$ remain constant throughout the iterations.

\subsubsection{Factor to stochastic node message parameter updates}

The updates for the parameters of factor to stochastic node messages require use of the VMP algorithms described in Subsection \ref{sec:VMP}. The following is a detailed explanation of their usage to obtain the remaining updates. 

Use Algorithm 2 of \cite{maestrini2021inverse} for the iterated Inverse G-Wishart Fragment with:
\begin{itemize}
	\item[] Graph Input: $G=G_{\tiny\mbox{full}}$.
	\item[] Shape Parameter Input: 1.
	\item[] Message Graph Input: $\GSUBaxTOpsigsqxax=G_{\tiny\mbox{diag}}$.
	\item[] Natural Parameter Inputs: $\etaSUBsigsqxTOpsigsqxax$, $\etaSUBpsigsqxaxTOsigsqx$, $\etaSUBaxTOpsigsqxax$, $\etaSUBpsigsqxaxTOax$.
	\item[] Graph Outputs: $\GSUBpsigsqxaxTOsigsqx$, $\GSUBpsigsqxaxTOax$.
	\item[] Natural Parameter Outputs: $\etaSUBpsigsqxaxTOsigsqx$, $\etaSUBpsigsqxaxTOax$.
\end{itemize}
Use Algorithm 2 of \cite{maestrini2021inverse} for the iterated Inverse G-Wishart Fragment with:
\begin{itemize}
	\item[] Graph Input: $G=G_{\tiny\mbox{full}}$.
	\item[] Shape Parameter Input: 1.
	\item[] Message Graph Input: $\GSUBaepsilonTOpsigsqepsilonaepsilon=G_{\tiny\mbox{diag}}$.
	\item[] Natural Parameter Inputs: $\etaSUBsigsqepsilonTOpsigsqepsilonaepsilon$, $\etaSUBpsigsqepsilonaepsilonTOsigsqepsilon$, $\etaSUBaepsilonTOpsigsqepsilonaepsilon$, $\etaSUBpsigsqepsilonaepsilonTOaepsilon$.
	\item[] Graph Outputs: $\GSUBpsigsqepsilonaepsilonTOsigsqepsilon$, $\GSUBpsigsqepsilonaepsilonTOaepsilon$.
	\item[] Natural Parameter Outputs: $\etaSUBpsigsqepsilonaepsilonTOsigsqepsilon$, $\etaSUBpsigsqepsilonaepsilonTOaepsilon$.
\end{itemize}
Use Algorithm \ref{alg:VMPalgo} of the current work with:
\begin{itemize}
	\item[] Data Inputs: $\by$, $\bK$.
	\item[] Natural Parameter Inputs: $\etaSUBxTOpxbsigsqx$, $\etaSUBpxbsigsqxTOx$, $\etaSUBsigsqxTOpxbsigsqx$, $\etaSUBpxbsigsqxTOsigsqx$.
	\item[] Natural Parameter Outputs: $\etaSUBpxbsigsqxTOx$, $\etaSUBpxbsigsqxTOsigsqx$.
\end{itemize}
Use the algorithm of Section 4.1.5 of \cite{wand2017fast} for the Gaussian Likelihood Fragment with:
\begin{itemize}
	\item[] Natural Parameter Inputs: $\etaSUBxTOpyxsigsqepsilon$, $\etaSUBpyxsigsqepsilonTOx$, $\etaSUBsigsqepsilonTOpyxsigsqepsilon$, $\etaSUBpyxsigsqepsilonTOsigsqepsilon$.
	\item[] Natural Parameter Outputs: $\etaSUBpyxsigsqepsilonTOx$, $\etaSUBpyxsigsqepsilonTOsigsqepsilon$.
\end{itemize}

\subsection{Approximating density functions}

After reaching convergence, the remaining task is deriving the optimal approximating densities. The densities of main interest are $q^*(\bx)$, $q^*(\sigma^2_\varepsilon)$ and $q^*(\sigma^2_x)$, and have the form expressed in \eqref{eq:qxDens}, \eqref{eq:qsigsqepsDens} and \eqref{eq:qsigsqxDens} that come from \eqref{eq:qStarVMP}. In particular,
\begin{equation*}
\begin{array}{c}
q^*(\bx)\propto\exp\left\{\left[\begin{array}{c}
\bx\\[1ex]
\vecof\big(\bx\bx^T\big)
\end{array}\right]^T\etaSUBqx\right\},\,\mbox{with}\,\,\etaSUBqx\equiv\etaSUBpxbsigsqxTOx + \etaSUBpyxsigsqepsilonTOx,\\[3ex]
q^*(\sigma^2_\varepsilon)\propto\exp\left\{\left[\begin{array}{c}
\log(\sigma^2_{\varepsilon})\\[1.2ex]
1/\sigma^2_{\varepsilon}
\end{array}\right]^T\etaSUBqsigsqepsilon\right\},\,\mbox{with}\,\,\etaSUBqsigsqepsilon\equiv\etaSUBpsigsqepsilonaepsilonTOsigsqepsilon + \etaSUBpyxsigsqepsilonTOsigsqepsilon\\[3ex]
\mbox{and}\,\,q^*(\sigma^2_x)\propto\exp\left\{\left[\begin{array}{c}
\log(\sigma^2_{x})\\[1.2ex]
1/\sigma^2_{x}
\end{array}\right]^T\etaSUBqsigsqx\right\},\,\mbox{with}\,\,\etaSUBqsigsqx\equiv\etaSUBpsigsqxaxTOsigsqx + \etaSUBpxbsigsqxTOsigsqx.
\end{array}
\end{equation*}
The $q$-density common parameters are then the following:
\begin{equation*}
	\begin{array}{c}
		\muqx=\Sigmaqx\left(\etaSUBqx\right)_{1:m},\,\,\,\Sigmaqx=-\frac{1}{2}\vecof^{-1}\left\{\left(\etaSUBqx\right)_{(m+1):m^2}\right\},\\[2.5ex]
		\kappaqsigsqepsilon=-2\left(1+\left(\etaSUBqsigsqepsilon\right)_1\right),\,\,\, \lambdaqsigsqepsilon=-2\left(\etaSUBqsigsqepsilon\right)_2,\\[2.5ex]
		\kappaqsigsqx=-2\left(1+\left(\etaSUBqsigsqx\right)_1\right),\,\,\, \lambdaqsigsqx=-2\left(\etaSUBqsigsqx\right)_2.
	\end{array}
\end{equation*}
Expressions for the other optimal approximating densities can be obtained in a similar manner.

\section{Removing $\bL$ from the variational algorithms}

\noindent We provide results that allow to simplify the variational algorithm updates involving the contrast matrix $\bL$ by means of simpler computational steps. The results are presented separately for one- and two-dimensional problems and make use of the following notation.
\begin{definition}
For vectors $\bv_1,\ldots,\bv_p$,
\begin{equation*}
\stack{i=1,\ldots,p}(\bv_i)\equiv
\left[
\begin{array}{c}
\bv_1\\
\vdots\\
\bv_p
\end{array}
\right].
\end{equation*}
\end{definition}
\begin{definition}
	For vectors $\bv$ and $\tilde{\bv}$, respectively of length $\dv$ and $\dv-1$,
	\begin{equation*}
		\tridiag\left(\bv,\tilde{\bv}\right)\equiv \left[\begin{array}{ccccc}
			v_{1} & \tilde{v}_{1} & 0 & \cdots & 0\\
			\tilde{v}_{1} & v_{2} & \tilde{v}_{2} & \ddots & \vdots\\
			0 & \tilde{v}_{2} & \ddots & \ddots & 0\\
			\vdots & \ddots & \ddots & v_{\dv-1} & \tilde{v}_{\dv-1}\\
			0 & \cdots & 0 & \tilde{v}_{\dv-1} & v_{\dv}
		\end{array}\right].
	\end{equation*}
\end{definition}
\begin{definition}
For vectors $\bv$ and $\tilde{\bv}$, respectively of length $\dv$ and $\left(\dv-c\right)$, with $\dv>c$ and $c\in\mathbb{N}$,
\begin{equation*}
\sparsetridiag\left(\bv,\tilde{\bv},c\right)\equiv \left[\begin{array}{cccccc}
v_{1} & \boldsymbol{0}_{c-1}^T & \tilde{v}_{1} & 0 & \cdots & 0\\
\boldsymbol{0}_{c-1} & v_{2} & & \tilde{v}_{2} & \ddots & \vdots\\
\tilde{v}_{1} & & & \ddots & \ddots & 0\\
0 & \tilde{v}_{2} & \ddots & \ddots & & \tilde{v}_{\dv-c}\\
\vdots & \ddots & \ddots & & & \boldsymbol{0}_{c-1}^T\\
0 & \cdots & 0 & \tilde{v}_{\dv-c} & \boldsymbol{0}_{c-1} & v_{\dv}
\end{array}\right].
\end{equation*}
\end{definition}
\noindent Note that if $c=1$, then $\sparsetridiag\left(\bv,\tilde{\bv},1\right)=\tridiag\left(\bv,\tilde{\bv}\right)$.

\subsection{One-dimensional case}

Consider a one-dimensional inverse problem analyzed via model \eqref{eq:linInvProbModel} and suppose $\bx$ has one-to-one correspondence with a sequence of hidden and equispaced locations on a line. Assume first nearest neighbor differences are modeled via the contrast matrix $\LoneD$ defined in \eqref{eq:L1D}. Then matrix $\bL$ can be removed from \eqref{eq:exp1} by making use of the following lemma. 
%
\begin{lemma}
Let $\bv$ be a vector of length $\dv$ and $\LoneD$ a $\left(\dv-1\right)\times \dv$ matrix having form \eqref{eq:L1D}. Then 
\begin{equation*}
\LoneD \bv = \left[\begin{array}{c}
v_2 - v_1\\
v_3 - v_2\\
\vdots\\
v_{\dv} - v_{\dv-1}
\end{array}\right],
\end{equation*}
which is a vector of length $\dv-1$.
\label{lem:lemmaA1d}
\end{lemma}
\noindent We can get rid of matrix $\bL$ from expressions having form \eqref{eq:exp2} through the next lemma.
%
\begin{lemma}
Let $\bM$ be a symmetric $\dM\times \dM$ matrix and $\LoneD$ a $\left(\dM-1\right)\times \dM$ matrix having form \eqref{eq:L1D}. Then for $i=1,\ldots,\dM-1$,
\begin{equation*}
\left(\LoneD\bM\LoneD^T\right)_{ii} = M_{i+1,i+1} - 2M_{i+1,i} + M_{i,i}
\end{equation*}
or, equivalently,
\begin{align*}
\diagonal\left(\LoneD\bM\LoneD^T\right) &= \diagonal\left(\bM\right)_{2:\dM} + \diagonal\left(\bM\right)_{1:\left(\dM-1\right)} \\
&\quad - 2\,\diagonal\left(\bM_{2:\dM,1:\left(\dM-1\right)}\right) ,
\end{align*}
which is a vector of length $\dM-1$.
\label{lem:lemmaB1d}
\end{lemma}
\noindent The following lemma simplifies the computation of \eqref{eq:exp3}.
%
\begin{lemma}
Let $\bw$ be a vector of length $\dw$ and $\LoneD$ a $\left(\dw-1\right)\times \dw$ matrix having form \eqref{eq:L1D}. Then
\begin{equation*}
	\LoneD^T\diag\left(\bw\right)\LoneD = \tridiag\left(\left[\begin{array}{c} w_1\\ \bw_{1:\left(\dw-1\right)} + \bw_{2:\dw}\\ w_{\dw} \end{array}\right],-\bw\right),
\end{equation*}
which is a matrix of size $\dw\times \dw$.
\label{lem:lemmaC1d}
\end{lemma}

In the \textsf{R} computing environment Lemmas \ref{lem:lemmaA1d}--\ref{lem:lemmaC1d} can be implemented with standard base functions. In particular, the function \texttt{diff(v)} automatically produces the result stated in Lemma \ref{lem:lemmaA1d}. The expressions originated by Lemmas \ref{lem:lemmaB1d} and \ref{lem:lemmaC1d} can be visualized through the examples provided in Section \ref{sec:proofLemmas1D}.

\subsection{Two-dimensional case}

Consider the study of bidimensional inverse problems through model \eqref{eq:linInvProbModel} with $\bx=\vecof(\bX)$ and $\bX$ being an $m_1\times m_2$ matrix whose entries correspond to a regular grid of locations. 
Then a first nearest neighbor contrast matrix, here denoted by $\LtwoD$, can be conveniently defined as
\begin{equation}
\LtwoD\equiv\left[\begin{array}{c}
\LH\bC\\
\LV
\end{array}\right],
\label{eq:L2D}
\end{equation}
which has size $d \times (m_1m_2)$, with $d=\dH+\dV$, $\dH=m_1(m_2-1)$ and $\dV=(m_1-1)m_2$. The single components of $\LtwoD$ are defined as follows:
\begin{equation}
\LH\equiv\bI_{m_1}\otimes\LzeroH\quad\mbox{and}\quad \LV\equiv\bI_{m_2}\otimes\LzeroV,
\label{eq:LHandLV}
\end{equation}
where $\LH$ and $\LV$ are matrices of size $\dH \times (m_1m_2)$ and $\dV \times (m_1m_2)$, respectively. The matrix $\bC$ is a $m_1\times m_2$ \textit{commutation matrix} such that
\begin{equation*}
\bC\vecof(\bX)=\vecof(\bX^T).
\end{equation*}
Lastly, matrices $\LzeroH$ and $\LzeroV$ have the form \eqref{eq:L1D} identified for the one-dimensional case, and dimensions $(m_2-1)\times m_2$ and $(m_1-1)\times m_1$, respectively. If for instance $\bX$ is of size $3\times4$ as in the example of Figure \ref{fig:pixelGrid}, then
\begin{equation}
\LzeroH=\left[\begin{array}{cccc}
-1 & 1 & 0 & 0\\
0 & -1 & 1 & 0\\
0 & 0 & -1 & 1
\end{array}\right]\quad\mbox{and} \quad
\LzeroV=\left[\begin{array}{ccc}
-1 & 1 & 0\\
0 & -1 & 1
\end{array}\right].
\label{eq:L0HandL0V}
\end{equation}
Superscripts H and V refer to differences between pairs of locations respectively computed in a horizontal and vertical fashion, given the row-wise and column-wise orientation identified by $\bX$.
The explicit expression of $\LtwoD$ for a $3\times4$ matrix of parameters $\bX$ is provided in Section \ref{sec:proofLemmas2D}. 
For clarity, the coefficients that define the dimensions of the $\LtwoD$ matrix components are summarized in Table \eqref{tab:coeffSummary}. 
\begin{table}
	\begin{center}
		\begin{tabular}{ll}
			\Xhline{3\arrayrulewidth}\\[-2.2ex]
			\textbf{Notation}   & \textbf{Description} \\[0.2ex]
			\Xhline{3\arrayrulewidth}\\[-2ex]
			$m_1$   & number of rows in $\bX$ (and $\bY$)\\
			$m_2$   & number of columns in $\bX$ (and $\bY$)\\
			$m=m_1\times m_2$ & total number of elements in $\bX$ (and $\bY$)\\
			$\dH=m_1(m_2-1)$   & number of horizontal differences in $\bX$ \\
			$\dV=(m_1-1)m_2$   & number of vertical differences in $\bX$ \\
			$d=\dH+\dV$   & total number of differences in $\bX$ \\
			\hline
		\end{tabular}
	\end{center}
	\caption{\it Notation used for two-dimensional inverse problems with first nearest neighbor differences and one-to-one correspondence between $\bX$ and $\bY$.}
	\label{tab:coeffSummary} 
\end{table}

Lemmas \ref{lem:lemmaA2d}--\ref{lem:lemmaC2d} are extensions of Lemmas \ref{lem:lemmaA1d}--\ref{lem:lemmaC1d} to the two-dimensional case, provided that $\LtwoD$ has the form identified by \eqref{eq:L2D}--\eqref{eq:L0HandL0V}. Vector and matrix dimensions are intentionally emphasized to highlight analogies with one-dimensional problems and set up a framework extendible to higher dimensions.

The next lemma simplifies the computation of expression \eqref{eq:exp1} and avoids calculating and storing matrix $\LtwoD$.
%
\begin{lemma}
Let $\bV$ be a matrix of size $m_1\times m_2$, $\bv=\vecof\left(\bV\right)$ a vector of length $\dv=m_1m_2$ and $\LtwoD$ a matrix of size $(\dH+\dV)\times \dv$ having the form defined in \eqref{eq:L2D}--\eqref{eq:L0HandL0V} with $\dH=m_1(m_2-1)$ and $\dV=(m_1-1)m_2$. Then
\begin{equation*}
\LtwoD\bv=\left[\begin{array}{c}
\tH\\
\tV
\end{array}\right],
\end{equation*}
where
\begin{equation*}
\tH=\stack{i=1,\ldots,m_1}\left(\left[\begin{array}{c}
u_{m_2(i-1)+2}-u_{m_2(i-1)+1}\\
u_{m_2(i-1)+3}-u_{m_2(i-1)+2}\\
\vdots\\
u_{m_2 i}-u_{m_2 i-1}
\end{array}\right]\right)
\end{equation*}
and
\begin{equation*}
\tV=\stack{i=1,\ldots,m_2}\left(\left[\begin{array}{c}
v_{m_1(i-1)+2}-v_{m_1(i-1)+1}\\
v_{m_1(i-1)+3}-v_{m_1(i-1)+2}\\
\vdots\\
v_{m_1 i}-v_{m_1 i-1}
\end{array}\right]\right),
\end{equation*}
are vectors of length $\dH$ and $\dV$, respectively, and $\bu=\vecof\left(\bV^T\right)$.
\label{lem:lemmaA2d}
\end{lemma}
\noindent Expression \eqref{eq:exp2} can be efficiently computed using the following lemma.
%
\begin{lemma}
Let $\bM$ be a symmetric $\dM \times \dM$ matrix with $\dM=m_1m_2$ and $\LtwoD$ a matrix of size $(\dH+\dV)\times\dM$ having the form defined in \eqref{eq:L2D}--\eqref{eq:L0HandL0V} with $\dH=m_1(m_2-1)$ and $\dV=(m_1-1)m_2$. Then
\begin{equation*}
\diagonal\left(\LtwoD\bM\LtwoD^T\right) = \left[\begin{array}{c}
\sH\\
\sV
\end{array}\right],
\end{equation*}
where $\sH$ and $\sV$ are vectors of length $\dH$ and $\dV$, respectively, such that
\begin{equation*}
\begin{array}{c}
\sH=\vecof\left\{\left(\vecof_{m_1,(m_2-1)}^{-1}\left(\szeroH\right)\right)^T\right\},\\[2ex]
\szeroH=\diagonal\left(\bM\right)_{(m_{1}+1):d_{M}}-2\,\diagonal\left(\bM_{(m_{1}+1):d_{M},1:(d_{M}-m_{1})}\right)+\diagonal\left(\bM\right)_{1:(d_{M}-m_{1})}
\end{array}
\end{equation*}
and
\begin{equation*}
\sV=\diagonal\left(\bM\right)_{(k_1+1):(k_{\dV}+1)}-2\,\diagonal\left(\bM_{(k_1+1):(k_{\dV}+1),k_1:k_{\dV}}\right)+\diagonal\left(\bM\right)_{k_1:k_{\dV}},
\end{equation*}
and $\bk$ is a vector of $\dV$ indeces obtained as follows:
\begin{equation*}
\bk=\stack{i=1,\ldots,m_2}\left(\left[\begin{array}{c}
m_1(i-1)+1\\
m_1(i-1)+2\\
\vdots\\
m_1 i -1
\end{array}\right]\right).
\end{equation*}
\label{lem:lemmaB2d}
\end{lemma}
\noindent The last lemma reduces the computational effort for implementing \eqref{eq:exp3}.
%
\begin{lemma}
Let $\bw$ be a vector of length $\dw=\dH+\dV$, with $\dH=m_1(m_2-1)$ and $\dV=(m_1-1)m_2$, and $\LtwoD$ a matrix of size $\dw\times (m_1m_2)$ having the form defined in \eqref{eq:L2D}--\eqref{eq:L0HandL0V}. Then
\begin{equation*}
\LtwoD^T\diag(\bw)\LtwoD=\bR - \diag\left(\left[\sum_{j=1}^{m_1m_2} R_{i,j}\right]_{i=1,\ldots,m_1m_2}\right),
\end{equation*}
where
\begin{equation*}
\bR=\sparsetridiag\big(\boldsymbol{0}_{m_1 m_2},\rH,m_1\big) + \sparsetridiag\big(\boldsymbol{0}_{m_1 m_2},\rV,1\big)
\end{equation*}
is a matrix of size $(m_1m_2)\times(m_1m_2)$, and
\begin{align*}
&\rH=-\vecof\left\{\left(\mbox{\rm vec}_{(m_2-1),m_1}^{-1}\left(\bw_{1:\dH}\right)\right)^T\right\}\\[1ex]
\mbox{and}\quad&\rV=-\left[\vecof\left(\left[\begin{array}{c}
\vecof_{(m_{1}-1),m_{2}}^{-1}\left(\bw_{(\dH+1):d_{w}}\right)\\[2ex]
\boldsymbol{0}_{m_{2}}^{T}
\end{array}\right]\right)\right]_{1:(m_1m_2-1)}
\end{align*}
are vectors of length $\dH$ and $m_1m_2-1$, respectively.
\label{lem:lemmaC2d}
\end{lemma}

The supplement provides heuristic arguments to prove Lemmas \ref{lem:lemmaA2d}--\ref{lem:lemmaC2d} and demonstrates the implementation of the two-dimensional case lemmas in \textsf{R}.


%

\subsection{Visualization of Lemmas \ref{lem:lemmaA1d}--\ref{lem:lemmaC1d}}\label{sec:proofLemmas1D}

Consider a simple one-dimensional inverse problem modeled through \eqref{eq:linInvProbModel} where $m=4$ and $d=m-1=3$. For this particular example the contrast matrix has size $d\times m$ and the following explicit expression:
\begin{equation*}
\LoneD=\left[\begin{array}{cccc}
-1 & 1 & 0 & 0\\
0 & -1 & 1 & 0\\
0 & 0 & -1 & 1
\end{array}\right].
\end{equation*}

\subsubsection{Visualization of Lemma \ref{lem:lemmaA1d}}

Vector $\bv$ has length $\dv=m=4$ and 
\begin{equation*}
\LoneD \bv = \left[\begin{array}{c}
v_2 - v_1\\
v_3 - v_2\\
v_{4} - v_{3}
\end{array}\right].
\end{equation*}

\subsubsection{Visualization of Lemma \ref{lem:lemmaB1d}}

Matrix $\bM$ is symmetric and has size $\dM\times \dM$, with $\dM=m=4$, and
\begin{equation*}
\diagonal\left(\LoneD\bM\LoneD^T\right)= \left[\begin{array}{c}
M_{2,2} - 2M_{2,1} + M_{1,1}\\
M_{3,3} - 2M_{3,2} + M_{2,2}\\
M_{4,4} - 2M_{4,3} + M_{3,3}
\end{array}\right].
\end{equation*}

\subsubsection{Visualization of Lemma \ref{lem:lemmaC1d}}

Vector $\bw$ has length $\dw=d=3$ and 
\begin{equation*}
\LoneD^T\diag\left(\bw\right)\LoneD = \left[\begin{array}{cccc}
w_1 & -w_1 & 0 & 0\\
-w_1 & w_1+w_2 & -w_2 & 0\\
0 & -w_2 & w_2+w_3 & -w_3\\
0 & 0 & -w_3 & w_3
\end{array}\right].
\end{equation*}

\subsection{Visualization and \textsf{R} implementation of Lemmas \ref{lem:lemmaA2d}--\ref{lem:lemmaC2d}}\label{sec:proofLemmas2D}

This subsection shows the results stated in Lemmas \ref{lem:lemmaA2d}--\ref{lem:lemmaC2d} through a simple two-dimensional inverse problem model where $\bX$ is a matrix of parameters of size $m_1\times m_2$ with $m_1=3$ and $m_2=4$. Referring to model \eqref{eq:linInvProbModel}, the length of $\bx=\vecof(\bX)$ is $m=m_1\times m_2$. For this particular case the contrast matrix $\LtwoD$ defined in \eqref{eq:L2D} is a matrix of size $d\times m$, hence of size $17\times12$, given that $d=\dH+\dV$, $\dH=m_1(m_2-1)=9$ and $\dV=(m_1-1)m_2=8$. The sub-components of $\LtwoD$ are
\begin{equation*}
\LH\bC=\left[\begin{array}{cccccccccccc}
-1 & 0 & 0 & 1 & 0 & 0 & 0 & 0 & 0 & 0 & 0 & 0\\
0 & 0 & 0 & -1 & 0 & 0 & 1 & 0 & 0 & 0 & 0 & 0\\
0 & 0 & 0 & 0 & 0 & 0 & -1 & 0 & 0 & 1 & 0 & 0\\
0 & -1 & 0 & 0 & 1 & 0 & 0 & 0 & 0 & 0 & 0 & 0\\
0 & 0 & 0 & 0 & -1 & 0 & 0 & 1 & 0 & 0 & 0 & 0\\
0 & 0 & 0 & 0 & 0 & 0 & 0 & -1 & 0 & 0 & 1 & 0\\
0 & 0 & -1 & 0 & 0 & 1 & 0 & 0 & 0 & 0 & 0 & 0\\
0 & 0 & 0 & 0 & 0 & -1 & 0 & 0 & 1 & 0 & 0 & 0\\
0 & 0 & 0 & 0 & 0 & 0 & 0 & 0 & -1 & 0 & 0 & 1
\end{array}\right]
\end{equation*}
and
\begin{equation*}
\LV=\left[\begin{array}{cccccccccccc}
-1 & 1 & 0 & 0 & 0 & 0 & 0 & 0 & 0 & 0 & 0 & 0\\
0 & -1 & 1 & 0 & 0 & 0 & 0 & 0 & 0 & 0 & 0 & 0\\
0 & 0 & 0 & -1 & 1 & 0 & 0 & 0 & 0 & 0 & 0 & 0\\
0 & 0 & 0 & 0 & -1 & 1 & 0 & 0 & 0 & 0 & 0 & 0\\
0 & 0 & 0 & 0 & 0 & 0 & -1 & 1 & 0 & 0 & 0 & 0\\
0 & 0 & 0 & 0 & 0 & 0 & 0 & -1 & 1 & 0 & 0 & 0\\
0 & 0 & 0 & 0 & 0 & 0 & 0 & 0 & 0 & -1 & 1 & 0\\
0 & 0 & 0 & 0 & 0 & 0 & 0 & 0 & 0 & 0 & -1 & 1
\end{array}\right],
\end{equation*}
where $\LH$ and $\LV$ are defined through \eqref{eq:LHandLV} by making use of matrices $\LzeroH$ and $\LzeroV$ shown in \eqref{eq:L0HandL0V}, and $\bC$ is a commutation matrix of appropriate size.

The objective of the following subsubsections is to visualize the results expressed in Lemmas \ref{lem:lemmaA2d}--\ref{lem:lemmaC2d} for the particular case under examination.

\subsubsection{Visualization of Lemma \ref{lem:lemmaA2d}}

Vector $\bv$ has length $\dv=m_1m_2=12$ and
\begin{equation*}
\LtwoD\bv=\left[\begin{array}{c}
\tH\\
\tV
\end{array}\right],
\end{equation*}
where
\begin{equation*}
\tH=\left[\begin{array}{c}
v_4-v_1\\
v_7-v_4\\
v_{10}-v_7\\
v_5-v_2\\
v_8-v_5\\
v_{11}-v_8\\
v_6-v_3\\
v_9-v_6\\
v_{12}-v_9
\end{array}\right],\quad\tV=\left[\begin{array}{c}
v_2-v_1\\
v_3-v_2\\
v_5-v_4\\
v_6-v_5\\
v_8-v_7\\
v_9-v_8\\
v_{11}-v_{10}\\
v_{12}-v_{11}
\end{array}\right].
\end{equation*}

\subsubsection{Visualization of Lemma \ref{lem:lemmaB2d}}

Matrix $\bM$ is symmetric and has size $\dM\times \dM$, with $\dM=m_1m_2=12$, and
\begin{equation*}
\diagonal\left(\LtwoD\bM\LtwoD^T\right) = \left[\begin{array}{c}\sH\\\sV\end{array}\right],
\end{equation*}
where
\begin{equation*}
\sH=\left[\begin{array}{c}
M_{4,4}-2M_{4,1}+M_{1,1}\\
M_{7,7}-2M_{7,4}+M_{4,4}\\
M_{10,10}-2M_{10,7}+M_{7,7}\\
M_{5,5}-2M_{5,2}+M_{2,2}\\
M_{8,8}-2M_{8,5}+M_{5,5}\\
M_{11,11}-2M_{11,8}+M_{8,8}\\
M_{6,6}-2M_{6,3}+M_{3,3}\\
M_{9,9}-2M_{9,6}+M_{6,6}\\
M_{12,12}-2M_{12,9}+M_{9,9}
\end{array}\right],\quad\sV=\left[\begin{array}{c}
M_{2,2}-2M_{2,1}+M_{1,1}\\
M_{3,3}-2M_{3,2}+M_{2,2}\\
M_{5,5}-2M_{5,4}+M_{4,4}\\
M_{6,6}-2M_{6,5}+M_{5,5}\\
M_{8,8}-2M_{8,7}+M_{7,7}\\
M_{9,9}-2M_{9,8}+M_{8,8}\\
M_{11,11}-2M_{11,10}+M_{10,10}\\
M_{12,12}-2M_{12,11}+M_{11,11}
\end{array}\right].
\end{equation*}

\subsubsection{Visualization of Lemma \ref{lem:lemmaC2d}}

Vector $\bw$ has length $\dw=d=17$ and 
\begin{equation*}
\LtwoD^T\diag\left(\bw\right)\LtwoD = \bR +
\diag\left(\left[\begin{array}{c}
	w_1+w_{10}\\
	w_4+w_{10}+w_{11}\\
	w_7+w_{11}\\
	w_1+w_2+w_{12}\\
	w_4+w_5+w_{12}+w_{13}\\
	w_7+w_8+w_{13}\\
	w_2+w_3+w_{14}\\
	w_5+w_6+w_{14}+w_{15}\\
	w_8+w_9+w_{15}\\
	w_3+w_{16}\\
	w_6+w_{16}+w_{17}\\
	w_9+w_{17}
	\end{array}\right]\right)\\
\end{equation*}
where
\begin{equation*}
\bR=\footnotesize{\left[\begin{array}{cccccccccccc}
	0 & -w_{10} & 0 & -w_1 & 0 & 0 & 0 & 0 & 0 & 0 & 0 & 0\\
	-w_{10} & 0 & -w_{11} & 0 & -w_4 & 0 & 0 & 0 & 0 & 0 & 0 & 0\\
	0 & -w_{11} & 0 & 0 & 0 & -w_7 & 0 & 0 & 0 & 0 & 0 & 0\\
	-w_1 & 0 & 0 & 0 & -w_{12} & 0 & -w_2 & 0 & 0 & 0 & 0 & 0\\
	0 & -w_4 & 0 & -w_{12} & 0 & -w_{13} & 0 & -w_5 & 0 & 0 & 0 & 0\\
	0 & 0 & -w_7 & 0 & -w_{13} & 0 & 0 & 0 & -w_8 & 0 & 0 & 0\\
	0 & 0 & 0 & -w_2 & 0 & 0 & 0 & -w_{14} & 0 & -w_3 & 0 & 0\\
	0 & 0 & 0 & 0 & -w_5 & 0 & -w_{14} & 0 & -w_{15} & 0 & -w_6 & 0\\
	0 & 0 & 0 & 0 & 0 & -w_8 & 0 & -w_{15} & 0 & 0 & 0 & -w_9\\
	0 & 0 & 0 & 0 & 0 & 0 & -w_3 & 0 & 0 & 0 & -w_{16} & 0\\
	0 & 0 & 0 & 0 & 0 & 0 & 0 & -w_6 & 0 & -w_{16} & 0  & -w_{17}\\
	0 & 0 & 0 & 0 & 0 & 0 & 0 & 0 & -w_9 & 0 & -w_{17} & 0
	\end{array}\right]}.
\end{equation*}
This involves the following two vectors:
\begin{equation*}
\brH=-\left[\begin{array}{c}
w_{1}\\
w_{4}\\
w_{7}\\
w_{2}\\
w_{5}\\
w_{8}\\
w_{3}\\
w_{6}\\
w_{9}
\end{array}\right]\,\,\mbox{and}\,\,\brV=-\left[\begin{array}{c}
w_{10}\\
w_{11}\\
0\\
w_{12}\\
w_{13}\\
0\\
w_{14}\\
w_{15}\\
0\\
w_{16}\\
w_{17}
\end{array}\right].
\end{equation*}

\subsection{Implementation of Lemmas \ref{lem:lemmaA2d}--\ref{lem:lemmaC2d} in \textsf{R}}\label{sec:implementationLemmas2D}

This subsection provides \textsf{R} code to implement Lemmas A, B and C for two-dimensional inverse problems. We make use of function \textit{invvec} from package \texttt{ks} \citep{duong2024}. Note that this function specifies matrix dimension with the number of columns preceding the number of rows. This is at odds with the definition of $\vecof^{-1}$ given in Section 1.2.

Consider, for instance, a two-dimensional dataset of size $50\times40$:
\begin{verbatim}
# Load required library:

library(ks)

# Obtain dimensions:

m1 <- 50   ;   m2 <- 40  
m <- m1*m2
dH <- m1*(m2-1)  ;  dV <- (m1-1)*m2
d <- dH + dV
\end{verbatim}

The result expressed in Lemma \ref{lem:lemmaA2d} can be computed in \textsf{R} with the following code:
\begin{verbatim}
# Create a vector of length m:

v <- rnorm(m)   

# Compute the result of Lemma 4:

vecVt <- vec(t(invvec(v,m2,m1))) 
indSetH <- seq(from=m2,to=dV,by=m2)
indSetV <- seq(from=m1,to=dH,by=m1)
tH <- diff(vecVt)[-indSetH]
tV <- diff(v)[-indSetV]
lemma4res <- c(tH,tV)
\end{verbatim}

The result expressed in Lemma \ref{lem:lemmaB2d} can be computed in \textsf{R} with the following code:
\begin{verbatim}
# Generate a square symmetric matrix M via vector v

M <- tcrossprod(v)
dM <- dim(M)[1]

# Compute the result of Lemma 5:

sH <- (diag(M)[(m1+1):dM] - 2*diag(M[(m1+1):dM,1:(dM-m1)]) 
       + diag(M)[1:(dM-m1)])
sH <- vec(t(invvec(sH,m2-1,m1)))
sV <- (diag(M)[setdiff(2:dM,indSetV+1)] 
       - 2*diag(M[setdiff(2:dM,indSetV+1),
                  setdiff(1:(dM-1),indSetV)])
       + diag(M)[setdiff(1:(dM-1),indSetV)])
lemma5res <- c(sH,sV)
\end{verbatim}

The result expressed in Lemma \ref{lem:lemmaC2d} can be computed in \textsf{R} as follows:
\begin{verbatim}
# Generate a vector w of lenght d

w <- rnorm(d)

# Compute the result of Lemma 6:

rH <- - vec(t(invvec(w[1:dH],m1,m2-1))) 
rV <- - vec(rbind(invvec(w[-(1:dH)],m2,m1-1),rep(0,m2)))[-m]
lemma6res <- matrix(0,m,m)
diag(lemma6res[1:(m-m1),(m1+1):m]) <- rH
diag(lemma6res[(m1+1):m,1:(m-m1)]) <- rH
diag(lemma6res[1:(m-1),2:m]) <- rV
diag(lemma6res[2:m,1:(m-1)]) <- rV
diag(lemma6res) <- - rowSums(lemma6res)
\end{verbatim}

\section{Additional biomedical data details}

\noindent Here we provide supplementary results to assess the performance of variational inference versus MCMC for the real biomedical data and additional details concerning the simulated biomedical data.

\subsection{Details on real biomedical data}

\begin{figure}[!ht]
	\centering
	{\includegraphics[width=.8\textwidth]{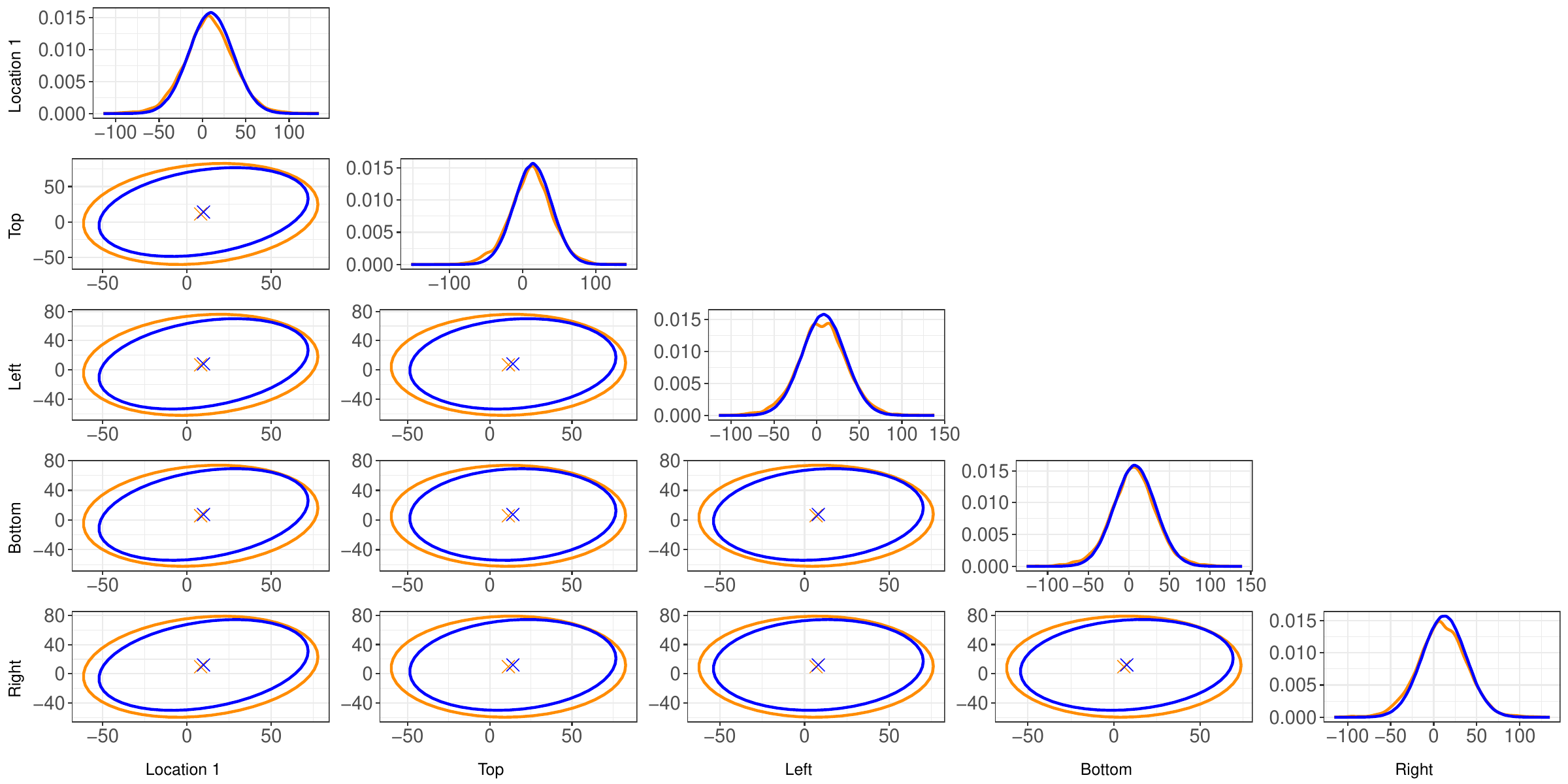}}
	\caption{\it Marginal and bivariate marginal posterior densities obtained via variational inference (blue) and MCMC (orange) for Location 1 of Figure \ref{fig:mouseResults} and the adjacent pixels on its top, left, bottom and right sides.}
	\label{fig:pixel1bivPlot} 
\end{figure}

\begin{figure}[!ht]
	\centering
	{\includegraphics[width=.8\textwidth]{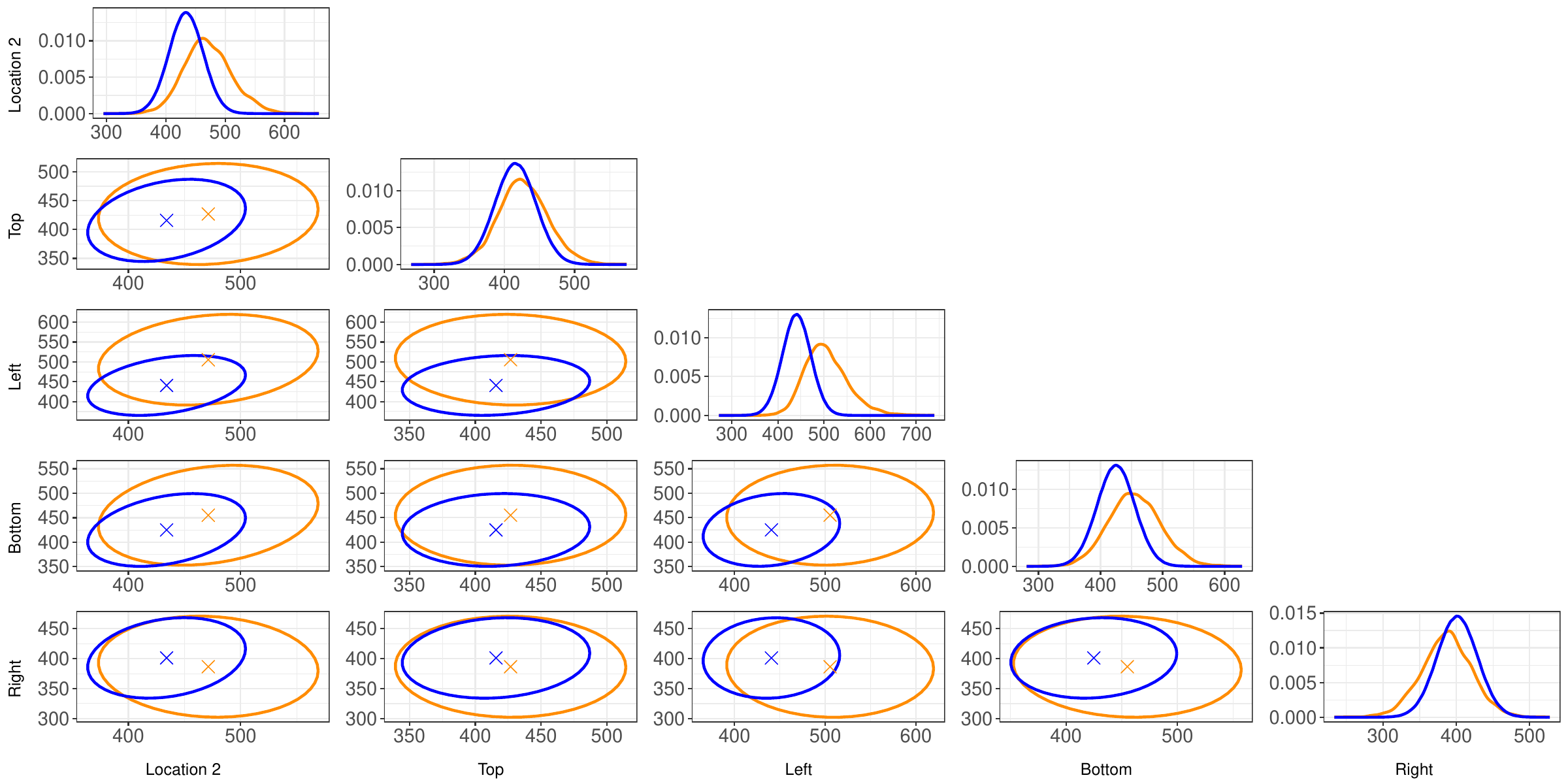}}
	\caption{\it Marginal and bivariate marginal posterior densities obtained via variational inference (blue) and MCMC (orange) for Location 2 of Figure \ref{fig:mouseResults} and the adjacent pixels on its top, left, bottom and right sides.}
	\label{fig:pixel2bivPlot} 
\end{figure}

\begin{figure}[!ht]
	\centering
	{\includegraphics[width=.8\textwidth]{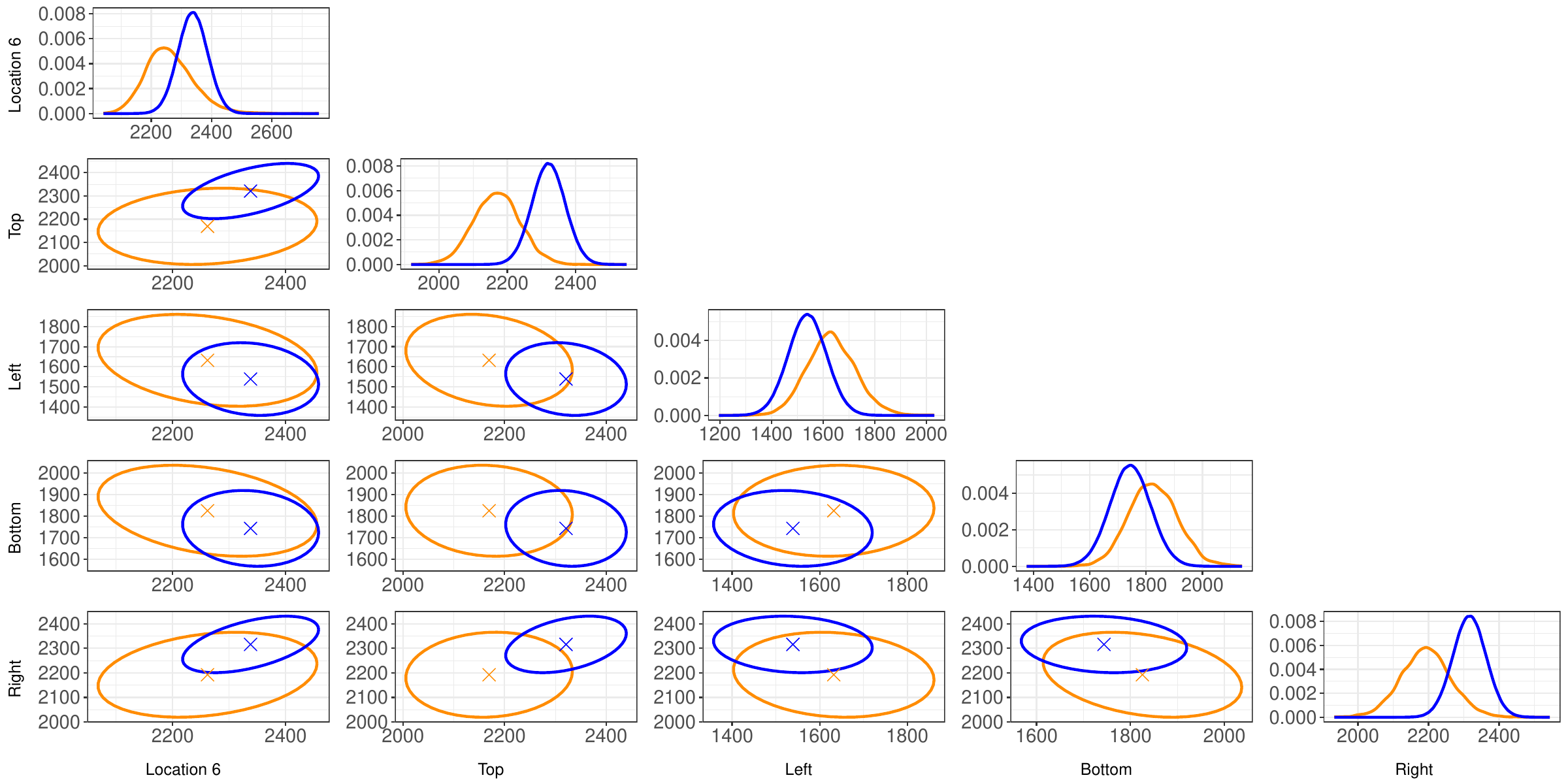}}
	\caption{\it Marginal and bivariate marginal posterior densities obtained via variational inference (blue) and MCMC (orange) for Location 1 of Figure \ref{fig:mouseResults} and the adjacent pixels on its top, left, bottom and right sides.}
	\label{fig:pixel6bivPlot} 
\end{figure}

\noindent Figures \ref{fig:pixel1bivPlot}--\ref{fig:pixel6bivPlot} provide a comparison between variational inference and MCMC in terms of marginal and bivariate marginal posterior densities for Location 1 (external to the mouse body), Location 2 (mouse thyroid) and Location 6 (mouse bladder) of Figure \ref{fig:mouseResults}. The bivariate marginal densities are 95\% ellipsoid obtained via the function \textit{geom\_density} of the \textsf{R} package \texttt{ggplot2} \citep{ggplot2}. While variational inference provides satisfactory approximations for Locations 1 and 2, its performance is less satisfactory for Location 6 corresponding to the mouse bladder and therefore to a location corresponding to higher technetium radioisotope emissions.


\subsection{Details on simulated biomedical data}

\begin{figure}[!ht]
	\centering
	{\includegraphics[width=.7\textwidth]{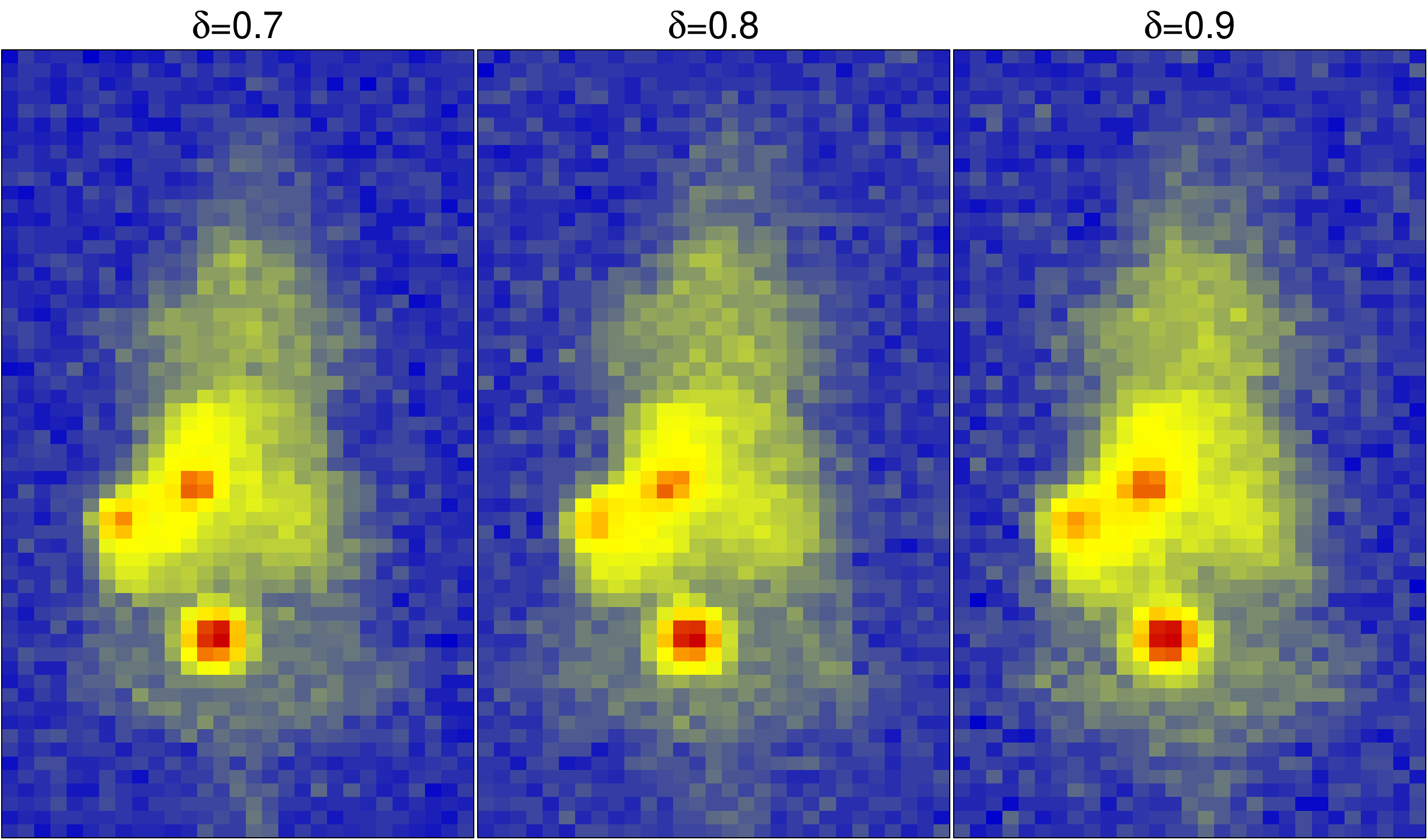}}
	\caption{\it Examples of data generated in the biomedical simulation study for different values of $\delta$.}
	\label{fig:MousePlotsSimStudyExamples} 
\end{figure}

Figure \ref{fig:MousePlotsSimStudyExamples} shows plots of three simulated biomedical datasets, one for each $\delta$ value considered in the simulation study. Higher $\delta$ values correspond to more blur in the images. 

%
\newpage
\section{Illustration for archaeological data \label{sec:archaeo}}

We here show how the message passing on factor graph fragments paradigm can be used to move from an inverse problem model analysis to another without deriving a variational inference algorithm from scratch. This is illustrated through data from archaeological magnetometry. In archaeology it is often required to investigate a potential site via geophysical remote sensing methods before any physical excavation is commenced. The model we consider includes a Skew Normal distribution for the response vector and a Horseshoe penalization in replacement to the Normal response and Laplace penalization of model \eqref{eq:linInvProbModel}.  

The data were collected from a mid Iron Age farmstead known as `The Park' through an archaeological exploration that took place in 1994 at Guiting Power in Gloucestershire, United Kingdom. After data collection, part of the area was also excavated and archaeologists drew an impression of the remains that were brought to light. 

The archaeological site was partitioned into a grid of 10m$\times$10m squares with the survey axes aligned in the directions of magnetic north and east. A fluxgate gradiometer FM18 with 0.1nT sensitivity was used to collect the data at 0.5m intervals. For each survey square, the gradiometer output was an array of $20\times20$ magnetic anomaly readings corresponding to the difference between the signals detected by the lower and upper sensors. The gradiometer lower sensor was held at 0.2m above the surface of the site and the upper sensor was fixed a further 0.5m higher. The Earth’s magnetic field is detected by both sensors, whereas the magnetic field from the buried feature is mostly detected by the lower sensor. Therefore the difference between the two sensors' readings  provides the magnetic anomaly due to the hidden feature, together with small random noise components. The random noise may be due to systematic causes such as machine-rounding errors, or non-systematic factors such as disturbance by small stones in the soil or interference from local magnetic sources \citep{scollar1970fourier}. 

We model the hidden surface through a single layer of rectangular prisms at a fixed burial depth. The scope is to estimate the prisms' magnetic \textit{susceptibility} in order to separate the constituent epochs of the site and locate relevant artifacts prior to excavation. The subsurface layer is assumed to be fixed at a burial depth of 0.3m, given that the topsoil across the excavated region of Guiting Power was found to be between 0.25 and 0.3m deep. According to the single layer subsurface model, each prism in the layer has the same constant extent, which we set to 0.5m. Although the vertical extent of each of the excavated features vary from 0.45m to 1.6m, the chosen value of prism vertical extent increases the chances of distinguishing low-susceptibility features.

All this information is relevant to design an appropriate matrix $\bK$ suitable for inverse problems concerning archaeological data.

\subsection{$\bK$ matrix for archaeological data \label{sec:Karchaeo}}

The matrix $\bK$ we adopt for the illustration on archaeological data has a more complicated definition and relies upon the \textit{spread function} defined in Section 2 of \cite{aykroyd2001bayesian}. The spread function models magnetic anomalies and depends on latitude and longitude of the archaeological site on the Earth's surface, the geometry of the gradiometer used to survey the area and the site physical properties. 

Suppose all the magnetic features are located at the same depth below the surface, have the same vertical thickness and the susceptibility is constant along any vertical line through the features, but may vary between horizontal locations.
We model the subsurface of the archaeological site as an ensemble of volume elements of equal size, called \textit{prisms}, each having uniform susceptibility, fixed vertical depth (0.3m) and extent (0.5m), and a square cross section in the horizontal plane.
Let $(x_1,y_1,z_1)$ and $(x_2,y_2,z_2)$ be the coordinates of opposite vertices of a prism with unit susceptibility, where the $x$, $y$ and $z$ axes point north, east and vertically downward, respectively. Then the vertical component of the anomaly due to the prism at a point with coordinates $(x,y,z)$ is 
\begin{equation*}
	\Delta Z(x,y,z)=\frac{B}{4\pi}\left\{\big[\Delta Z^{(1)}+\Delta Z^{(2)}+\Delta Z^{(3)}\big]^{\zeta=z-z_1}_{\zeta=z-z_2}\right\},
\end{equation*}
where $B\approx4.8\times10^4$nT (nanoteslas) is the magnitude of the magnetic flux density due to the Earth's field. The three additive components of $\Delta Z(x,y,z)$ are:
\begin{align*}
	&\Delta Z^{(1)}=\left[-\sin I\tan^{-1}\left(\frac{\xi\eta}{\zeta(\xi^2+\eta^2+\zeta^2)^{1/2}}\right)\right]^{\xi=x-x_1,\eta=y-y_1}_{\xi=x-x_2,\eta=y-y_2},\\
	&\Delta Z^{(2)}=\left[\frac{1}{2}\cos I\cos\theta\log\left(\frac{(\xi^2+\eta^2+\zeta^2)^{1/2}+\eta}{(\xi^2+\eta^2+\zeta^2)^{1/2}-\eta}\right)\right]^{\xi=x-x_1,\eta=y-y_1}_{\xi=x-x_2,\eta=y-y_2},\\
	&\Delta Z^{(3)}=\left[\frac{1}{2}\cos I\sin\theta\log\left(\frac{(\xi^2+\eta^2+\zeta^2)^{1/2}+\xi}{(\xi^2+\eta^2+\zeta^2)^{1/2}-\xi}\right)\right]^{\xi=x-x_1,\eta=y-y_1}_{\xi=x-x_2,\eta=y-y_2},
\end{align*}
where $I$ is the inclination of the Earth's magnetic field and $\theta$ is the angle between the direction of magnetic north and the $x$ axis. In our application $I=65$\textdegree and $\theta=0$\textdegree. 

The difference between two simultaneous readings from two sensors is recorded. One sensor is mounted vertically above the other at a distance of 0.5m. Then the recorded reading originated by a prism identified by coordinates $(x,y)$ is 
\begin{equation}
	h(x,y)=\Delta Z(x,y,z_{\tiny\mbox{U}})-\Delta Z(x,y,z_{\tiny\mbox{L}}),
	\label{eq:hfunc}
\end{equation}
where $z_{\tiny\mbox{U}}$ is the vertical coordinate of the upper sensor and $z_{\tiny\mbox{L}}$ is that of the lower sensor, which are held at 0.7m and 0.2m above the surface throughout the survey. 

Suppose that a vector $\by$ of $n$ readings is recorded over a rectangular site at coordinates $\bs_j$, $j=1,\ldots,n$. Assume that the subsurface is divided into a rectangular assemblage of $m$ prisms having coordinates $\bt_i$, $i=1,\ldots,m$, and producing susceptibilities that are collected in an $m\times1$ vector $\bx$. Then the influence of the prism $i$ at surface location $j$ is
\begin{equation}
	K_{ij}=h(\bt_i-\bs_j)
	\label{eq:Karchaeo}
\end{equation}
where $h$ is the function defined in \eqref{eq:hfunc}. The linear relationship between $\by$ and $\bx$ is then modeled as $E(\by) = \bK\bx$, where each element of the $n\times m$ matrix $\bK$ is given by \eqref{eq:Karchaeo}.

\subsection{Model and VMP Implementation}

As affirmed in \cite{aykroyd2001bayesian}, the assumptions used to model the hidden surface provide a realistic basis for modeling the data but can be quite restrictive. The recorded magnetic anomaly may include both positive and negative values, generate shifts in the apparent location of features, be asymmetric in shape and vary across different sampling regions. For this reason we model the outcome variable $\by$, i.e. the anomaly detected by the gradiometer, through a Skew Normal distribution and impose a Horseshoe penalization to the difference between hidden susceptibility values $\xDelta$. The Horseshoe prior shrinks the small signals and enhances the big ones, highlighting the contrast between buried features and plain soil.

The model we fit is the following:
\begin{equation}
	\begin{array}{c}
		y_i\vert \bx,\sigma^2_{\varepsilon},\lambda,c_i\simind N\left(\left(\bK\bx\right)_i+\frac{\sigma^2_{\varepsilon}\lambda|c_i|}{\sqrt{1+\lambda^2}},\frac{\sigma^2_{\varepsilon}}{1+\lambda^2}\right),\quad c_i\simind N(0,1),\quad i=1,\ldots,m,\\[1ex]
		(\xDelta)_j\vert b_j,\sigma^2_x\simind N\left(0,\sigma^2_x/b_j\right),\quad b_j\simind\pi^{-1}b_j^{-1/2}\left(1+b_j\right)^{-1},\quad j=1,\ldots,d,\\[1ex]
		\sigma^2_\varepsilon\sim\mbox{Inverse-}\chi^2\left(A_\varepsilon,B_\varepsilon\right),\quad\lambda\sim N\left(0,S^2_\lambda\right),\\[1ex]
		\sigma^2_x\vert a_x\sim \mbox{Inverse-$\chi^2$}\left(1,1/a_x\right),\quad a_x\sim\mbox{Inverse-$\chi^2$}\left(1,1/A^2_x\right).
	\end{array}
	\label{eq:linInvProbModel_SNHS}
\end{equation}
The first line of this model gives rise to a Skew Normal likelihood and is equivalent to
\begin{equation*}
		y_i\vert \bx,\sigma^2_{\varepsilon},\lambda\simind \mbox{Skew-Normal}\left(\left(\bK\bx\right)_i,\sigma^2_{\varepsilon},\lambda\right).	
\end{equation*}
Here the density function of a random variable $z$ having a $\mbox{Skew-Normal}\left(\mu,\sigma^2,\lambda\right)$ distribution is $p(z)=(2/\sigma)\phi\{(z-\mu)/\sigma\}\}\Phi\{\lambda(z-\mu)/\sigma\}$, with scale parameter $\sigma>0$, skewness parameter $\lambda$, and $\phi$ and $\Phi$ denoting the Standard Normal density and cumulative distribution functions.
The second line of \eqref{eq:linInvProbModel_SNHS} specifies a Horseshoe penalization on the $(\xDelta)_j$'s, as indicated by Table \ref{tab:bDistrib}. The priors at line three give rise to conjugate message passing updates for the stochastic nodes of $\sigma_\varepsilon$ and $\lambda$. As in model \eqref{eq:linInvProbModel}, $\sigma_x$ is assigned a Half-Cauchy$\left(A_x\right)$ prior through the auxiliary variable $a_x$.

\begin{figure}[!ht]
	\centering
	{\includegraphics[width=.6\textwidth]{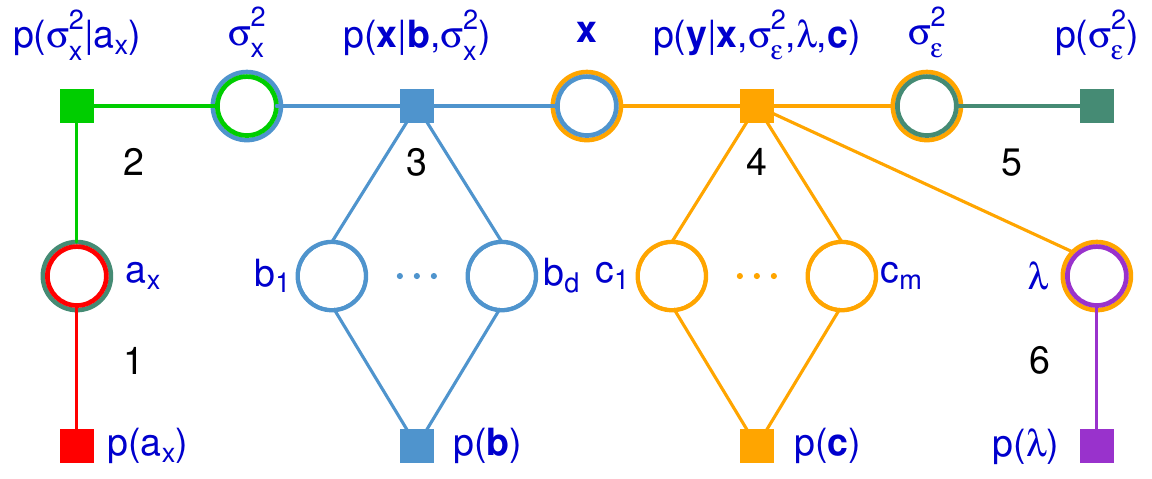}}
	\caption{\it Factor graph representation of the Skew Normal response model with Horseshoe penalization in \eqref{eq:linInvProbModel_SNHS}, where the square nodes correspond to the density functions, or factors, on the right-hand side of \eqref{eq:modelFactoriz_SNHS}. The circular nodes correspond to stochastic nodes of the $q$-density factorization in \eqref{eq:mfRest_SNHS}. Numbers are used to show the distinction between fragments, whereas colors identify different fragments types.}
	\label{fig:InvProbFacGraph_SNHS} 
\end{figure}

The joint density function of model \eqref{eq:linInvProbModel_SNHS} is given by 
\begin{align}
	\begin{split}
	p\left(\by,\bx,\bc,\bb,\sigma^2_\varepsilon,\lambda,\sigma_x^2,a_x\right)&=p\left(\by\vert\bx,\bc,\sigma^2_\varepsilon,\lambda\right)p\left(\bx\vert\bb,\sigma^2_x\right)p\left(\bc\right)p\left(\bb\right)\\
	&\quad\times  p\left(\sigma^2_\varepsilon\right)p\left(\lambda\right)p\left(\sigma^2_x\vert a_x\right)p\left(a_x\right).
	\end{split}
	\label{eq:modelFactoriz_SNHS}
\end{align}
Steps similar to those presented for the base model can be used to implement VMP for the model with Skew Normal responses and Horseshoe prior. The starting point is again the choice of a factorization for the approximating $q$-densities.  We impose the following mean field approximation to the joint posterior:
\begin{equation}
	p\left(\bx,\bc,\bb,\sigma^2_\varepsilon,\lambda,\sigma_x^2,a_x\vert\by\right)\approx q\left(\bx\right)q\left(\sigma^2_\varepsilon\right)q\left(\lambda\right)q\left(\sigma^2_x\right)q\left(a_x\right)\prod_{i=1}^m q_i\left(c_i\right)\prod_{j=1}^d q_j\left(b_j\right).
	\label{eq:mfRest_SNHS}
\end{equation}
The right-hand sides of \eqref{eq:modelFactoriz_SNHS} gives rise to the factor graph representation displayed as
Figure \ref{fig:InvProbFacGraph_SNHS}.
In this factor graph, four of the seven fragments arising from the base model \eqref{eq:linInvProbModel}, those numbered 1--3, are preserved. Fragment 4 is now the \textit{Skew Normal likelihood fragment} studied in Section 3.4 of \cite{mclean2019variational}. The message passed from this fragment to $\sigma^2_\varepsilon$ takes the form
\begin{equation}
	m_{p(\by\vert\bx,\sigma^2_{\varepsilon},\lambda,\bc)\rightarrow\sigma_\varepsilon^2}(\sigma_\varepsilon^2)=\exp\left\{\left[\begin{array}{c}
		\log(\sigma_{\varepsilon}^{2})\\[1ex]
		1/\sigma_{\varepsilon}\\[1ex]
		1/\sigma_{\varepsilon}^{2}
	\end{array}\right]^{T}\biggerbdeta_{p(\by\vert\bx,\sigma^2_{\varepsilon},\lambda,\bc)\rightarrow\sigma_\varepsilon^2}\right\}
\label{eq:messTOsigsqeps}
\end{equation}
and is proportional to an \textit{Inverse Square Root Nadarajah} density function, whereas the message to $\lambda$ is
\begin{equation}
	m_{p(\by\vert\bx,\sigma^2_{\varepsilon},\lambda,\bc)\rightarrow\lambda}(\lambda)=\exp\left\{\left[\begin{array}{c}
		\log(1+\lambda^{2})\\[1ex]
		\lambda^{2}\\[1ex]
		\lambda\sqrt{1+\lambda^{2}}
	\end{array}\right]^{T}\biggerbdeta_{p(\by\vert\bx,\sigma^2_{\varepsilon},\lambda,\bc)\rightarrow\lambda}\right\}
\label{eq:messTOlambda}
\end{equation}
and is part of the \textit{Sea Sponge} family. These two families of distributions are defined in Sections S.2.3 and S.2.5 of the supplementary material of \cite{mclean2019variational}. Priors on $\sigma_\varepsilon^2$ and $\lambda$ that are conjugate to these messages must have the form
\begin{align}
	&m_{p(\sigma_\varepsilon^2)\rightarrow\sigma_\varepsilon^2}(\sigma_\varepsilon^2)=\exp\left\{\left[\begin{array}{c}
		\log(\sigma_{\varepsilon}^{2})\\[1ex]
		1/\sigma_{\varepsilon}\\[1ex]
		1/\sigma_{\varepsilon}^{2}
	\end{array}\right]^{T}\biggerbdeta_{p(\sigma_\varepsilon^2)\rightarrow\sigma_\varepsilon^2}\right\}\label{eq:mpsigsqeps}\\
	\mbox{and}\,\,&	m_{p(\lambda)\rightarrow\lambda}(\lambda)=\exp\left\{\left[\begin{array}{c}
	\log(1+\lambda^{2})\\[1ex]
	\lambda^{2}\\[1ex]
	\lambda\sqrt{1+\lambda^{2}}
\end{array}\right]^{T}\biggerbdeta_{p(\lambda)\rightarrow\lambda}\right\},\label{eq:mplambda}
\end{align}
for some $3\times 1$ vectors $\biggerbdeta_{p(\sigma_\varepsilon^2)\rightarrow\sigma_\varepsilon^2}$ and $\biggerbdeta_{p(\lambda)\rightarrow\lambda}$. Priors $\sigma^2_\varepsilon\sim\mbox{Inverse-}\chi^2\left(A_\varepsilon,B_\varepsilon\right)$ and $\lambda\sim N\left(0,S^2_\lambda\right)$ of model \eqref{eq:linInvProbModel_SNHS} are respectively conjugate to \eqref{eq:messTOsigsqeps} and \eqref{eq:messTOlambda}, since for this choice of priors the messages $m_{p(\sigma_\varepsilon^2)}(\sigma_\varepsilon^2)$ and $m_{p(\lambda)\rightarrow\lambda}(\lambda)$ can be written as \eqref{eq:mpsigsqeps} and \eqref{eq:mplambda} with
\begin{equation*}
	\biggerbdeta_{p(\sigma_\varepsilon^2)\rightarrow\sigma_\varepsilon^2}=[-(A_\varepsilon/2)-1\,\,\,\,0\,\,\,-B_\varepsilon/2]^T\,\,\mbox{and}\,\,\biggerbdeta_{p(\lambda)\rightarrow\lambda}=[0\,\,\,-1/(2S^2_\lambda)\,\,\,\,0]^T.
\end{equation*}
We impose diffuse priors using $A_\varepsilon=B_\varepsilon=10^{-2}$ and $S_\lambda=10^5$.

\begin{figure}
	\centering
	\begin{subfigure}
		\centering
		\raisebox{-0.5\height}{\includegraphics[width=.525\linewidth]{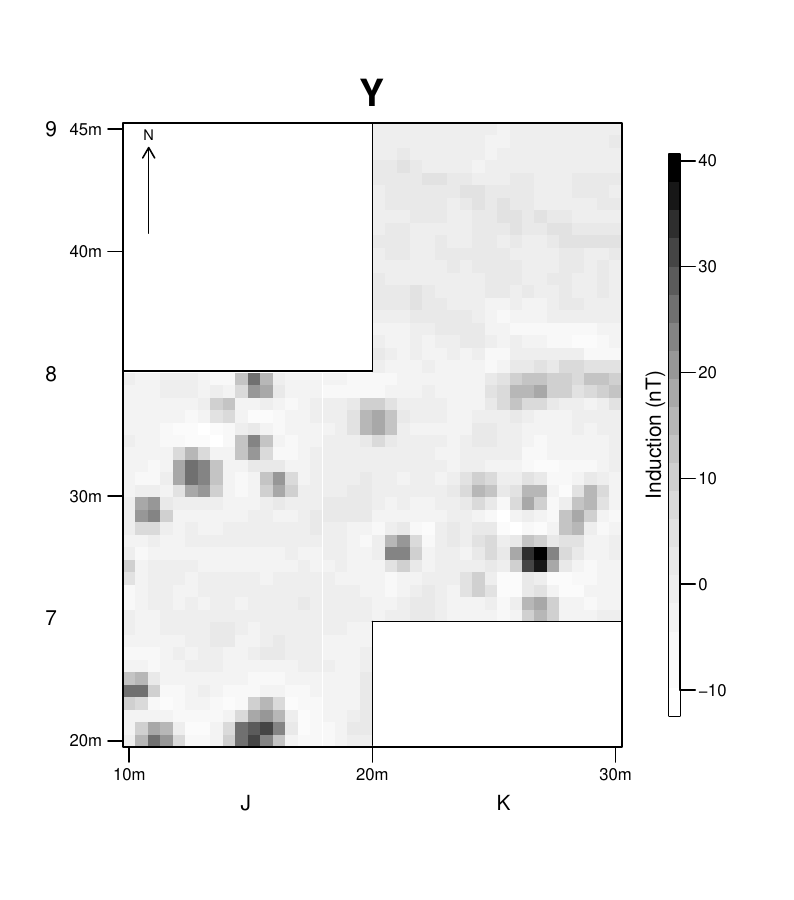}}
	\end{subfigure}
	\hspace{-1.2cm}
	\begin{subfigure}
		\centering
		\raisebox{-0.5\height}{\includegraphics[width=.525\linewidth]{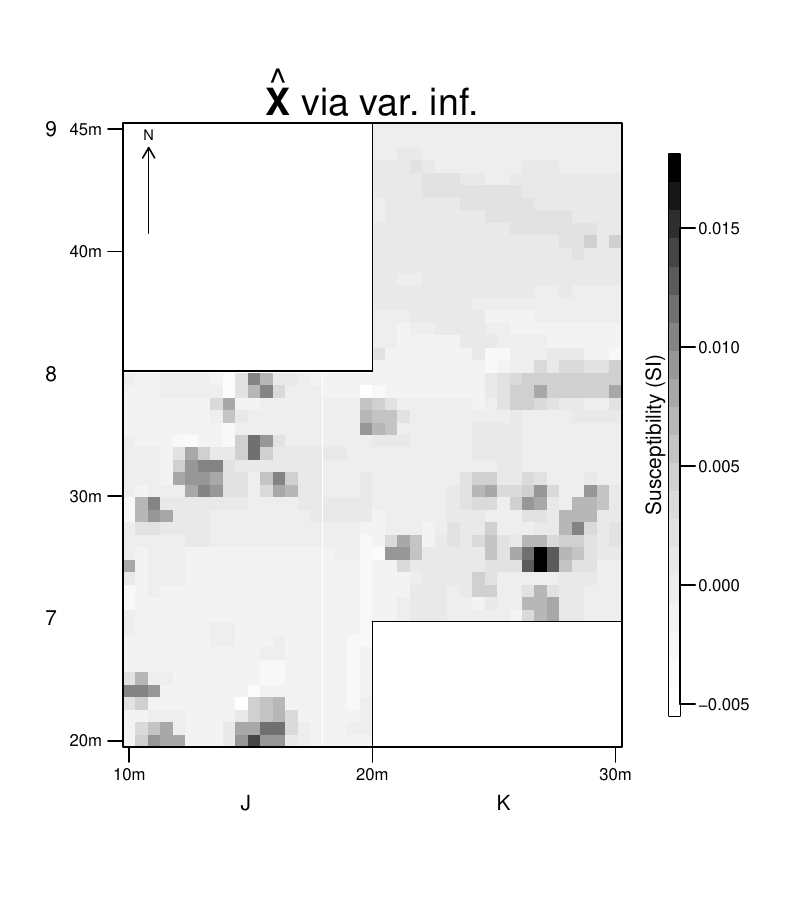}}
	\end{subfigure}\\\vspace{-0.9cm}
	\hspace{0.2cm}
	\begin{subfigure}
		\centering
		\raisebox{-0.49\height}{\includegraphics[width=.395\linewidth]{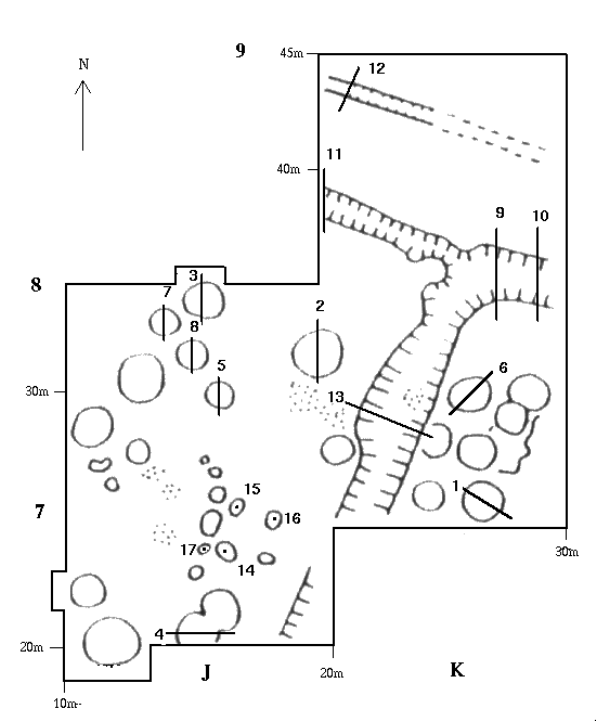}}
	\end{subfigure}
	\hspace{.8cm}
	\begin{subfigure}
		\centering
		\raisebox{-0.51\height}{\includegraphics[width=.47\linewidth]{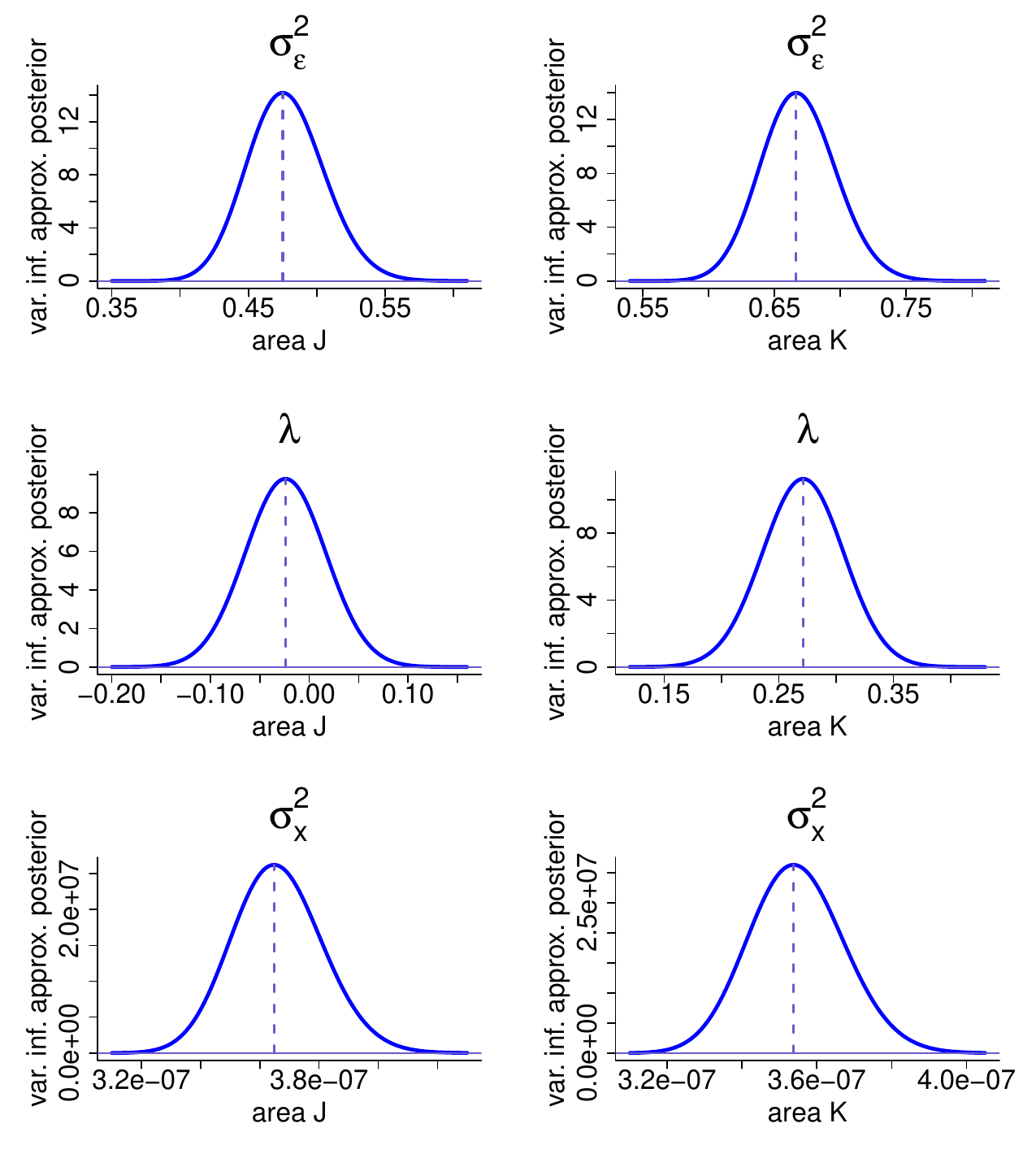}}
	\end{subfigure}
	\caption{\it Results of the archaeological data study conducted via model \eqref{eq:linInvProbModel_SNHS} and variational inference. The top-left image displays the archaeological dataset denoted by $\bY$. Its variational inference reconstruction, denoted by $\widehat{\bX}$, is shown in the top-right image, whereas the bottom-left image corresponds to the archaeologist’s impression of the 1994 excavation of `The Park'. The plots in the bottom-right side show the variational approximate marginal posterior densities of the Skew Normal variance and skewness parameters for areas J and K of the archaeological site.}
	\label{fig:RealArchaeoData}
\end{figure}

Full implementation of VMP  is based on iteratively updating, for each factor graph fragment depicted in Figure \ref{fig:InvProbFacGraph_SNHS}: (i) the parameter vectors of messages passed from the fragment’s neighboring stochastic nodes to the fragment’s factors; (ii) the parameter vectors of the messages passed
from the fragment’s factors to their neighboring stochastic nodes. The first step is very simple and entails application of (7) of \cite{wand2017fast}. The factor to stochastic node updates of the second step can be performed through various VMP procedures:
\begin{itemize}
	\item Fragment 1 is an Inverse G-Wishart prior fragment and the factor to stochastic node parameter vector updates can be performed using Algorithm 1 of \cite{maestrini2021inverse} with inputs $G_{\Theta}=G_{\tiny\mbox{full}}$, $\xi_{\Theta}=1$ and $\bLambda_{\Theta}=(A_x^2)^{-1}$.
	\item Fragment 2 is an iterated Inverse G-Wishart prior fragment and the factor to stochastic node parameter vector updates can be performed using Algorithm 2 of \cite{maestrini2021inverse} with inputs $G=G_{\tiny\mbox{full}}$, $\xi=1$ and $G_{\bA\rightarrow p(\bSigma\vert \bA)}=G_{\tiny\mbox{diag}}$.
	\item The factor to stochastic node parameter vector updates of fragment 3 can be performed through Algorithm \ref{alg:VMPalgo}.
	\item Fragment 4 is the Skew Normal likelihood fragment and the factor to stochastic node parameter vector updates can be performed using Algorithm 4 of \cite{mclean2019variational} with $\by$ and $\bK$ as data inputs. 
	\item Fragment 5 corresponds to the imposition of an Inverse Square Root Nadarajah prior distribution on the variance parameter  $\sigma^2_{\varepsilon}$. The output of VMP applied to this fragment is the natural parameter vector of the prior density function, that is, $\biggerbdeta_{p(\sigma_\varepsilon^2)\rightarrow\sigma_\varepsilon^2}$ from \eqref{eq:mpsigsqeps}. 
	\item Fragment 6 corresponds to the imposition of a Sea Sponge prior distribution on the skewness parameter $\lambda$. The output of VMP applied to this fragment is the natural parameter vector of the prior density function, that is, $\biggerbdeta_{p(\lambda)\rightarrow\lambda}$ from \eqref{eq:mplambda}.
\end{itemize}
We restrict our attention to the portion of the archaeological dataset corresponding to the excavated area, which enables a qualitative performance assessment of our fitting method. Figure \ref{fig:RealArchaeoData} displays the data under examination ($\bY$) together with the results of the application of VMP to model \eqref{eq:linInvProbModel_SNHS} and the impression drawn by archaeologists.
The data were handled in a disjoint way by separating the two rectangular areas corresponding to indices J and K of the archaeologists' impression. It is standard practice to examine grids as soon as they are collected and this division of the surveyed area allows to partition the data into two full matrices $\bY_J$ for area J and $\bY_K$ for area K of size $30\times20$ and $40\times20$, respectively. Model \eqref{eq:linInvProbModel_SNHS} can be used setting $\by=\vecof(\bY_J)$ or $\by=\vecof(\bY_K)$ and the reconstruction $\widehat{\bX}$ is simply obtained as the inverse vectorization of the VMP estimate of $\bx$. We employ the mean of the optimal approximating density $q^*(\bx)$, which is a $N\big(\muqx,\Sigmaqx\big)$ density function, to estimate $\bx$. Expressions for $q^*(\bx)$ and the other $q$-densities of interest are provided in the supplement.

If compared to the original dataset, the VMP reconstruction of Figure \ref{fig:RealArchaeoData} shows greater contrast between background and features, and some weak features are more evident in the posterior mean reconstruction. Despite the data being treated in a disjoint way, discontinuity in the estimate of $\bX$ between areas J and K is not very apparent. Careful inspection of the reconstruction shows the locations of some reconstructed features are shifted if compared to their apparent position in the original survey data image. This is important information, considering that each pixel corresponds to a square area of 0.5m side and that archaeological excavations require intensive manual work. The approximate posterior densities of $\lambda$ seem to indicate that the data from area J is symmetric, whereas that from area K is positively skewed.

\subsection{Approximating density functions}

From (10) of \cite{wand2017fast}, the densities of main interest, $q^*(\bx)$, $q^*(\sigma^2_\varepsilon)$, $q^*(\lambda)$ and $q^*(\sigma^2_x)$, have the following forms:
\begin{equation*}
	\begin{array}{c}
		q^*(\bx)\propto\exp\left\{\left[\begin{array}{c}
			\bx\\[1ex]
			\vecof\big(\bx\bx^T\big)
		\end{array}\right]^T\etaSUBqx\right\},\,\mbox{with}\,\,\etaSUBqx\equiv\etaSUBpxbsigsqxTOx + \biggerbdeta_{p(\by\vert\bx,\sigma^2_{\varepsilon},\lambda,\bc)\rightarrow\bx},
	\end{array}
\end{equation*}
\begin{equation*}
	\begin{array}{c}
		q^*(\sigma^2_\varepsilon)\propto\exp\left\{\left[\begin{array}{c}
			\log(\sigma^2_{\varepsilon})\\[1.2ex]
			1/\sigma_{\varepsilon}\\[1.2ex]
			1/\sigma^2_{\varepsilon}
		\end{array}\right]^T\etaSUBqsigsqepsilon\right\},\,\mbox{with}\,\,\etaSUBqsigsqepsilon\equiv\biggerbdeta_{p(\sigma^2_{\varepsilon})} + \biggerbdeta_{p(\by\vert\bx,\sigma^2_{\varepsilon},\lambda,\bc)\rightarrow\sigma^2_\varepsilon},
	\end{array}
\end{equation*}
\begin{equation*}
	\begin{array}{c}
		q^*(\lambda)\propto\exp\left\{\left[\begin{array}{c}
			\log(1+\lambda^{2})\\[1ex]
			\lambda^{2}\\[1ex]
			\lambda\sqrt{1+\lambda^{2}}
		\end{array}\right]^T\biggerbdeta_{q(\lambda)}\right\},\,\mbox{with}\,\,\biggerbdeta_{q(\lambda)}\equiv\biggerbdeta_{p(\lambda)} + \biggerbdeta_{p(\by\vert\bx,\sigma^2_{\varepsilon},\lambda,\bc)\rightarrow\lambda}
	\end{array}
\end{equation*}
\begin{equation*}
	\begin{array}{c}
		\mbox{and}\,\,q^*(\sigma^2_x)\propto\exp\left\{\left[\begin{array}{c}
			\log(\sigma^2_{x})\\[1.2ex]
			1/\sigma^2_{x}
		\end{array}\right]^T\etaSUBqsigsqx\right\},\,\mbox{with}\,\,\etaSUBqsigsqx\equiv\etaSUBpsigsqxaxTOsigsqx + \etaSUBpxbsigsqxTOsigsqx.
	\end{array}
\end{equation*}
Expressions for the other optimal approximating densities can be obtained in a similar manner. The full list of optimal approximating density functions respecting restriction \eqref{eq:mfRest_SNHS} is the following:
\begin{align*}
	&q^*(\bx)\,\text{is a}\, N\big(\muqx,\Sigmaqx\big)\,\text{density function},\\
	&q^*(c_i)\propto\exp\left\{\left[\begin{array}{c}
		\vert c_i\vert\\[1ex]
		c_i^2\\
	\end{array}\right]^{T}\biggerbdeta_{c_i}\right\},\,\text{for a}\,2\times1\,\text{natural paramter vector}\,\biggerbdeta_{c_i}\,\text{and}\, i=1,\ldots,m,\\
	&q^*(b_j)\,\text{is an}\, \text{Inverse-Gaussian}\big(\mu_{q(b_j)},\lambda_{q(b_j)}\big)\,\text{density function, for}\, j=1,\ldots,d,\\
	&q^*(\sigma^2_\varepsilon)\,\mbox{is an}\, \mbox{Inverse-Square-Root-Nadarajah}\left(\alpha_{q(\sigma^2_\varepsilon)},\beta_{q(\sigma^2_\varepsilon)},\gamma_{q(\sigma^2_\varepsilon)}\right)\,\text{density function},\\
	&q^*(\lambda)\,\mbox{is a}\, \mbox{Sea-Sponge}\left(\alpha_{q(\lambda)},\beta_{q(\lambda)},\gamma_{q(\lambda)}\right)\,\text{density function},\\
	&q^*(\sigma^2_x)\,\mbox{is an}\,\mbox{Inverse-}\chi^2\big(\kappaqsigsqx,\lambdaqsigsqx\big)\,\text{density function}\\
	\text{and }&q^*(a_x)\,\mbox{is an}\,\mbox{Inverse-}\chi^2\big(\kappaqax,\lambdaqax\big)\,\text{density function}.
\end{align*}
The common parameters of the $q$-densities of interest are:
\begin{equation*}
	\begin{array}{c}
		\muqx=\Sigmaqx\left(\etaSUBqx\right)_{1:m},\,\,\,\Sigmaqx=-\frac{1}{2}\vecof^{-1}\left\{\left(\etaSUBqx\right)_{(m+1):m^2}\right\},\\[2.5ex]
		\alpha_{q(\sigma^2_{\varepsilon})}=-2\left(1+\left(\etaSUBqsigsqepsilon\right)_1\right),\,\,\, \beta_{q(\sigma^2_{\varepsilon})}=-\left(\etaSUBqsigsqepsilon\right)_3,\,\,\,\gamma_{q(\sigma^2_{\varepsilon})}=-\left(\etaSUBqsigsqepsilon\right)_2,\\[2.5ex]
				\alpha_{q(\lambda)}=\left(\etaSUBqlambda\right)_1,\,\,\, \beta_{q(\lambda)}=-\left(\etaSUBqlambda\right)_2,\,\,\,\gamma_{q(\lambda)}=\left(\etaSUBqlambda\right)_3,\\[2.5ex]
		\kappaqsigsqx=-2\left(1+\left(\etaSUBqsigsqx\right)_1\right),\,\,\, \lambdaqsigsqx=-2\left(\etaSUBqsigsqx\right)_2.
	\end{array}
\end{equation*}
To obtain the common parameters of $q(\sigma^2_{\varepsilon})$ and $q(\sigma^2_{\lambda})$ from their natural parameters we make use of the results from Sections S.2.3 and S.2.5 of the supplementary material of \cite{mclean2019variational} concerning the Inverse Square Root Nadarajah and Sea Sponge distributions.

\end{document}